\documentclass[preprint2,twocolappendix]{aastex631}
\usepackage{xfrac}
\usepackage{amsmath,amssymb}
\usepackage{upgreek}
\usepackage{isomath}

\newcommand\Rm{{\rm Rm} }
\newcommand\Rey{{\rm Re} }
\newcommand\Pm{{\rm Pm} }

\newcommand\SNr{\dot\sigma_{\rm Sn}}
\newcommand\OSN{\Omega_{\rm Sn}}
\newcommand\ESK{E_{\rm m}}
\newcommand\EST{E_{\rm th}}

\newcommand\Ms{{\cal{M}} }
\newcommand{\vect}[1]{{{\mbox{\boldmath $#1$}}}}
\newcommand{\mathbfss}[1]{\textbf{\textsf{#1}}}
\newcommand\K{~ {\rm K}}
\newcommand\cmcube{~ {\rm cm^{-3}}}
\newcommand\cs{ c_{\rm s}}
\newcommand\cplocal{ c_{\rm p}}

\newcommand\cv{ c_{ v}}

\newcommand\kpc{~ {\rm kpc}}
\newcommand\pc{~ {\rm pc}}
\newcommand\dx{ {\delta x}}
\newcommand\dxx{ {\delta \vect{x}}}
\newcommand\Myr{~ {\rm Myr}}
\newcommand\Gyr{~ {\rm Gyr}}
\newcommand\erg{~ {\rm erg}}
\newcommand\kms{~ {\rm km~s}^{-1}}
\newcommand\eB{{\langle e_{ B}\rangle}}
\newcommand\eK{{\overline{e_{ K}}}}
\newcommand\eKt{{\overline{e_{ K}^\prime}}}

\newcommand\mB{{\langle\vect{B}\rangle}}
\newcommand\mU{{\langle\vect{u}\rangle}}
\newcommand\Div{{\cal{C}}}
\newcommand\EMF{{\vect{\cal{E}}}}

\newcommand\LA{{L2}}

\newcommand\LB{{L2-cl}}
\newcommand\LC{{L0-B0}}
\newcommand\LD{{L2-B0}}
\newcommand\HA{{H2}}

\definecolor{midblue}{rgb}{0.0,0.4,0.7}
\definecolor{midgreen}{rgb}{0.1,0.6,0.3}
\definecolor{mypurple}{rgb}{0.7,0.3,0.8}

\received{June 10, 2023}
\revised{\today}
\accepted{}
\submitjournal{ApJ}

\shorttitle{Small-scale vs large-scale dynamo}
\shortauthors{Gent et al.}

\begin{document}

\title{Transition from small-scale to large-scale dynamo in a supernova-driven, multiphase medium}

\correspondingauthor{Maarit J. Korpi-Lagg}
\email{Email: frederick.gent@aalto.fi, mordecai@amnh.org,\\ maarit.korpi-lagg@aalto.fi}

\author[0000-0002-1331-2260]{Frederick A. Gent}
\affiliation{
Astroinformatics, Department of Computer Science, Aalto University, P.O. Box 15400, FI-00076 Espoo, Finland
 }
\affiliation{
    Nordic Institute for Theoretical Physics,
      SE-106 91 Stockholm, Sweden
}
\affiliation{
    School of Mathematics, Statistics and Physics,
      Newcastle University, NE1 7RU, UK
 }

\author[0000-0003-0064-4060]{Mordecai-Mark {Mac Low}}
\affiliation{
  Department of Astrophysics, American Museum of Natural History,
  New York, NY 10024, USA
}

\author[0000-0002-9614-2200]{Maarit J. {Korpi-Lagg}}
\affiliation{
Astroinformatics, Department of Computer Science, Aalto University, P.O. Box 15400, FI-00076 Espoo, Finland
}
\affiliation{
Max Planck Institute for Solar System Research,
D-37707 G\"ottingen, Germany
}
\affiliation{
    Nordic Institute for Theoretical Physics,
      SE-106 91 Stockholm, Sweden
}


\begin{abstract}
Magnetic fields are now widely recognised as critical at many scales to
galactic dynamics and structure, including multiphase pressure balance, dust
processing, and star formation. Using imposed magnetic fields cannot reliably
model the interstellar medium’s (ISM) dynamical structure nor phase
interactions. Dynamos must be modelled. ISM models exist of turbulent magnetic
fields using small-scale dynamo (SSD). Others model the large-scale dynamo
(LSD) organising magnetic fields at scale of the disc or spiral arms.
Separately, neither can fully describe the galactic magnetic field dynamics
nor topology. We model the LSD and SSD together at sufficient resolution to
use the low explicit Lagrangian resistivity required. The galactic
SSD saturates within 20 Myr. We show that the SSD is quite insensitive to the
presence of an LSD and is even stronger in the presence of a large-scale shear
flow. The LSD grows more slowly in the presence of SSD, saturating after 5 Gyr
vs. 1–2 Gyr in studies where the SSD is weak or absent. The LSD primarily
grows in warm gas in the galactic midplane. Saturation of the LSD occurs due
to $\alpha$-quenching near the midplane as the growing mean field produces a
magnetic $\alpha$ that opposes the kinetic $\alpha$.  The magnetic energy in
our models of the LSD shows slightly sublinear response to increasing
resolution, indicating that we are converging towards the physical solution at
1 pc resolution. Clustering supernovae in OB associations increases the growth
rates for both the SSD and the LSD, compared to a horizontally uniform
supernova distribution.
\end{abstract}

\section{Introduction}\label{sec:intro}

Magnetic fields have a major impact on our observation and interpretation of
astrophysical entities, their structure and dynamics. The intensity and
polarization of emission that we measure to observe the universe are affected
by the location, orientation, and strength of the magnetic field both in the
Solar neighbourhood \citep{TEEMP23, HHFRE23} and the more distant universe
where these emissions originate or propagate \citep{BSSW03}.  \citet{Dickey22}
compare observations of Faraday rotation measures of diffuse Galactic
synchrotron emission against extragalactic sources. They show that
measurements are highly sensitive to the local structure of the magnetic field
by comparing correlations between the surveys across the sky \citep[see
also,][]{Girichidis21}. Measurement of this magnetic field relies on
inferences, based on existing models.  Simulations of the magnetic field can
be used to test such models and inferences \citep[e.g.,][]{Betti19},
particulary if dynamo models can generate  realistic turbulent and large-scale
structures.

Magnetic fields influence the dynamics, morphology, and evolution of galaxies.
Presence of a turbulent magnetic field improves by 30--50\% \citep{KMG22} the
dust survival rates against supernova (SN) blast waves, a critical factor in
the recycling of the interstellar medium (ISM) and subsequent star formation
processes. The large-scale magnetic field also affects the gas scale height in
galactic discs \citep{Gressel08b, Hill:2012a, Gent:2012} and the multiphase
structure of the ISM \citep{Hill:2012a, EGSFB16, EGSFB19}. Observations of
external spiral galaxies {often} show magnetic fields aligned along the
disc and parallel to the spiral arms{, or otherwise ordered over kpc
scales \citep[e.g.,][their Table 4]{BCEB19}}. The total field is often
stronger in the spiral arms \citep{FBSBH11, AKKWBD13, Beck15}, although
occasionally in the interarm regions \citep{BH96, Harnett04}.

However, some inconsistencies between measurements tracing differing
properties of the magnetic field raise questions about its strength and
structure as a function of galactic radius \citep{Beck15a}, in different
thermal phases, or specific environments such as star-forming regions
\citep{Sofia21}. We seek to explore numerically the extent differences in the
action of dynamos at various scales and in different thermal phases account
for the observations.

Simulations of the ISM in disc galaxies at kiloparsec scales, which include
magnetic fields by imposing a background \citep{deAvillez:2005, DP08, KO15B},
or an initially strong \citep{Hill:2012a, IH17} horizontal field, evolve SN
driven turbulence to compare the multiphase structure of the ISM with and
without magnetic fields.  However, to produce magnetic fields with realistic
structure and energetics, the large-scale dynamo (LSD) must be modelled
explicitly.

\begin{figure*}
\centering
\includegraphics[trim=2.2cm 0.0cm 0.0cm 0.0cm,clip=true,height=1.18\columnwidth]{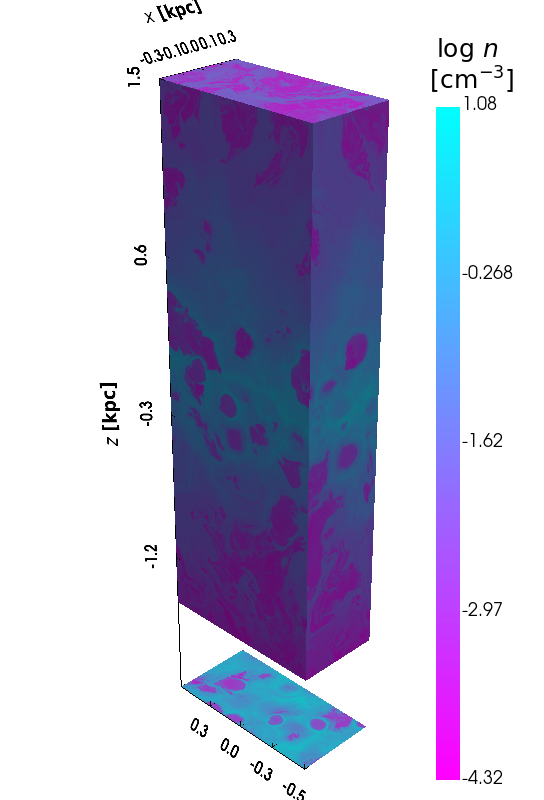}\hspace{-0.15cm}
\includegraphics[trim=2.2cm 0.0cm 0.0cm 0.0cm,clip=true,height=1.18\columnwidth]{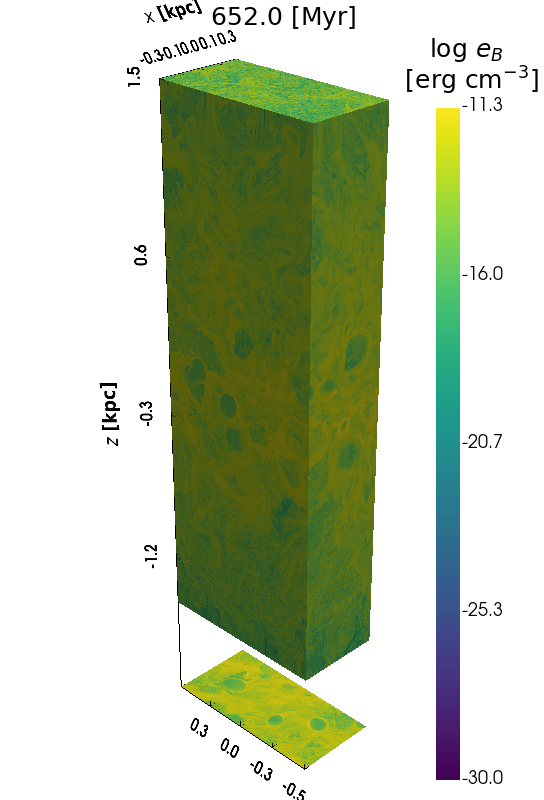}\hspace{-0.15cm}
\includegraphics[trim=2.2cm 0.0cm 0.0cm 0.0cm,clip=true,height=1.18\columnwidth]{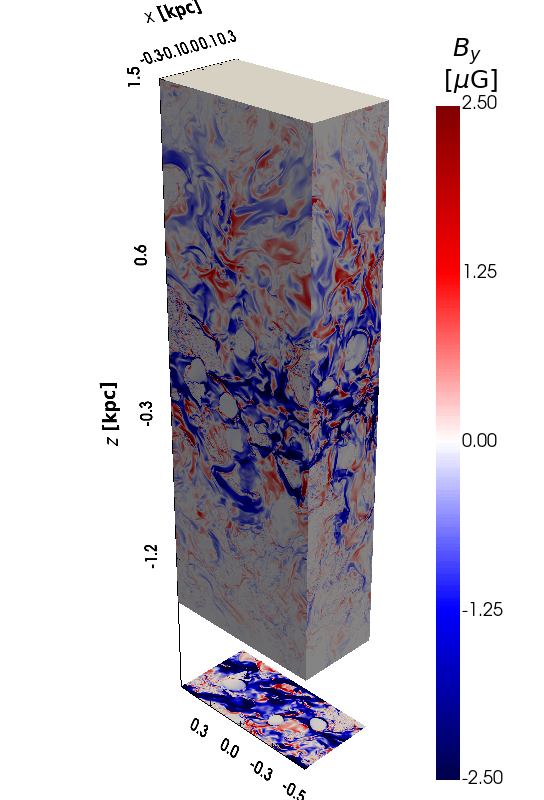}\hspace{-0.15cm}
\caption{
Structure of high resolution ($\delta x = 1$~pc) Model~\HA\, with properties
given in Table~\ref{tab:models}. Figure shows (left to right) gas number
density, magnetic energy density, and azimuthal magnetic field during the
nonlinear stage of the dynamo. The lower offset planes show horizontal midplane
slices. The online version of this article includes {a 52 second} animation of
this Figure, demonstrating the rapid transition to turbulence, the saturation
of the turbulent magnetic energy within about 30~Myr and the much slower
organisation into large scales of the azimuthal component of the magnetic field
up to 652~Myr. The breathing and sloshing modes of the disc are also apparent.
The vertical striping evident in the magnetic energy occurs from inflows
through the boundary which has a vertical magnetic field condition. A high
resolution version is included in the data repository \citep{GentAMNH}.
\label{fig:ism}
}
\end{figure*}

Some global models of galactic dynamos that span much larger scales
\citep[e.g.,][]{OLLS97, HWK09, Kotarba10, Pakmor13, BZKYA17} do not resolve
the turbulent scales of the local magnetic field so that the small-scale
dynamo (SSD) is absent or only present at kiloparsec scale
\citep[e.g.,][]{RT16, MDST18}.

Models in spiral galaxies with SN-driven turbulence on length scales of at
most a few kiloparsecs \citep[e.g.,][]{Korpi:1999b, Gressel:2008, Gent:2013a,
Bendre2018} show the LSD saturates on time scales of 1--2 Gyr near equipartion
with the ISM kinetic energy density, depending on the rates of galactic
rotation and SN explosions. However, the underlying turbulent structure of the
magnetic field produced by the SSD is either absent or under-resolved.

The parameters and properties of the SSD in SN-driven turbulence are studied
by \citet{BKMM04, BalKim05, GMKS21} and \citet{GMKS22}. \citet{GMKS21}
demonstrate that SSD models numerically converge at cell size $\lesssim 1$\,pc
and that the SSD is present if the microphysical {magnetic diffusivity}
$\eta \lesssim 5 \cdot 10^{-3}$\,kpc\,km\,s$^{-1}$.

In this work, we model the combination of LSD and SSD in a stratified galactic
disc containing SN-driven turbulence, including high enough grid resolution to
resolve the SSD. We compare to solutions with the LSD suppressed to identify
the effect the LSD has on the SSD; we consider the effect of clustering of SN
in OB associations; and we examine the effect of resolution on the solution of
the LSD. In Section~\ref{sec:models} we explain the numerical setup, model
parameters and notation conventions. In Section~\ref{sec:results} we describe
how the saturated SSD is affected by the LSD{, some insights into the
growth rates and drivers of the LSD,} and the effect of SN clustering{. In
Section~\ref{sec:robust} we consider} the effect of resolution {and the
robustness of the numerical setup}.  In {Section~\ref{sec:disc}} we
discuss how the results relate to previous understanding of the galactic
magnetic field {and in Section~\ref{sec:conc} we summarise our main
conclusions}. We provide supplementary illustrative material in
Appendix~\ref{sec:supp}{, including a reference table of notation employed
in this manuscript,} and outline in Appendix~\ref{sec:code} what versions of
the Pencil Code were used during the months and years over which the models
were evolved.  Data from the models presented in this paper are available in
{a repository at} the American Museum of Natural History Digital Library \citep{GentAMNH}.

\section{Model description}\label{sec:models}

\begin{table}
\begin{center}
\caption{
Models included in analysis of LSD \& SSD interactions. {Model names
include H (L) for  high (low) resolution, then the rotation velocity in Solar
neighborhood units. OB clustering is labelled ``cl''. ``B0'' denotes zero mean
magnetic field.}
\label{tab:models}
}
\begin{tabular}{lcccc}
\hline\hline
Model&$\dxx$&$\Omega$&Dynamo          &$\nu_6,\chi_6,\eta_6$  \\
     &[pc]      &[$\OSN$]&                &[kpc$^5$ km s$^{-1}$]  \\\hline
\HA  &1.0       &  2     &LSD+SSD         &1~(-15)                \\
\LA  &4.0       &  2     &LSD+SSD         &5~(-12)                \\
\LB  &4.0       &  2     &LSD+SSD         &5~(-12)                \\
\LC  &4.0       &  0     &SSD             &5~(-12)                \\
\LD  &4.0       &  2     &SSD             &5~(-12)                \\\hline
\end{tabular}
\end{center}
\end{table}

\subsection{Model parameters}\label{sec:pars}

We use the shearing sheet approximation to model a cylindrical wedge segment
of a galactic disc using Cartesian coordinates $(x, y, z)$ aligned with
galactocentric radius, azimuth, and the normal to the plane of the disc. We
use dimensions of ${\vect{L}}=$(0.512, 1.024, 3.072) kpc, centered
vertically on the midplane. Grid cells in our models are cubical with size
$\delta x=4\pc$, aside from one high-resolution model with $\delta x=1\pc$.
The SSD has been shown to numerically converge at these resolutions
\citep{GMKS21}.  Model \LA\ has been initialized with a random field of a few
nanogauss, while all other models start with a random seed field of a few
picogauss.  This difference evidently has no effect on the growth rates, once
an eigenmode of the SSD has been established.  The density and magnetic field
structure of the high-resolution model are illustrated in
Figure~\ref{fig:ism}.

The set of models included are listed in Table\,\ref{tab:models}.  Models \LA\
and \HA\ apply our standard set of parameters, differing only by having low or
high resolution.  Models otherwise differ from Model \LA\ as follows. SNe
occur in OB associations in Model \LB. To isolate the properties of the SSD
from the LSD in this stratified system, we numerically exclude the mean field
by subtracting at every time step the averages of each magnetic field
component in each horizontal plane for Model \LC\ without rotation and Model
\LD\ with the same differential rotation as Model \LA.

The radial domain size is sufficient to contain the typical scale of
superbubbles that evolve \citep{TB88,MM88,NI89,FMZ91} from multiple SNe, while
the azimuthal domain size of 1 kpc supports multiple independent eddies of the
largest turbulent structures.  The vertical scale is sufficient to support two
scale heights of the warm gas, above and below the plane. Open vertical
boundaries allow flows in and out of the domain, but our grid has insufficient
vertical extent to fully model a galactic fountain.  Hence, to sustain the
simulation against gas losses due to vertical advection out of the domain, we
continuously maintain the total gas mass by compensating total gains or losses
each time step with a density adjustment in every cell proportional to the
local gas density.  The natural rate of gas outflows yields mass loss of
around 10\% per gigayear, so the instantaneous adjustments are very small, and
relative gradients in momentum are conserved locally in space and time.

The initial condition for all models has a vertical distribution of gas
density and temperature in {nonadiabatic hydrostatic} equilibrium.  This
is solved numerically in a one-dimensional vertical column of the ISM.  The
total mass in the initial state is approximately equivalent to the combined
mass of the cold and warm neutral medium and the ionized medium, as summarised
in \citet{Ferriere01} Equation (2).

We choose the {magnetic diffusivity} $\eta = 8\cdot 10^{-4}$~kpc km s$^{-1}$
and the viscosity $\nu = 4\cdot 10^{-3}$~kpc km s$^{-1}$ to be low enough to
support a vigorous SSD, although numerical diffusivity remains dominant in our
low resolution models \citep[see][for details of how this was
demonstrated]{GMKS21}. Thus, the nominal magnetic Prandtl number in our models
is Pm = 5.  We omit explicit thermal conductivity, which acts at time and
length scales below our model resolution.

\subsection{Dynamical equations}\label{sec:eq}
We use the Pencil Code \citep{brandenburg2002,Pencil-JOSS} to model SN-driven
turbulence as described previously in \citet{Gent:2012} and \citet{GMKSH20}.

Variations of the system of non-ideal, compressible, non-adiabatic,
magnetohydrodynamics, Equations~\eqref{eq:mass}--\eqref{eq:ent}, have been used
to simulate SN-driven turbulence in \citet{Gent:2013b, Gent:2013a, KGVS18,
GMKS21} and \citet{GMKS22}.  \citet{GMKS21} and \citet{GMKS22} omit the
gravitational potential ${\Phi}$, the angular velocity $\vect{\Omega}$, or the
shear $S$.  The other three studies omit hyperdiffusion, have explicit
diffusivities proportional to sound speed, and include thermal diffusivity
$\chi$.  Hydrodynamic models of \citet{Gent:2013a} and \citet{KGVS18} exclude
Equation~\eqref{eq:ind}, Ohmic heating and Lorentz force.  \citet{Gent:2013b}
include shock resistivity for Equation~\eqref{eq:ind}.

In this study we solve
  \begin{eqnarray}
  \label{eq:mass}
    \frac{\text{D}\rho}{\text{D}t} &=&
    -\rho \nabla \cdot \vect{u}
    +\nabla \cdot\zeta_D\nabla\rho,\\[0.5cm]
  \label{eq:mom}
    \rho\frac{\text{D}\vect{u}}{\text{D}t} &=&
    {\ESK\dot\sigma}h^{-1}
    -\rho\cs^2\nabla\left({s}/{\cplocal}+\ln\rho\right)-\rho\nabla{\Phi}\nonumber\\
    &-&\rho Su_x\vect{\hat{y}}-2\rho\vect{\Omega}\times\vect{u}
    +\mu_0^{-1}\nabla\times\vect{B}\times\vect{B}
    \nonumber\\
    &+&\nabla\cdot \left(2\rho\nu{\mathbfss W}\right)
    +\rho\nabla\left(\zeta_{\nu}\nabla \cdot \vect{u} \right)
    \nonumber\\
    &+&\nabla\cdot \left(2\rho\nu_6{\mathbfss W}^{(5)}\right)
  -\vect u\vect{\nabla}\cdot\left(\zeta_D\vect{\nabla}\rho\right),\\[0.5cm]
  \label{eq:ind}
    \frac{\partial \vect{A}}{\partial t} &=&
    \vect{u}\times\vect{B}-S A_y\vect{\hat{y}}-Sx\frac{\partial \vect{A}}{\partial y}\nonumber \\
    &+&\eta\nabla^2\vect{A}
    +\eta_6\nabla^6\vect{A},\\[0.5cm]
  \label{eq:ent}
    \rho T\frac{\text{D} s}{\text{D}t} &=&
    \EST\dot\sigma h^{-1} +\rho\Gamma
    -\rho^2\Lambda +\eta\mu_0^{-1}|\nabla\times\vect{B}|^2
    \nonumber\\
    &+&2 \rho \nu\left|{\mathbfss W}\right|^{2}
    +\rho~\zeta_{\nu}\left(\nabla \cdot \vect{u} \right)^2
    \nonumber\\
    &+&\nabla\cdot\left(\zeta_\chi\rho T\nabla s\right)
    +\rho T\chi_6\nabla^6 s
    \nonumber\\
    &-& \cv~T \left(
    \zeta_D\nabla^2\rho + \nabla\zeta_D\cdot\nabla\rho\right).
  \end{eqnarray}
The ideal gas equation of state closes the system, assuming an adiabatic index
(ratio of specific heats) $\cplocal/\cv=5/3$.  Treating the ISM as a monatomic,
fully ionized plasma we apply a mean molecular weight of 0.531.  Most variables
take their usual meanings; a list of notations used is tabulated within
Appendix~\ref{sec:code}.

\subsection{Gravity and shear}\label{sec:gravity}

We include an external gravity field with stellar and dark halo contributions
following \citet{KG89}, such that the vertical gravitational acceleration
\begin{equation}
\label{KG89grav}
g{_z} = {-}\frac{\partial {\Phi}}{\partial z} =
-\frac{a_sz}{\sqrt{z_s^2+z^2}}-\frac{a_hz}{z_h},
\end{equation}
with $a_s=1.4\kms\Myr^{-1}$, $z_s=0.2\kpc$, $a_h=0.5\kms\Myr^{-1}$ and
$z_h=1\kpc$.  {In the local shearing periodic box approximation
\citep[e.g.,][]{GL65,brandenburg1995,yang09} the angular velocity $\Omega$ is
treated as constant.  Coriolis forces are included to first order, while
centrifugal forces are assumed to balance radial gravity, and thus neglected
\citep[see Eq.~{[76]} of][]{Balbus03}.}

We do not follow the self-gravity of the gas.  Taking the mean midplane warm
gas density of 0.1 cm$^{-3}$ and a mean sound speed of $12\kms$ in the warm
gas, {the Jeans length} $\lambda_{\rm J} \gg 1\kpc$ and more so for the
diffuse hot gas. As we do not allow the cold gas to cool below  $\sim100$\,K at
maximal number density of a few tens cm$^{-3}$, star-forming structures cannot
form.  In reality, star-forming molecular clouds indeed have a filling fraction
of only around 5\% \citep[e.\ g.][]{Ferriere01} and so can be neglected for the
purposes of modelling the dynamo.

In Equation~\eqref{eq:mom} $\vect{\Omega}=(0,0,\Omega)$ with angular momentum
$\Omega\propto\OSN$ and the Solar neighborhood value $\OSN=30\kms\kpc^{-1}$,
assuming an orbital speed for the Sun of $255\kms$ at galactocentric radius
$8.5\kpc$. The advective derivative,
\begin{equation}\label{eq:adv}
\frac{\text{D}~}{\text{D}t}=\frac{\partial~}{\partial t} + (\vect{U}+\vect{u})\cdot\nabla,
\end{equation}
includes transport by an imposed shear flow $\vect{U}=(0,Sx,0)$, in which the
rate of shear $S=-\Omega$ assuming a flat rotation curve, and a perturbation
velocity $\vect{u}$ that represents the local deviation from the overall
rotational velocity $\vect{U}$.

\subsection{Cooling and heating}\label{sec:cooling}

The uniform far ultraviolet heating follows the functional form given by
\citet{Wolfire:1995} with temperature dependence approximated as
\begin{equation}\label{eq:Gamma}
\Gamma_{\rm UV}(T) = \frac{\Gamma_0}{2}\left(1+\tanh\left[\frac{
     2\cdot 10^4\,{\rm K} - T}{2000\,{\rm K}}\right]\right),
\end{equation}
with $\Gamma_0 = 0.0147$\,erg\,g$^{-1}$\,s$^{-1}$.  However,  we apply an
enhanced heating rate $\Gamma = 3.5\,\Gamma_{\rm UV}$ in order to support the
thickness of the disc in the absence of ionization and cosmic ray energy
\citep{HMGI18}. The cooling, applies a piecewise power-law temperature
dependence $\Lambda (T) = \Lambda_k T^{\beta_k}$ within the range $T_k < T <
T_{k+1}$ based on \citet{Wolfire:1995} for cold and warm gas phases and
\citet{Sarazin:1987} for the hot phase, as parameterized previously \citep[see
e.g,][]{SVG02,Gressel:2008} and see \citet[][their Table 1]{Gent:2013a}, but
with $\Lambda=0$ at $T<90$~K.  This cooling law includes thermally unstable
branches at high and low temperatures as expected for the ISM.

\subsection{Artificial diffusivities}\label{sec:shocks}

Terms in the dynamical equations that contain $\zeta_D=2\Div,~\zeta_\nu=5\Div$
and $\zeta_\chi=2\Div$ resolve shock discontinuities with artificial diffusion
of mass, momentum, and thermal energy, respectively.  They depend on the local
convergence $\Div=-\nabla\cdot\vect{u}|_{-\rm ve}$, effectively quadratic in
Equation~\eqref{eq:mom}, as described in detail in  \citet{GMKSH20}.
Equations~\eqref{eq:mom} and \eqref{eq:ent} include momentum and energy
conserving corrections for the artificial mass diffusion $\zeta_D$  applied in
Equation~\eqref{eq:mass}.  Following \citet{GMKS21}, resistive shock diffusion
is omitted from Equation~\eqref{eq:ind}, where hyperdiffusion is sufficient to
resolve grid-scale instabilities without suppressing the important physical
process of field compression in SN remnant shells.

Terms containing $\nu_6,~\chi_6$ and $\eta_6$ apply sixth-order hyperdiffusion
to resolve grid-scale instabilities \citep[see, e.g.,][]{ABGS02,HB04}, with
mesh Reynolds number set to be $\simeq1$ at the scale of the zone size $\dx$.
Hyperdiffusivity coefficients are $\nu_6=\chi_6=\eta_6=$ $5\times10^{-12}$ and
$1\times10^{-15}$ for resolution at 4 pc and 1 pc, respectively.

Due to continual code development in response to the performance and evolution
of the simulations over a period exceeding two years in real time, some run
parameters and the version of the code being executed vary.  The versions and
key adjustments to the Pencil Code applied to each model at the times indicated
are tabulated in Appendix~\ref{sec:code}.

\subsection{Supernova explosions}\label{sec:SN}
For each SN explosion thermal energy $\EST = 10^{51}\erg$ is instantaneously
added to a source region with spherical Gaussian distribution of radial scale
$r_0$.  For models with $\delta x=4\pc$, $r_0\geq20\pc$, while with
$\delta\vect{x}=1\pc$, $r_0\geq12\pc$ in order to adequately approximate a
sphere on a Cartesian grid.  Where the ambient ISM is very diffuse the radius
is adjusted such that the total mass in the ambient ISM is approximately
180\,${\rm M}_\odot$ to avoid excessive temperatures negatively impacting the
timestep.

In dense regions at $\delta x=4\pc$ up to 5\% of the energy is instead added
via momentum injection, $\ESK$ to compensate for excessive cooling
\citep[see][]{KO15,GMKSH20}.  This applies an axially symmetric diverging flow,
$\Delta\vect{u}$, with speed profile following the same Gaussian distribution
as the thermal energy. The pre-existing ambient ISM is not modified at the
explosion site, so the injected thermal energy, and velocity field when used,
is added to the existing ambient turbulent ISM.

In most cases no ejecta mass is added, as this is negligible relative to the
ambient ISM mass, but in cases where the temperature exceeds about $10^8$\,K,
some mass is included, with the same Gaussian profile as the energy, to cool
the gas and maintain the timestep \citep{OSU05,joung2006}.  Cumulatively, this
added ejecta is $\sim2M_\odot$ per SN and $\lesssim40M_\odot$ for any single
remnant.

SN Type II and I are distinguished only by their Poisson rates of
$\dot\sigma_{\rm II} = 15\kpc^{-2}\Myr^{-1}$ and $\dot\sigma_{\rm I} =
2\,\kpc^{-2}\Myr^{-1}$, respectively.  The model SN rate is therefore
$\dot\sigma=\dot\sigma_{\rm II}+\dot\sigma_{\rm I}=0.5\SNr$
\citep{BT91,Mannucci05}. The rate half that of the Solar neighbourhood is
adopted to reduce mass losses via galactic wind, while maintaining a multiphase
structure relevant to disc galaxies.

Both types of SNe are uniform randomly distributed horizontally, with Gaussian
distributions in $z$ of scale height $h_{\rm II}=90\pc$ and $h_{\rm I}=325\pc$,
respectively.  The SN rate $\dot\sigma$ applies the galactic area rate of SNe
independently of the vertical size of the domain $h$, which is included in
Equations~\eqref{eq:mom} and \eqref{eq:ent} to express the average volume rate.
The number of SNe does not change significantly with domain height, provided it
is $\gtrsim 6h_{\rm I}$, so $h=3.072$~kpc is not a model parameter.

Breathing and sloshing oscillations of the disc \citep{WC01} grow castastrophic
if SNe are located independent of the movement of the disc.  Therefore, we
track the vertical centre of the Type~II SN probability distribution to the
most dense layer of the gas $n_{\rm peak}={\rm max}(\langle
n\rangle_{xy}){/(1\cmcube)}$.  The Type~II SN scale height also varies in
response to the thickness of the disc. It is assumed that $n_{\rm peak}$ would
tend to be 1.  The scale height of the probability distribution is adjusted to
$h_{\rm II}/n_{\rm peak}$.  The model total mass is conserved, so ${n_{\rm
peak}}<(>)1$ suggests a higher (lower) scale height for the gas.  The stellar
population is not modelled, so this assumes a simple correlation between star
formation distribution and gas disc position and thickness.

In Model \LB, 70\% of Type II SNe are also clustered within associations of
radius 40 pc. We assume a Solar neighbourhood OB area rate
$\lesssim0.5\kpc^{-2}\Myr^{-1}$. With the Model Type~II SN area
rate~$\simeq15\kpc^{-2}\Myr^{-1}$ the average cluster includes about 22 SNe.

When each cluster is initiated the time to the next cluster is determined using
a Poisson rate.  The random location of a Type~II SN remnant identifies the
centre of the new cluster.  There is a 70\% probability that a remnant will
occur within the cluster.  The cluster has a spherical normal probability
distribution of radial scale 40~pc.  Otherwise an SN is located using the
parameters standard to all models.

For the small area of our model it is reasonable to assume that there are no
concurrent clusters, and that consecutive clusters follow immediately.

 \subsection{Averaging and normalization conventions}

Angular brackets without a subscript indicate the quantity is averaged over the
entire volume, while the subscript $xy$ limits the average to horizontal
planes, and the subscript $\ell$ indicates that the quantity has been Gaussian
smoothed over a kernel of length $\ell$ prior to averaging.  An overbar
indicates averaging over a domain {\em and} an interval of time, which may
vary, as specified in the text.

The velocity deviation from the shear flow $\vect{u}$, solved for in
Equation~\eqref{eq:mom}, excludes the imposed galactic shear velocity
$\vect{U}$. However, the system includes a stratified disc and so is
susceptible to the anisotropic {kinetic} $\alpha$-effect (AKA), dependent
on the rate of shear \citep[see, e.g.,][]{BR01,KGVS18}.  Hence, the deviation
velocity $\vect{u}$ includes mean vertical winds and horizontal vortical flows.
These are excluded to determine the turbulent velocity $\vect{u}^\prime$, which
is most relevant to the SSD.  The turbulent velocities are derived from the
horizontally averaged velocities as
\begin{equation}
\label{eq:uturb}
u_j^\prime =
u_j - \langle u_j\rangle_{xy}
.
\end{equation}

Similarly the kinetic energy
\begin{equation}\label{eq:eK}
	\eK=\frac{1}{2}\overline{\langle\rho u^2\rangle}
\end{equation}
also includes mean flows, so to isolate the contribution from the turbulent
kinetic energy from the horizontally averaged quantities we use
\begin{equation}
\label{eq:ekinturb}
\langle e_{ K}^\prime\rangle_{xy} =
\langle e_{ K}\rangle_{xy} - \frac{1}{2}\langle \rho\rangle_{xy}
\sum_{j=1}^3\langle u_j^2\rangle_{xy},
\end{equation}

To present the solutions for the magnetic energy growth with respect to the
kinetic energy density these are normalised by time- and volume-averaged
quantities.  In the case of $\eK$ or $\eKt$ only the time interval after the
magnetic field has passed its minimum energy is averaged to exclude initial
transients in the kinetic quantities.

\subsection{Mean electromotive force}\label{sec:EMF}

For the LSD the large-scale or mean-field evolution version of
Equation~\eqref{eq:ind} can be derived in terms of the mean magnetic field
$\mB$ as
\begin{equation}
\label{eq:meanB}
\frac{\partial\mB}{\partial t} = \nabla\times(\mU\times\mB+\EMF - \eta\nabla\times\mB)
\end{equation}
in which $\EMF$ contributes the electromotive force (EMF) from averaging
turbulent fluctuations $\langle\vect{u}^\prime\times\vect{B}^\prime\rangle$.
Applying a second order correlation approximation to an inhomogeneous,
anisotropic turbulence as found in the ISM of a spiral galaxy, a general
expression for the EMF has the form
\begin{eqnarray}
\label{eq:meanEMF}
\EMF &=& \tensorsym\alpha\cdot\mB + \vect\gamma\times\mB - \tensorsym\beta\cdot(\nabla\times\mB)\nonumber\\
&-&\vect\delta\times(\nabla\times\mB) -\tensorsym\kappa\cdot(\nabla\mB)^{(s)},
\end{eqnarray}
in which tensors $\tensorsym\alpha$ and $\tensorsym\beta$ are second order and
$\tensorsym\kappa$ third order, and each term represents a physical process
acting on the dynamo.  The $\tensorsym\alpha$ tensor applies effects from
small-scale helicity, $\tensorsym\beta$ turbulent diffusivity, $\vect\delta$
turbulent pumping, $\vect\gamma$ shear or rotation current effects, and
$\tensorsym\kappa$ includes residual effects, dependent on the symmetric part
of the magnetic gradient tensor $\nabla\mB)^{(s)}$.  A comprehensive study of
each of these effects will be required to explain fully the LSD, but that is
beyond the scope of the current investigation.

However, some indication of the EMF properties may be extracted from examining
a simplified analog of the $\tensorsym\alpha$ term.  {The action of the
Coriolis force on the sheared,} vertically stratified disc will tend to inject
helicity into the systemic vertical flows of opposite sign on either side of
the midplane.  The dynamo amplifies a magnetic field with opposing
{small-scale} helicity, which will quench the dynamo, if it cannot be
removed. {\citep{BS05,SS21}}

Under the assumption of isotropic, homogeneous turbulence $\tensorsym\alpha$
can be reduced to a scalar
\begin{equation}\label{eq:alpha}
{\alpha = \alpha_K + \alpha_M.}
\end{equation}
The kinetic
helicity contributes to the LSD as
\begin{equation}
\label{eq:alphaK}
\alpha_K \approx -\frac{1}{3}\tau\langle\vect{u}^\prime\cdot\vect{\omega}^\prime\rangle,
\end{equation}
where $\tau$ is the correlation time of the turbulence  and
$\vect{\omega^\prime} = \nabla \times \vect{u^\prime}$ is its vorticity
{\citep{Moffatt78,KR80}}.  Due to the conservation of magnetic helicity,
for the LSD to exist some small-scale helicity flux is required
{\citep{PFL76,SB04}}. The magnetic helicity contributes to the LSD as
\begin{equation}
\label{eq:alphaM}
\alpha_M \approx \frac{1}{3}\tau\left\langle{\frac{\left(\nabla\times\vect{B}\right)^\prime\cdot\vect{B}^\prime}{\rho}}\right\rangle.
\end{equation}
While these simplified expressions are insufficient to describe the full range
of turbulent transport motions driving the LSD, they are relatively easy to
obtain and offer some indication of how the LSD changes over time and differs
between models.  The Lorentz force acts against the flow as the dynamo
approaches saturation, where opposite sign of $\alpha_M$ can lead to
$\alpha$-quenching \citep{KR80} of the LSD.

\section{Results}\label{sec:results}

\subsection{Small-scale dynamo}\label{sec:SSD}

\begin{figure}
\centering
\includegraphics[trim=0.5cm 0.5cm 0.0cm 0.2cm,clip=true,width=\columnwidth]{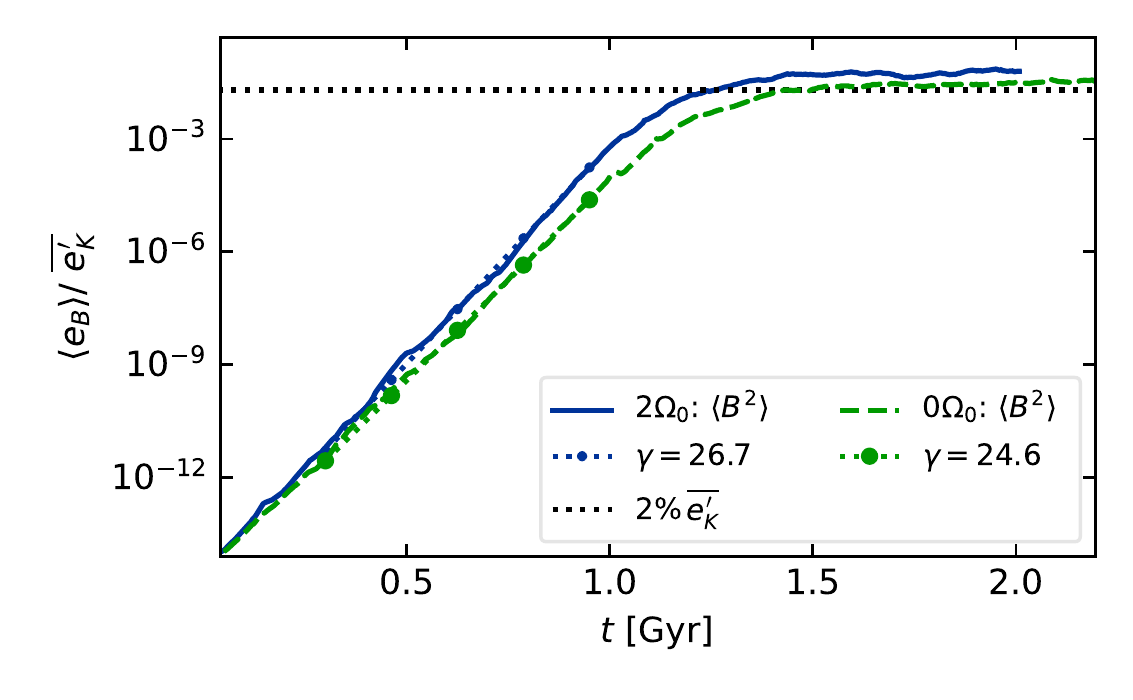}
\caption{
Volume-averaged magnetic energy density for models \LD\ (\emph{blue}) with
$\Omega=2\OSN$ and \LC\ (\emph{green}) with $\Omega=0$, in which LSD is
suppressed by continuously removing the horizontally averaged magnetic field
components to ensure only SSD is present. Exponential fits for the growth rates
measured during the dotted intervals are listed in the legend.  Normalisation
is by the time-averaged turbulent kinetic energy $\eKt$.  The horizontal dotted
line indicates 2\% equipartition.
\label{fig:eB0-gamma}
}
\end{figure}

To isolate the properties of the SSD in the stratified and differentially
rotating ISM we exclude the presence of a mean field in models \LC\ and \LD\ by
continuously subtracting the averages of each component of the magnetic field
on each horizontal plane.  The energy density of the turbulent magnetic field
generated by the SSD is plotted in Figure~\ref{fig:eB0-gamma}.

{Over a range of models and regimes the SSD is constrained to saturate at
or below equipartition with the kinetic energy density, but has so far been
found to actually saturate well below equipartition.  \citet{ADN07} obtain an
intermittently superequipartition solution with a smooth flow, but it reduces
to $\sim10\%$ in the turbulent regime.  \citet{FCSBKS11} report saturation
levels of only 0.8\%--5\%of equipartition for isothermal forced turbulence at
sonic Mach number $\Ms=2$, reaching the higher value with solenoidal forcing.
This drops to 2\% for $\Ms\geq10$.  As high as 60\% is possible only with
incompressible flows and solenoidal forcing.  \citet{SSFBK15} derive a
semi-analytic model prediction of 1.3\% to 43.8\% at $\Pm>1$, with the lower
end applying for highly compressible flows.}

The saturation of the magnetic field for Model\,\LC\  occurs at about 2\% of
equipartition with the time-averaged turbulent kinetic energy density $\eKt$ of
Equation~\eqref{eq:ekinturb}.  {The saturation ratio is therefore a
signature of the compressibility at high Rm \citep[see also,
e.g.,][]{SBK02,GMKS21} and is likely to evolve rapidly in the ISM to a minimum
of 1--2\% equipartition.  }

The SSD in Model \LC, without shear, grows with exponential growth rate
$\gamma= 24.6\Gyr^{-1}$, while Model \LD\ grows slightly faster, with
$\gamma=26.7\Gyr^{-1}$, and saturates at a higher level.  Large-scale shear and
rotation, therefore, slightly enhance the SSD.

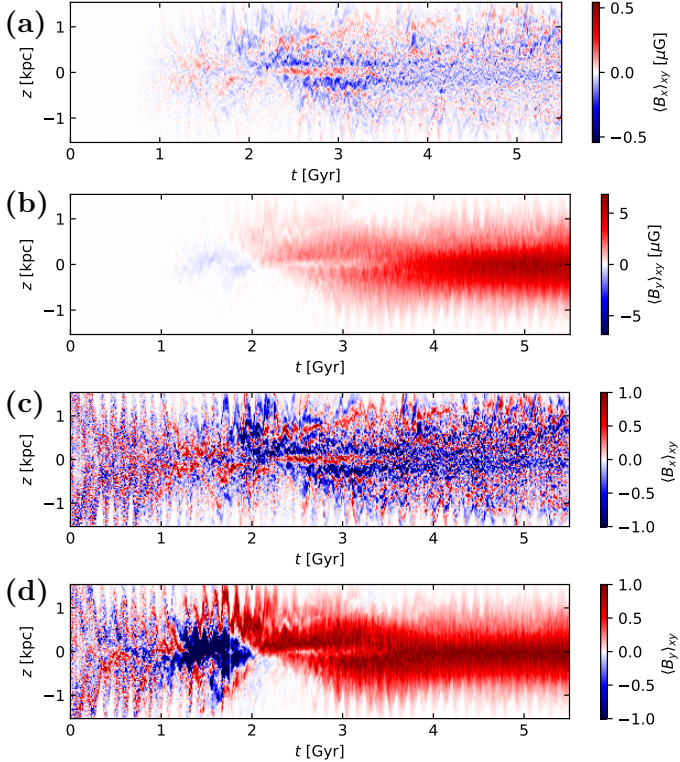
\begin{figure}
\centering
\includegraphics[trim=0.45cm 0.45cm 0.45cm 0.25cm,clip=true,width=1.05\columnwidth]{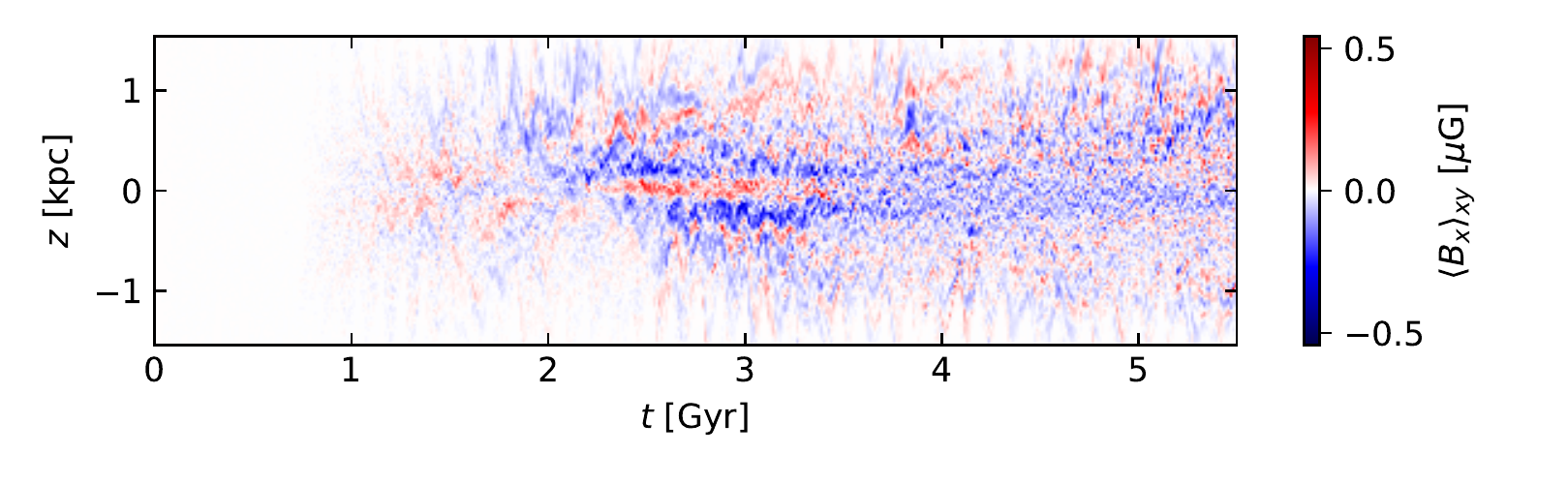}
\includegraphics[trim=0.45cm 0.45cm 0.45cm 0.25cm,clip=true,width=1.05\columnwidth]{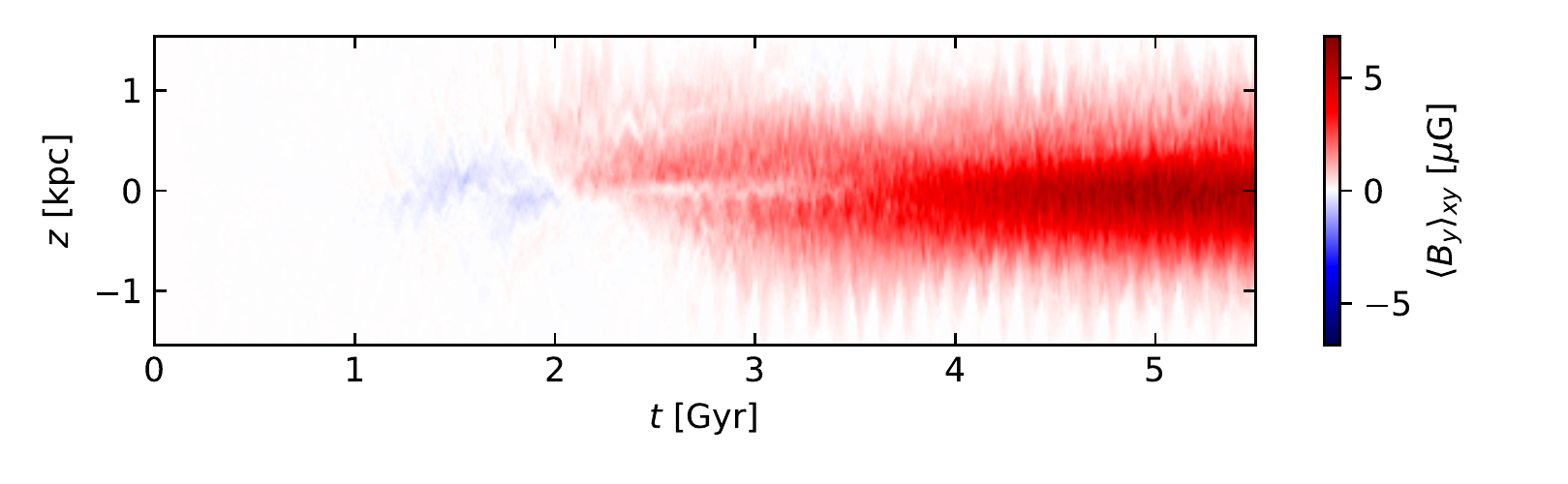}
\includegraphics[trim=0.45cm 0.45cm 0.45cm 0.25cm,clip=true,width=1.05\columnwidth]{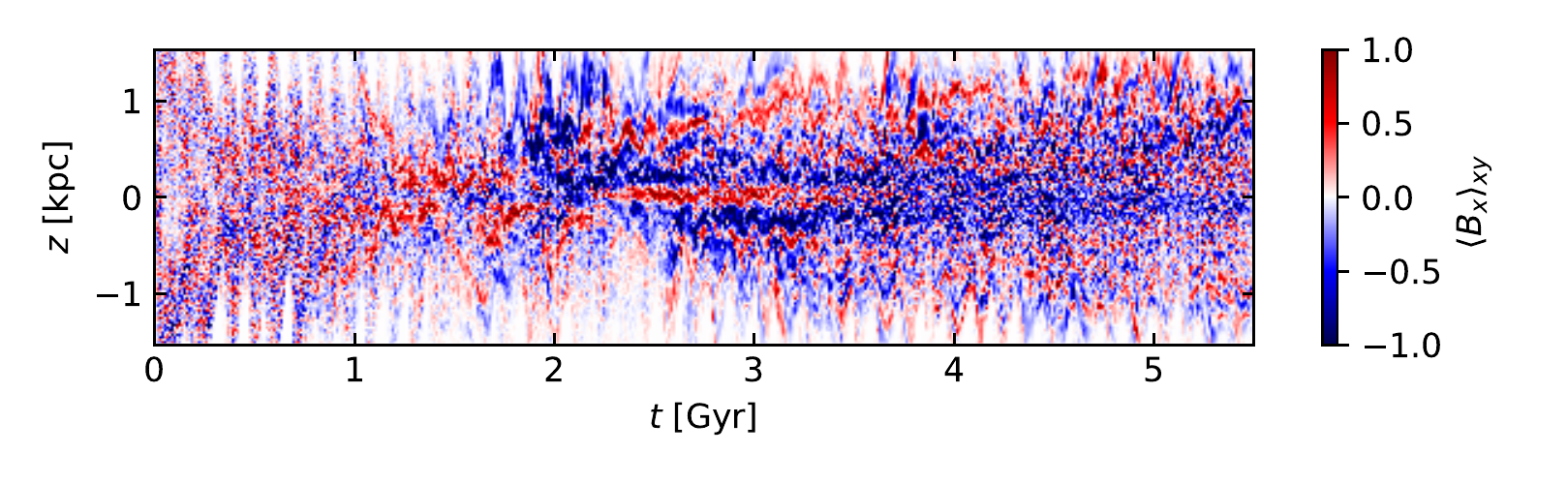}
\includegraphics[trim=0.45cm 0.45cm 0.45cm 0.25cm,clip=true,width=1.05\columnwidth]{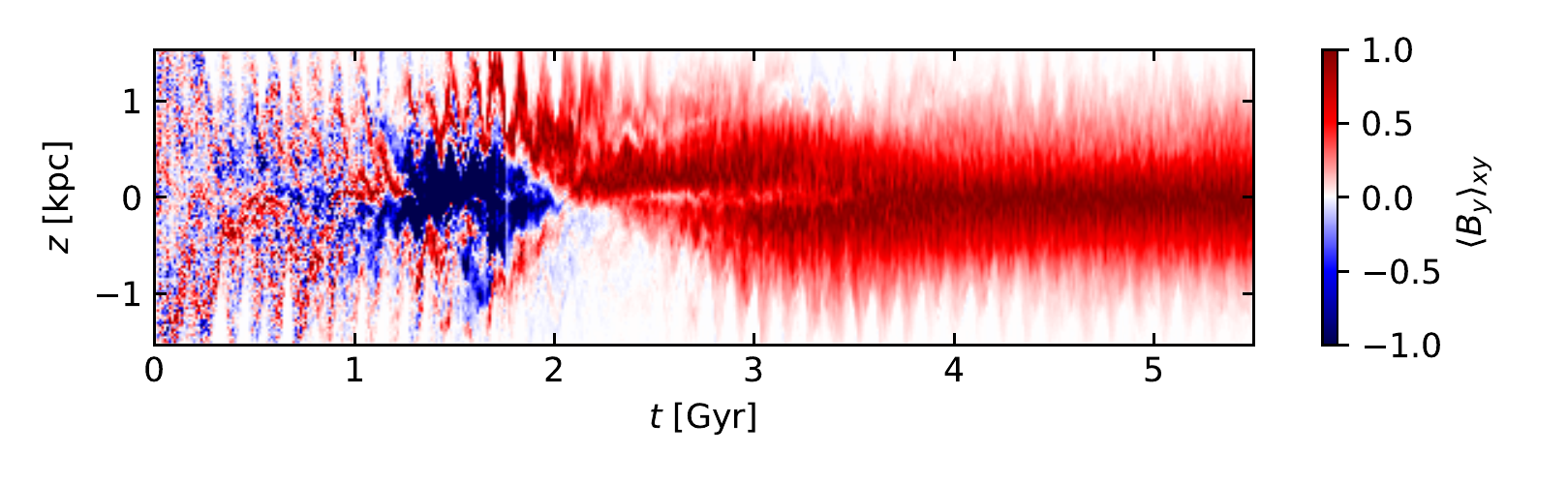}
 \begin{picture}(0,0)
    \put(-125,290){{\sf\bf{(a)}}}
    \put(-125,222){{\sf\bf{(b)}}}
    \put(-125,146){{\sf\bf{(c)}}}
    \put(-125, 75){{\sf\bf{(d)}}}
  \end{picture}
\caption{
Time-latitude diagrams of the horizontally averaged magnetic field \emph{(a)}
$\langle B_x \rangle_{xy}$  and \emph{(b)} $\langle B_y \rangle_{xy}$, for the
fiducial model \LA\ in microgauss and, respectively, \emph{(c)} and \emph{(d)}
normalised by the time-dependent maxima to highlight the evolving structure and
scales of the LSD.  \label{fig:tav-L2O}
}
\end{figure}

\subsection{Large-scale dynamo}\label{sec:LSD}

Having examined the characteristics of the SSD in a rotating shear flow with
Model \LD, we now apply the same parameters, but do not suppress the growth of
the large scale field in Model \LA.  Figure~\ref{fig:tav-L2O} displays the
evolution of the mean magnetic field as horizontal averages of the magnetic
field components $\langle B_x \rangle_{xy}$ and $\langle B_y \rangle_{xy}$ up
to saturation of the LSD.  Relative to the horizontal components, $B_z$ has
stronger local fluctuations driven by the vertical flows.  However, due to
periodic boundary conditions on $A_x$ and $A_y$, its horizontal averages must
vanish.

The azimuthal shear favours a steep pitch angle of the mean magnetic field,
which ultimately primarily aligns azimuthally.  The late time value of $\langle
B_y \rangle_{xy}$ is strongly positive and an order of magnitude stronger than
mean $\langle B_x \rangle_{xy}$.  In the period beyond 3.9\,Gyr, after which
the mean field remains quite steady, along the midplane $\overline{\langle
B_x\rangle_{xy}}=-0.17\pm0.20\,\upmu $G, while $\overline{\langle
B_y\rangle_{xy}}=9.16\pm 3.56\,\upmu $G, with errors given by the standard
deviation.

In Figures~\ref{fig:tav-L2O}\emph{(c)} and \emph{(d)}, horizontal averages of
$B_x$ and $B_y$ are shown normalised by their time-dependent maximal values, so
that their structure at early times can be easily compared to the saturated
state.  It is evident that the mean of both components during the kinematic
stage of the SSD fluctuates on short time and spatial scales.  More coherent
and persistent structures emerge after 1\,Gyr, when the SSD has already
saturated.  At first these are typically of opposite sign to the eventual
state, with more positive $B_x$ and negative $B_y$.  Although the sign
reversals persist in $B_x$, these fluctuations remain much weaker than in
$B_y$. The vertical topology of the field remains steady after about 3.75\,Gyr,
although the strength of the field continues to grow as seen in panel
\emph{(b)}.

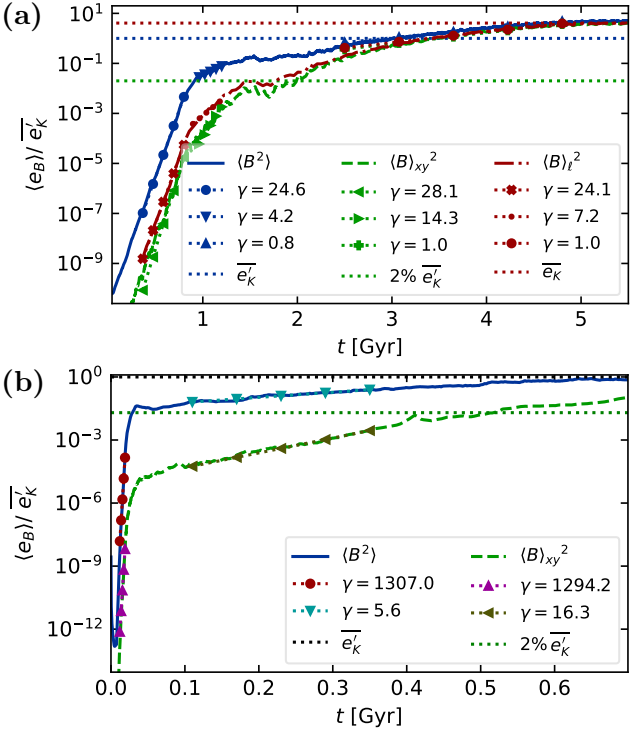
\begin{figure}
\centering
\includegraphics[trim=0.3cm 0.5cm 0.0cm 0.0cm,clip=true,width=\columnwidth]{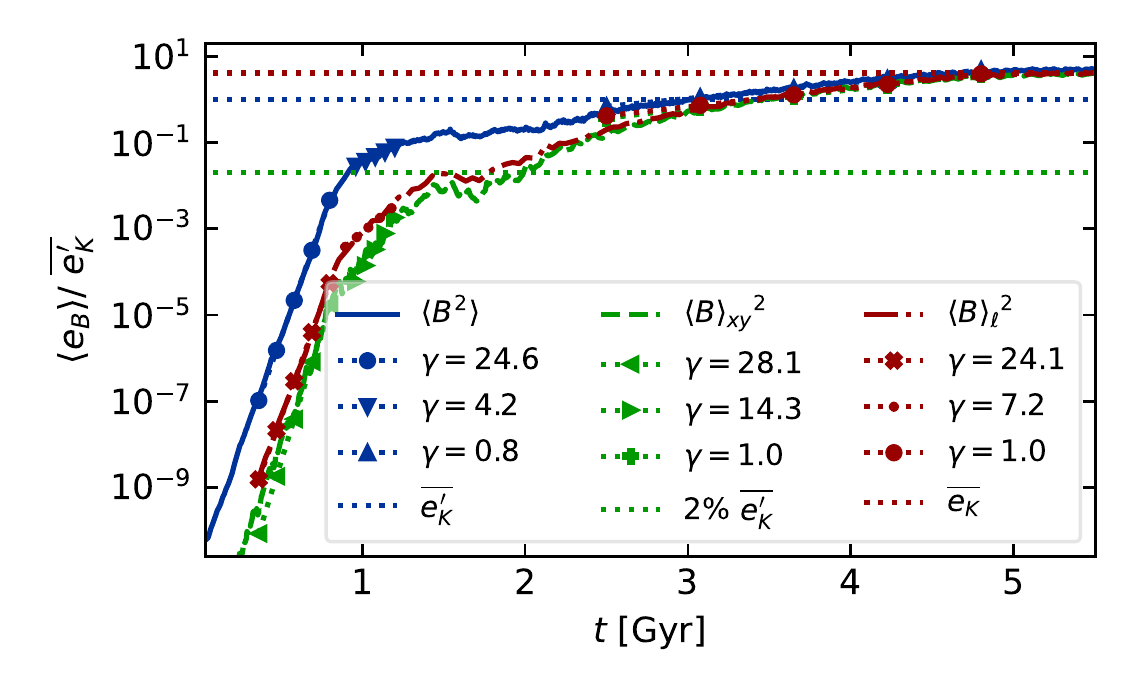}
\includegraphics[trim=0.5cm 0.5cm 0.0cm 0.3cm,clip=true,width=\columnwidth]{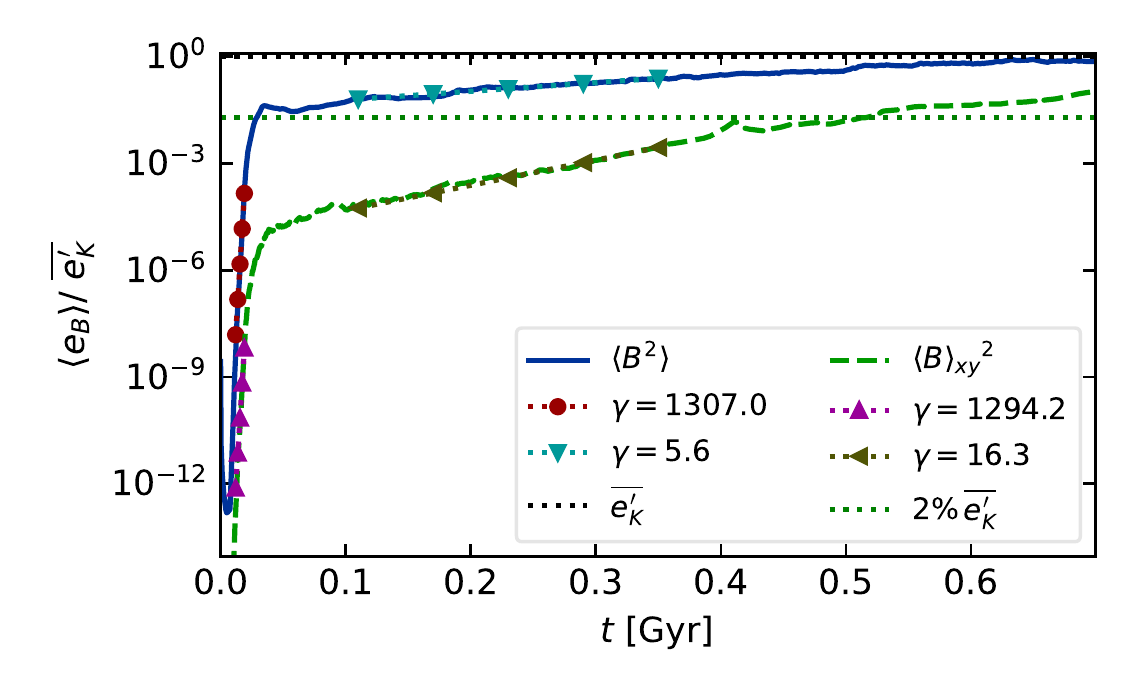}
 \begin{picture}(1,0)
    \put(-122,280){{\sf\bf{(a)}}}
    \put(-122,141){{\sf\bf{(b)}}}
  \end{picture}
\caption{
{Energy density of total magnetic field $\langle B^2\rangle(8\uppi)^{-1}$
(\emph{blue}) and horizonally averaged mean-field ${\langle B
\rangle_{xy}}^2(8\uppi)^{-1}$ (\emph{green}) for \emph{(a)} Fiducial LSD
Model~\LA\ and \emph{(b)} Model \HA.  For \LA\ Gaussian smoothed mean-field is
also plotted ${\langle B \rangle_{\ell}}^2(8\uppi)^{-1}$ (\emph{red}), with
$\ell=50\pc$. (Factors of $(8 \uppi)^{-1}$ are dropped in the legend for
readability in this and following figures showing magnetic energy evolution.)
Normalisation is by the time-averaged turbulent kinetic energy density $\eKt$
(\emph{blue dotted}), while the total kinetic energy is given by $\eK$
(\emph{red dotted}).  Exponential fits for the growth rates are listed in the
legends, applying to intervals in \emph{(a)} $365\Myr<t<800\Myr$,
$0.9\Gyr<t<1.2\Gyr$ and $2.5\Gyr<t<4.8\Gyr$, respectively, and for \emph{(b)}
spanning $12\Myr<t<19\Myr$ and $0.11\Gyr<t<0.35\Gyr$, respectively.
}
\label{fig:eB-gamma}
\label{fig:eB1}
}
\end{figure}

All profiles of Figure~\ref{fig:tav-L2O} exhibit oscillatory structure.  These
correspond to the breathing and sloshing modes predicted by \citet{WC01}, due
to {the} gravitational {interaction of the stellar disc} with the gas
outflows.  (This is also shown in gas number density in
Appendix~\ref{subsec:appen-density} for a zoomed-in interval.)  The period of
100--150~Myr supports the model of \citet{WC01} well for Solar neighbourhood
parameters.

In Figure~\ref{fig:eB-gamma}{\emph{(a)}} we plot the growth of the total
magnetic {energy} (blue) for Model \LA.  The mean field {energy} is
also plotted as horizontal averages $\langle B\rangle_{xy}$ and Gaussian
smoothed $\langle B\rangle_{\ell}$, with a smoothing length $\ell=50\pc$, as
described previously \citep{Gent:2013a}.  Fits for the exponential growth rate
$\gamma$ are listed in the legend for intervals spanning the SSD, a
transitional stage, and the LSD.

The growth rate of the total {magnetic energy} during the SSD matches that
of Model \LC\ at $24.6\Gyr^{-1}$ and lies about 10\% below that of Model \LD,
with otherwise equal parameters to this model.  The kinematic SSD therefore
appears to be mildly inhibited by the action of the LSD.  During the SSD
{kinematic stage}, the mean field {magnetic energy} grows almost
identically with the SSD, and, counter to the finding of \citet{Gent:2013a}, is
near insensitive to the definition of the mean field.

Before the total {magnetic energy} reaches 2\% $\eKt$, {the level} at
which the SSD saturates for Model \LC\ and \LD\ (Figure~\ref{fig:eB0-gamma}),
the mean-field {energy} growth slows down.  The SSD {energy} does not
saturate {at this point}, but continues to grow at a smaller rate $\gamma =
4.2\Gyr^{-1}$ above 2\% $\eKt$.  The strength of the growth of the fluctuating
field {energy} is unexpected. The mean field {energy} remains at least
two orders of magnitude weaker than the turbulent field {energy}. Tangling
must therefore be capable of generating turbulent field at rates substantially
larger than the source.  The mean field {energy} grows at a rate $\gamma$
between 7 and $14\Gyr^{-1}$, but this slows down {to $1\Gyr^{-1}$} as it
approaches 2\% $\eKt$.

{Measuring the growth rate of the magnetic field {magnitude} $\gamma_B
= \gamma/2$, \citet{GEZR08} obtain $4\Gyr^{-1}$ for $\dot\sigma=0.5\SNr$, with
a shear $S=4\OSN$ twice as large as our models. At a comparable epoch, our
Model~\LA\ has $\gamma_B=2.1\Gyr^{-1}$.  With SN rate $\dot\sigma=\SNr$ for
$S=2\OSN,\,4\OSN$, and $8\OSN$, \citet{GEZR08} yield $\gamma_B=6.8,\,9.8$ and
$19.2\Gyr^{-1}$, respectively.}

Thus, for $S\gtrsim\OSN${, when there is a flat rotation curve as applies
here}, effectively $\gamma_B \propto S$, consistent with the theoretical
prediction for a {supercritical} mean-field galactic dynamo summarised by
\citet[][Eqs.~{[{11.17--11.18}]}]{BS05}. At a galactocentric radius of 4
kpc, \citet[][Chapter 11.2]{SS21} predict $\gamma_B=6.25\Gyr^{-1}$, which is
remarkably close to the numerical solutions for $S=2\OSN$ and
$\dot\sigma=\SNr$.

During this transition the definition of the mean field does substantially
affect the measured growth rate, which agrees with the analysis of
\citet{Gent:2013a}, who omitted the kinematic phase and considered only this
transition. Later, as the LSD becomes dominant, the definition makes little
difference to evolution of the mean-field energy, although it affects very much
its topology \citep[e.g., see Figure~7.6][]{SS21}.

In Figure~\ref{fig:eB1}{\emph{(b)}} we show the evolution of magnetic
energy of Model~\HA. For comparison with Model~\LA\ {the} intervals
{relevant} to the {kinematic SSD have $\gamma=1307\Gyr^{-1}$, while the
transition phase has $\gamma=5.6\Gyr^{-1}$. (The high resolution model H2 has
not yet completed the transition to the LSD.)  The extremely rapid kinematic
growth at high resolution is} quite consistent with our previous results for
SSD in the multiphase ISM {\citep{GMKS21,GMKS22}}.  {The near
self-similar growth rates during the kinematic SSD within each model} suggest
the growth of the mean field during this phase is the result of winding up by
the large scale shear of the growing random field. This is reflected in the
small scale structure of the horizontal averages during this period as shown in
Figure~\ref{fig:tav-L2O}{\emph{(c)--(d)}}.

In modelling of SSD and LSD evolution at high Rm with an $\alpha^2$ dynamo with
helical forcing, \citet{SB14} and \citet{BSB16} find no LSD during the
kinematic phase of the SSD, with single eignemode growth at all scales.  Only
in the nonlinear stage of the SSD do the growth rates at large scales exceed
growth at small scales{. Even when shear is included \citet[][their figures
1 and 2]{BSB19} show no distinction between the growth rates at each scale
until the SSD becomes nonlinear. With the same presentation (see our Appendix)
we could infer the same conclusion. However, this would be incorrect since for
Model \LA\ we do find the LSD enhances the growth rate at large scales
throughout} the kinematic SSD.

\begin{figure}
\centering
\includegraphics[trim=0.32cm 1.64cm 0.30cm 0.1cm,clip=true,width=\columnwidth]{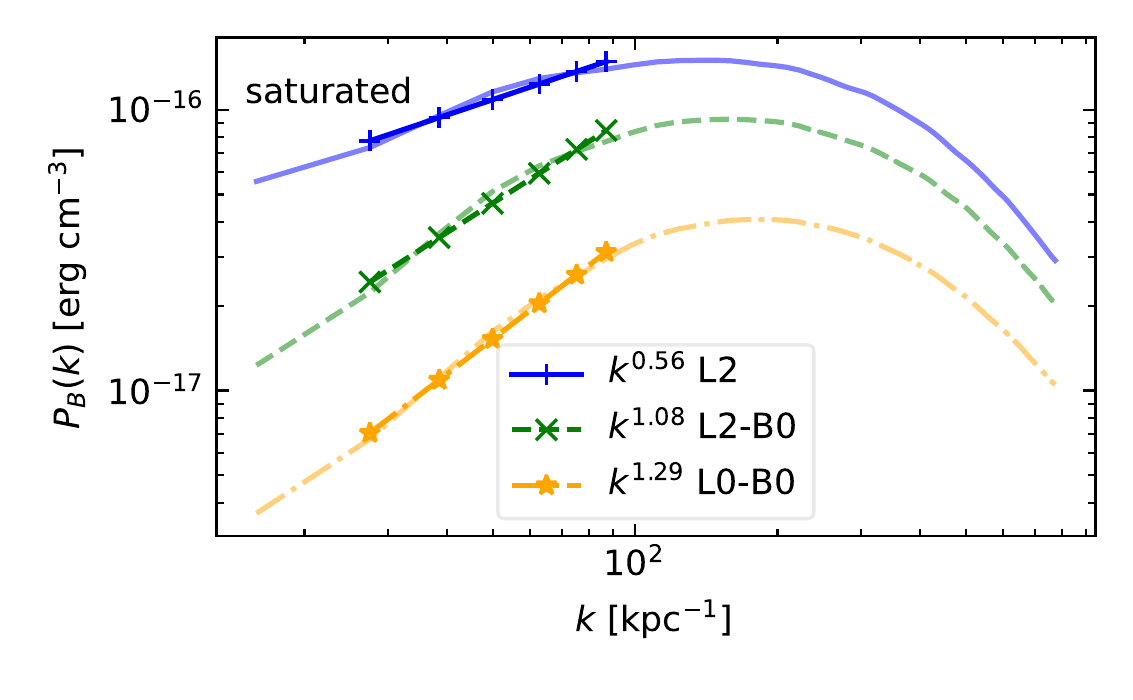}
\includegraphics[trim=0.32cm 1.64cm 0.30cm 0.1cm,clip=true,width=\columnwidth]{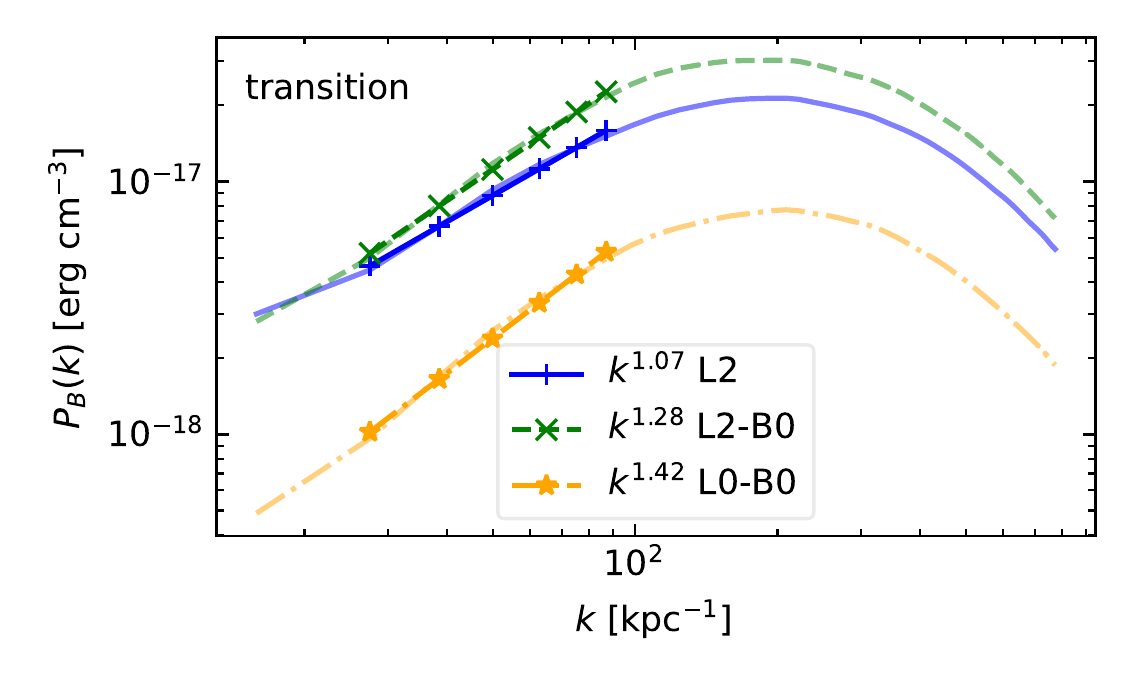}
\includegraphics[trim=0.32cm 0.55cm 0.31cm 0.1cm,clip=true,width=\columnwidth]{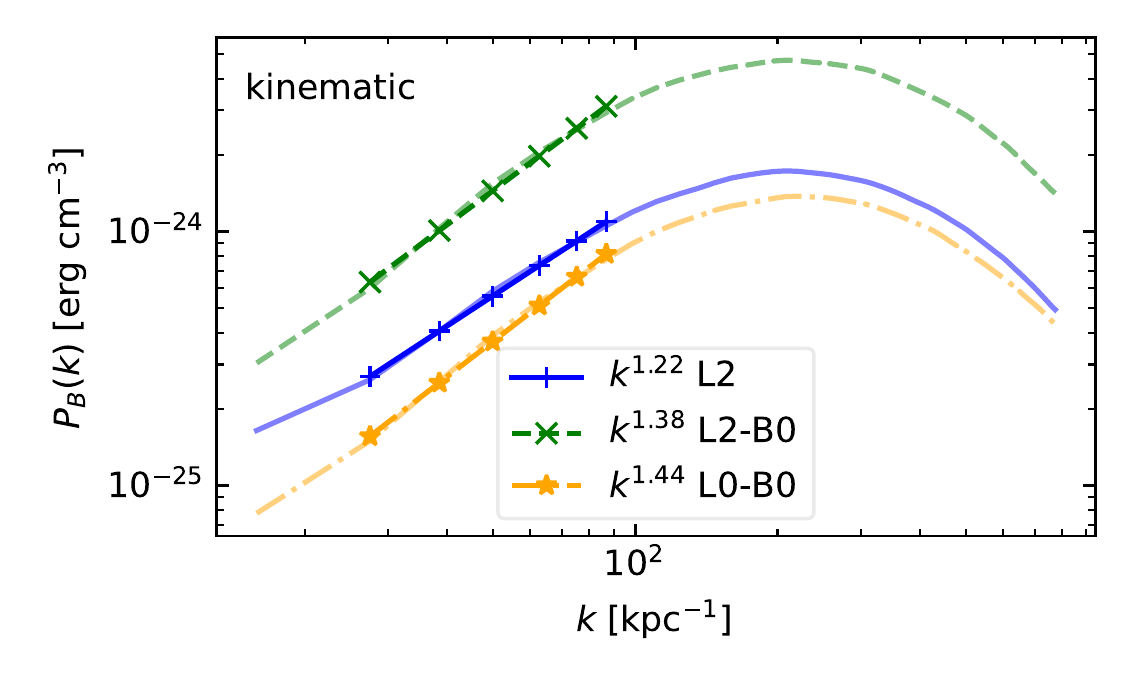}
 \begin{picture}(0,0)
    \put(-118,380){{\sf\bf{(a)}}}
    \put(-118,260){{\sf\bf{(b)}}}
    \put(-118,145){{\sf\bf{(c)}}}
  \end{picture}
\caption{
Magnetic energy spectra during epochs of the SSD as indicated in each panel
(\LA\ is shifted 365 Myr, as discussed in Sect.~\ref{sec:convergence}).  Models
listed in the legends have power law fits for comparison with Kazantsev
$k^{3/2}$ scaling. Spectra have been smoothed with a Gaussian filter of length
5\,Myr {centered about $t$} in time and 1.5\,$k_1$ {about $k$}. 
\label{fig:lsd-ssd-kspec}
}
\end{figure}
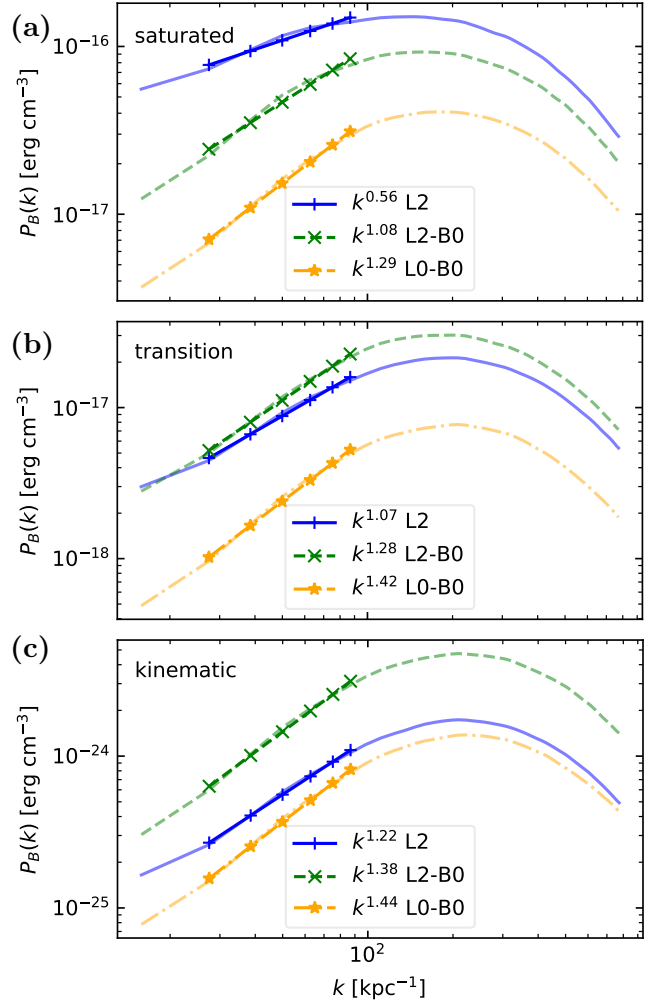

This {enhancement} is evident by inspection of
Figure~\ref{fig:lsd-ssd-kspec}. We plot the magnetic energy spectra together
for Models~\LA, \LC, and \LD, at three characteristic stages of the SSD.  To
compare equivalent stages of the evolution of the SSD between the three models
the times selected for Model~\LA\ are shifted by 365~Myr. In the early
kinematic stage the Models \LC\ and \LD\ have a power law approximating
$k^{3/2}$ \citep{K68} at small wavenumbers, consistent with the expected
Kazantsev theory. {An expanded examination of the kinematic stage in
\citet[][figure 10]{BSB19} might therefore show a similar trend.}

Even at these early times the power law is somewhat more shallow for Model \LD,
due to the differential rotation, and yet more so for \LA, due to the inclusion
of LSD, increasing energy at larger scales.  Throughout the kinematic stage,
however, all have the peak energy at $k\simeq17k_1$.  In common with
\citet{HBD04,CD09,BS13} and \citet{Eyink13}, all models show a shift in energy
toward lower $k$ as the SSD saturates, with the power law $\rightarrow k^1$
scaling at low $k${. O}nly for Model~\LA\ with LSD does the power law
{approach} $k^0$ at late times.

\subsection{The \texorpdfstring{$\alpha$-effect}{alpha effect}}\label{sec:alpha}

\begin{figure}
\centering
\includegraphics[trim=0.35cm 0.45cm 0.59cm 0.25cm,clip=true,width=\columnwidth]{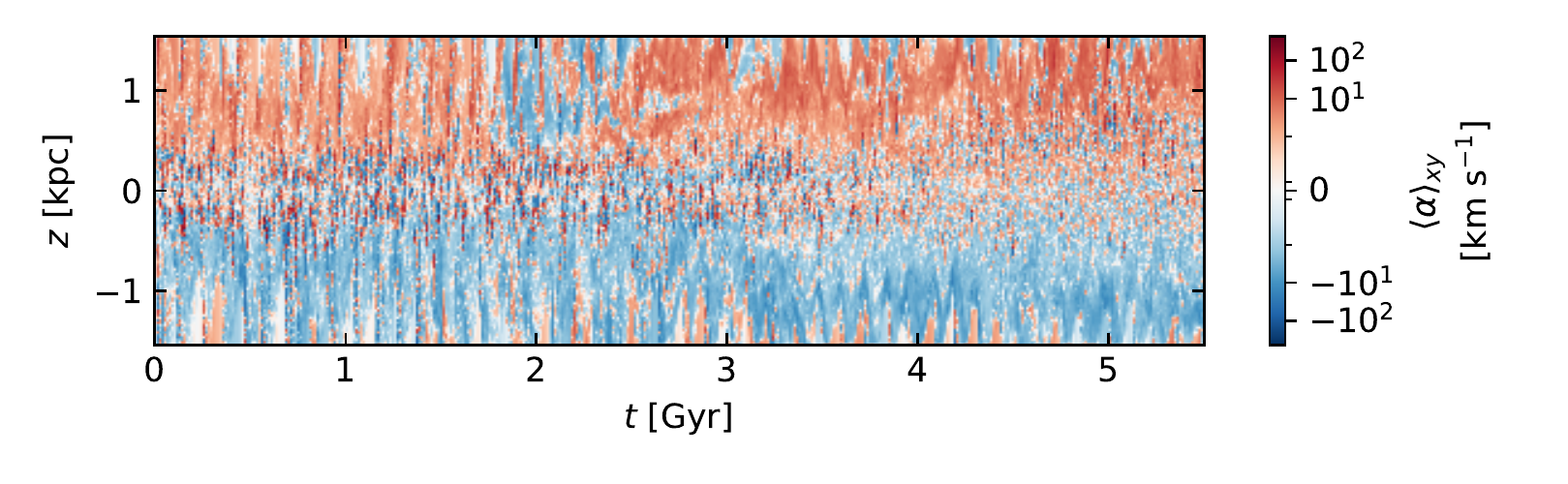}
\includegraphics[trim=0.35cm 0.45cm 0.59cm 0.25cm,clip=true,width=\columnwidth]{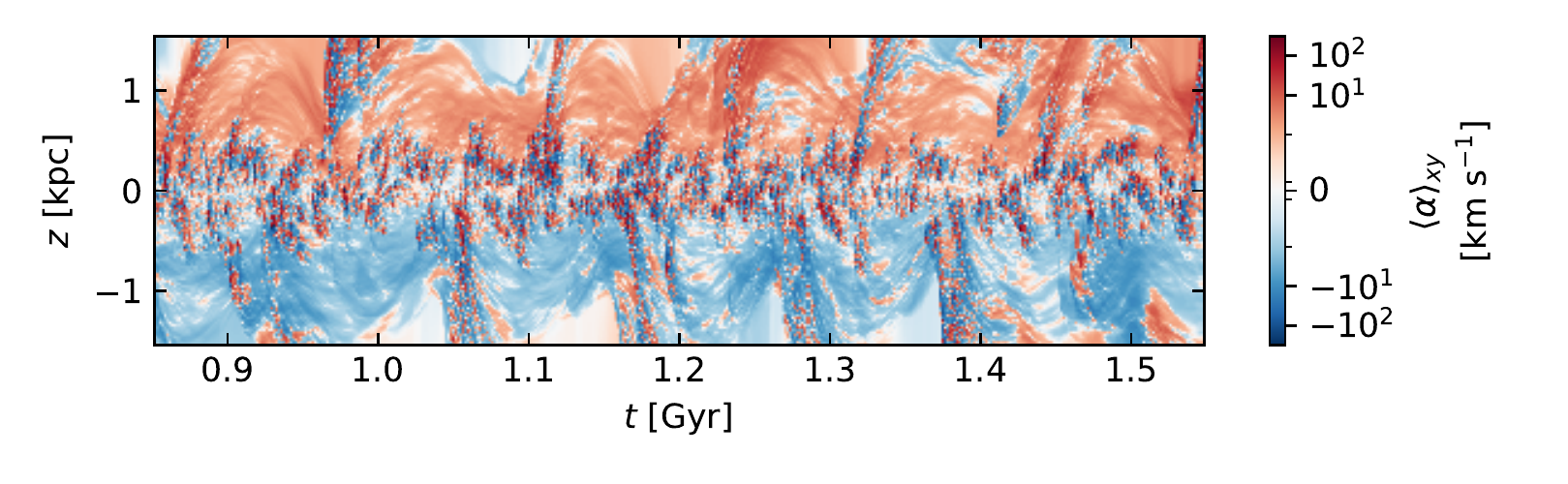}
\includegraphics[trim=0.35cm 0.45cm 0.69cm 0.25cm,clip=true,width=\columnwidth]{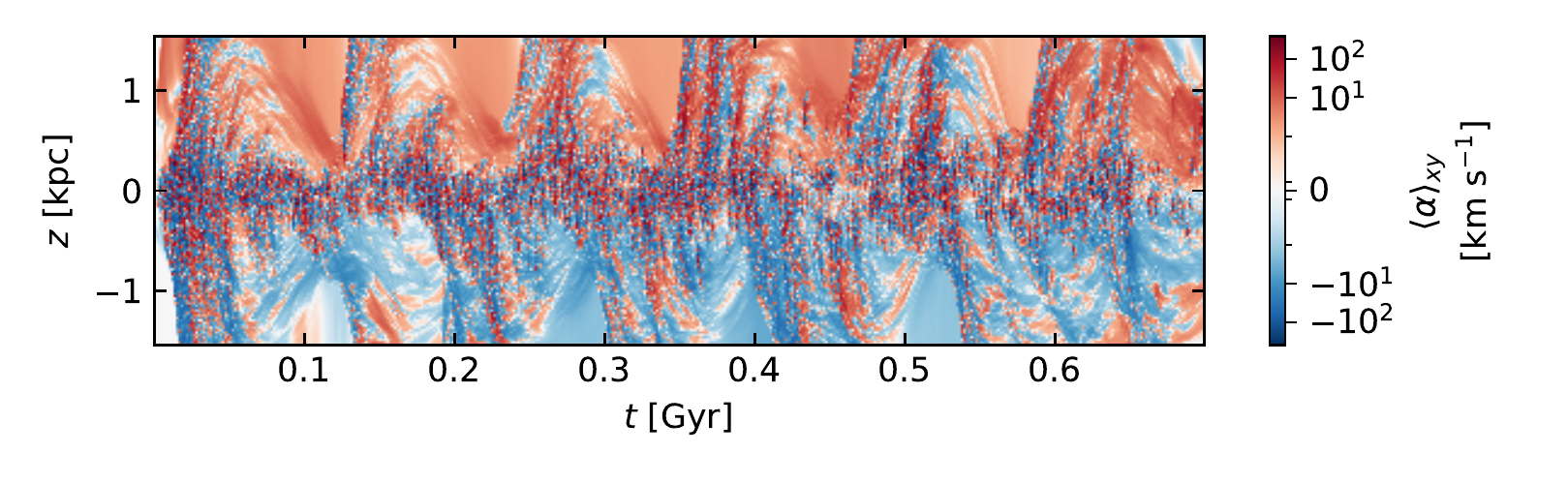}
 \begin{picture}(0,0)
    \put( -90,208){{\sf\bf{\LA}}}
    \put( -90,140){{\sf\bf{\LA}}}
    \put( -90, 70){{\sf\bf{\HA}}}
    \put(-123,208){{\sf\bf{(a)}}}
    \put(-123,140){{\sf\bf{(b)}}}
    \put(-123, 70){{\sf\bf{(c)}}}
  \end{picture}
\caption{
Time-latitude diagrams of horizontally averaged $\alpha$-effect {from
Equation~\eqref{eq:alpha}}  for Model~\LA\ \emph{(a)} throughout and \emph{(b)}
during a restricted interval comparable with \emph{(c)} Model~\HA.  A linear
colour-scale is applied in the range $-5\kms<\alpha<5\kms$ and symmetric
logarithmic colour-scale for larger magnitudes.
\label{fig:av-alpha}
}
\end{figure}
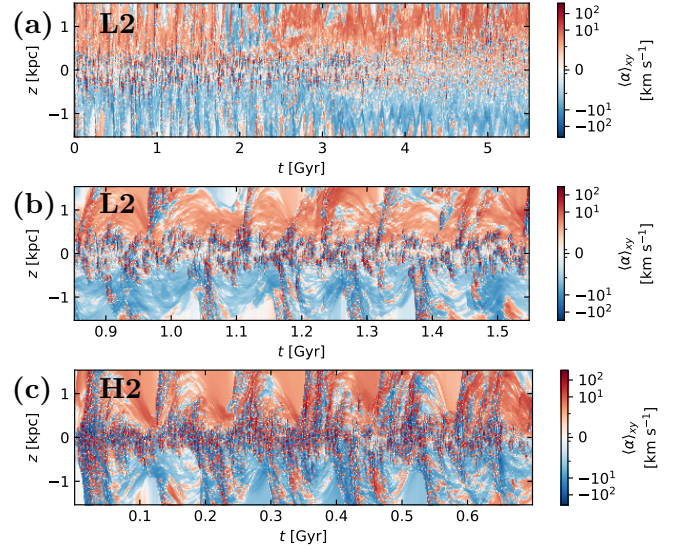

We approximate the proxy for $\alpha=\alpha_K+\alpha_M$ in
Eqs.~\eqref{eq:alphaK} and \eqref{eq:alphaM} from the time series of horizontal
averages with
\begin{equation}
\label{eq:alphaKxy}
\langle\vect{u}^\prime\cdot\vect{\omega}^\prime\rangle_{xy}=
\langle\vect{u}\cdot\vect{\omega}\rangle_{xy}-\sum_{k=1}^3
\langle u_k\rangle_{xy}\langle \omega_k\rangle_{xy}
\end{equation}
and
\begin{align}
\label{eq:alphaMxy}
\left<\frac{\vect{B}^\prime\cdot{\left(\nabla\times\vect{B}\right)}^\prime}{\rho}\right>_{xy}&=
\frac{\langle\vect{B}\cdot{\nabla\times\vect{B}}\rangle_{xy}}{\langle\rho\rangle_{xy}}\\
&-
\frac{\langle B_k\rangle_{xy}{\langle \epsilon_{ijk}\partial_iB_j\rangle_{xy}}}{\langle\rho\rangle_{xy}},\nonumber
\end{align}
where the first expression on the right hand side is used because
$\langle(\vect{B}\cdot{\nabla\times\vect{B}})\rho^{-1}\rangle_{xy}$ is
unavailable{, and in the second $\epsilon_{ijk}$ denotes the Levi-Cevita
symbol and summation over repeated indices $i,j,k$ is assumed}.  In
Figure~\ref{fig:av-alpha} this $\alpha$ is illustrated as time-latitude
diagrams for Models~\LA\ and \HA, during their entire evolution,
and also during a period of Model~\LA\ comparable to Model~\HA\ to assist
comparison.

We assume a correlation time of $\tau=10\Myr$ {in
Equation~\eqref{eq:alphaK} to calculate both $\alpha_K$ and $\alpha_M$} for
both models.  This is consistent with an application of the test-field method
\citep{Bendre2018} and yields rms values for $\alpha$ in Models \LA, \HA\ and
\LB\ of 15, 30, and $89\kms$, which correspond reasonably to calculations of
$u^\prime$ included in Figure~\ref{fig:eBall}(c).  {This value of $\tau$}
is slightly higher than the $5\Myr$ determined by \citet{HSSFG17}, the
$1.9\Myr$ of \citet{BGKKMR13}, or the order of magnitude estimate of 3--7 Myr
of \citep{CS20}.  Whether $\tau$ differs between magnetic and kinetic $\alpha$,
or overall, a factor of 1/5 would not qualitatively alter the dynamics.

Considering the turbulent $\alpha$ displayed for Model~\LA\ in
Figure~\ref{fig:av-alpha}(a) the northern latitudes are predominantly positive,
while the southern latitudes are negative, which is consistent with the action
of the Coriolis force on each side of the midplane.  We provide a breakdown of
each $\alpha$ contribution in Appendix~\ref{subsec:appen-alpha}.  This shows
that the evolution of $\alpha$ is dominated by the contribution of kinetic
$\alpha_K$ in the first 2 Gyr. The negative sign in the northern latitudes near
2 Gyr is due to current helicity, as are the strong high latitude signatures
after 2.5 Gyr.  However, the high altitude $\alpha$-effect where the magnetic
field is quite weak likely has little effect on the LSD.

The reversal in sign of the mean field after 2 Gyr corresponds to damping of
$\alpha$ in the dynamo active warm gas near the midplane, but also with growth
of magnetic $\alpha_M$ of opposite sign to $\alpha_K$ in this region (see
Appendix~\ref{subsec:appen-alpha}).  The $\alpha$-quenching near the midplane,
due to reducing $\alpha_K$ and growing magnetic $\alpha$ of opposite sign
saturates the LSD.

Comparing  Models~\LA\ and \HA\ in Figures~\ref{fig:av-alpha}(b) and (c) during
corresponding 700~Myr intervals, their $\alpha$ topologies are qualitatively
similar.  At high resolution, $\alpha$ is typically stronger, and also has
reversals in sign, which are more ubiquitous and on smaller spatial scales and
time scales. In both models helicity of opposite sign crosses the midplane and
propagates away from the disc.  The small-scale structure of the helicity
occupies the outflows, {which are evident as small-scale turbulent
structures that appear first at the midplane and propagate quickly towards the
halo,} while large-scale, same-sign helicity populates the inflows{,
appearing at the boundary and propagating more slowly towards the midplane}.
Thus, the small-scale helicity {at sufficiently high Rm \citep{Mitra10}} is
sporadically removed from the disc allowing the LSD to grow {and for
sufficiently high helicity fluxes to saturate even at levels above
equipartition \citep{CSSS14}}.

\subsection{Effect of clustering}\label{sec:cluster}

Figure~\ref{fig:eBall}\emph{(a)} {plots total magnetic energy density
evolution for all models. It} shows that Model \LB, with OB clustering
included, has growth of magnetic energy driven by the SSD over four times
faster than Model \LA, without clustering.  (Note that, although they start
with somewhat different seed fields, the growth rates are unaffected by the
difference.) Clustering increases the volume filling fraction of hot
superbubbles in the midplane.  The increased effective sound speed supports
rapid growth of the {turbulent magnetic field,} twice that of Model \LA,
and 4/3 faster after the SSD has saturated.

{The turbulent kinetic energy density in Figure~\ref{fig:eBall}\emph{(b)}
is similar with and without clustering, while the turbulent velocity in
Figure~\ref{fig:eBall}\emph{(c)} is much higher due to clustering.  W}hen OB
clustering is included, the highest turbulent velocities are in hot regions of
low density and thus contribute little to the kinetic energy.  \citet{GMKS22}
have demonstrated that it is the flow velocity, rather than the kinetic energy,
that is critical to the SSD growth rate in the multiphase ISM, while saturation
strength is related to the kinetic energy and is thus independent of the
inclusion of clustering.

\section{{Robustness of Results}}\label{sec:robust}
\subsection{Numerical convergence} \label{sec:convergence}
\begin{figure}
\centering
\includegraphics[trim= 0.5cm 1.35cm 0.0cm 0.2cm,clip=true,width=\columnwidth]{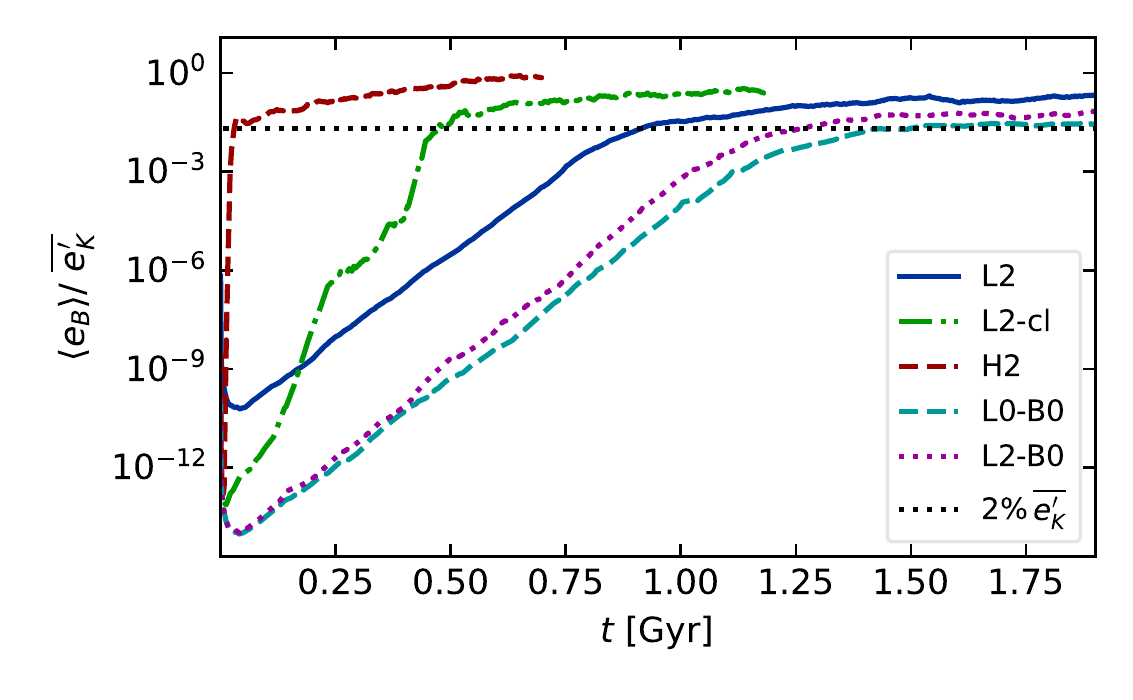}
\includegraphics[trim=-0.3cm 1.35cm 0.0cm 0.2cm,clip=true,width=\columnwidth]{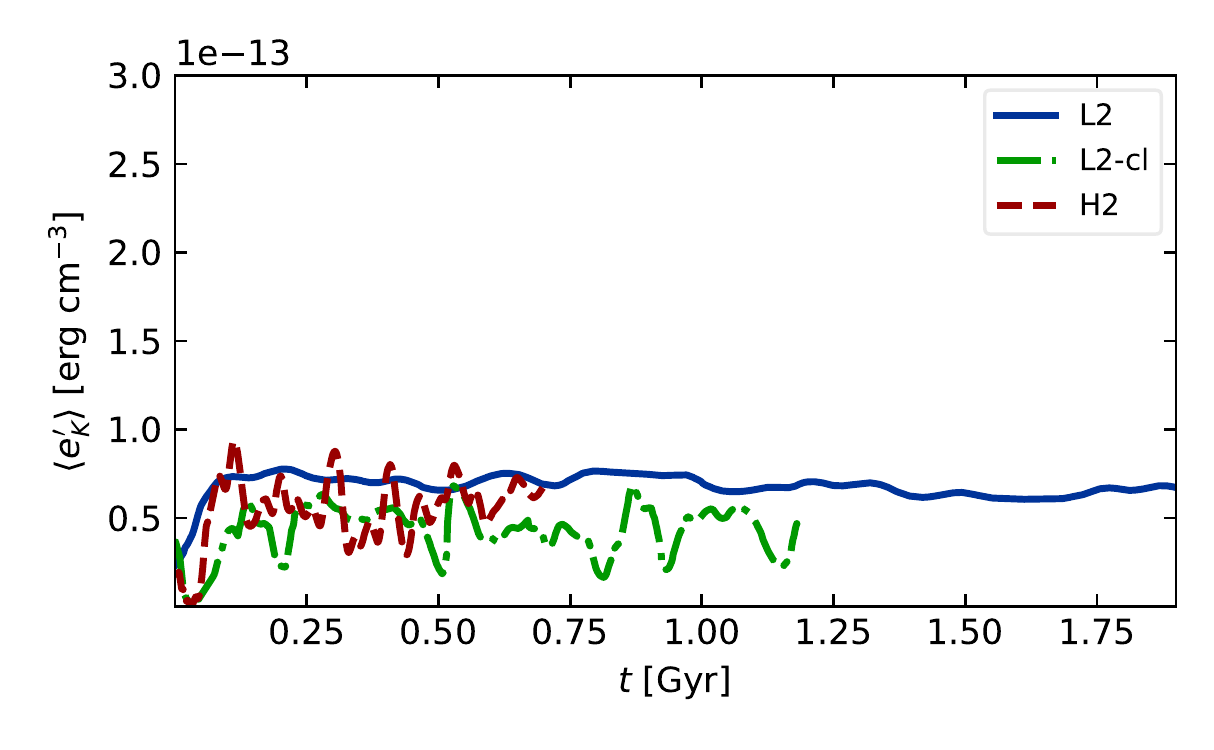}
\includegraphics[trim=-0.2cm 0.50cm 0.0cm 0.2cm,clip=true,width=\columnwidth]{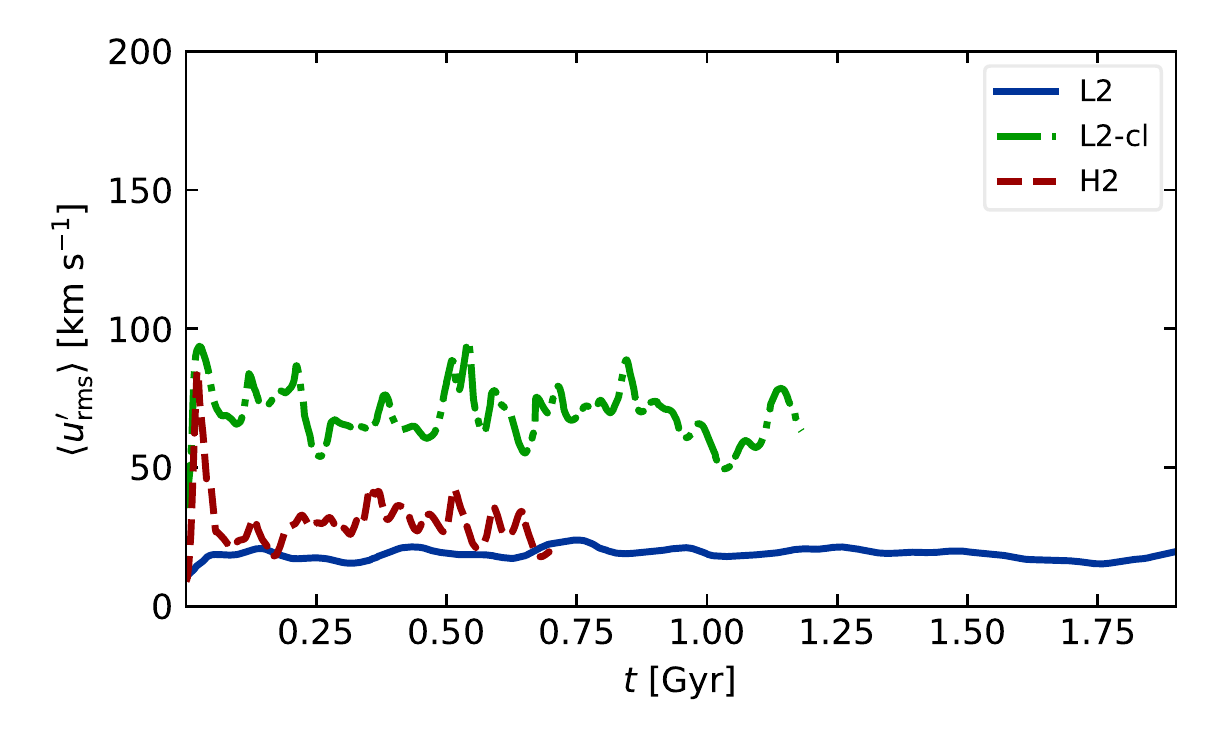}
 \begin{picture}(1,0)
    \put(-120,375){{\sf\bf{(a)}}}
    \put(-120,245){{\sf\bf{(b)}}}
    \put(-120,135){{\sf\bf{(c)}}}
  \end{picture}
\caption{
\emph{(a)} Volume-averaged magnetic energy density for all models listed in the
legend, normalised by the time-averaged turbulent kinetic energy density
$\eKt$.   \emph{(b)} Volume-averaged turbulent kinetic energy density for the
LSD models, and \emph{(c)} volume-averaged root-mean-square turbulent velocity
$u^\prime$.
\label{fig:eBall}
}
\end{figure}
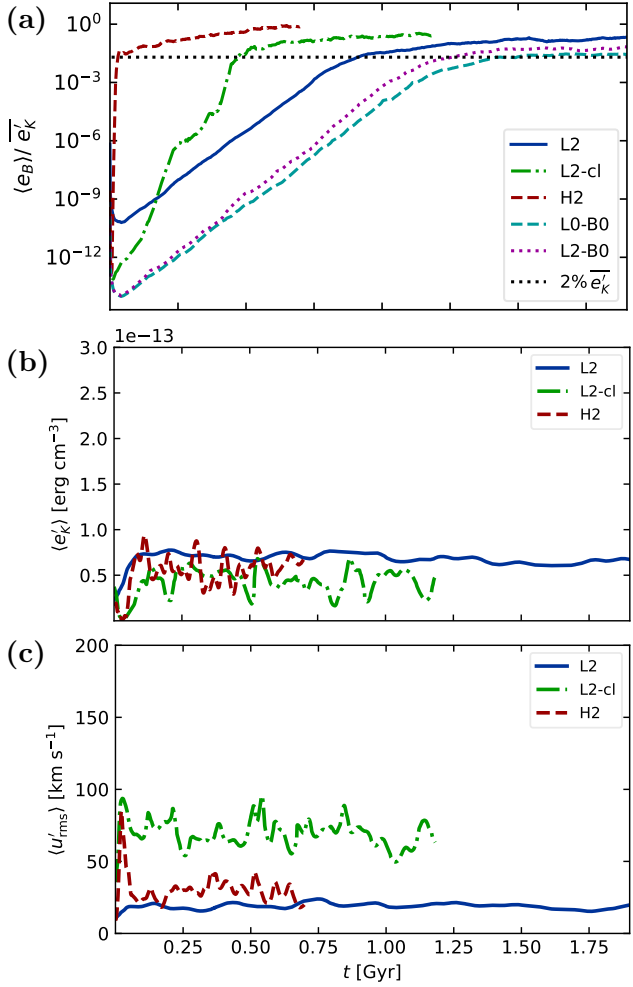

We next study the numerical convergence of the combined dynamos.  We first
consider the convergence behavior of the SSD.  Figure~\ref{fig:eBall}\emph{(a)}
shows the extreme sensitivity of the growth rate of the SSD to numerical
resolution {\citep[as reported][]{GMKS21,GMKS22}. In} the high-resolution
Model~\HA\ {an eigenmode is established at 7~Myr,} saturating {26} Myr
{later.  By contrast, the low-resolution models \LA\ and \LD\ take up to
40~Myr to establish the eigenmode, then 900 and 500 Myr, respectively, to
saturate.}

The SN rate in models \LA\ and \HA\ are 2.5 times higher than modelled in
\citet{GMKS22} and the density is stratified, facilitating even higher velocity
and sound speed.  {In \citet{GMKS22} the high and ultra high resolution SSD
established an eigenmode at 40~Myr and then saturated within 17 Myr. More
realistic FUV heating allows the warm gas to cool, which increases the
fractional volume of hot gas. It is likely therefore, given higher SN rates
that the duration of the SSD would reduce further to a few megayears or even
shorter.}

{{The kinematic growth rate $\gamma_B$ differs by a factor of about 50
between model \HA\ and \LA.  Theory predicts $\gamma_B\propto\Rey$ with some
additional weak dependence on $\ln(\Rm)$ \citep{KRS86,Kandu97} and inverse
dependence on $\Ms$ \citep{MossS96,F16}.}}

{{
Without explicit diffusivities, using a second order solver, \citet{TFD06} find
that their effective $\Rey$ scales inversely with the square of grid
resolution.  For high resolution \citet{GMKS21} show the Lagrangian
diffusivities used here dominate, while at low resolution numerical
diffusivities dominate.  Their direct comparison to second order solutions of
\citet{BKMM04} demonstrates that numerical diffusivity reduces substantially at
sixth order.  The ratio of numerical $\Rey$ between \HA\ and \LA\ could
therefore be as high as $4^6=4096$, which would more than account for a factor
of 50 in $\gamma_B$.  }}

{{ISM turbulence is multi-scale due to random explosions and variation
in remnant location and size. It is strongly affected by nonadiabatic cooling.
To test the applicability of theoretical predictions a dedicated parameter
study of SSD at resolution $\delta x\leq 1\pc$ is required.}}

In order to study the numerical convergence of the LSD {solution} after
saturation of the SSD, the large variation in time of SSD saturation needs to
be taken into account{.  The} saturation energy of the magnetic field
{is anyway well} converged.  {For} the period that the LSD dominates
field growth, Model~\HA\ {is} shifted by 970~Myr to align the LSD
solutions.  (Similarly, shifting the dynamo solution for Model~\LA\ illustrated
in Figure~\ref{fig:eBall}(a) by 365 Myr aligns it with the SSD solutions
{of} Models~\LC\ and \LD\ for {Figure~\ref{fig:lsd-ssd-kspec} in}
Section~\ref{sec:LSD} above.)

The turbulent velocity of model \HA\ (Fig.~\ref{fig:eBall}\emph{(c)}) does
slightly exceed that of \LA, but the main driver of the high rate of SSD at
high resolution is {that the filamentary dense structures occupy a lower
fractional volume, reducing net cooling and increasing the fractional volume of
hot diffuse} turbulent structures.  The {total} turbulent kinetic energy
density (Fig.~\ref{fig:eBall}\emph{(b)}) is well converged at the resolution of
these models, as is the saturation strength of the SSD.

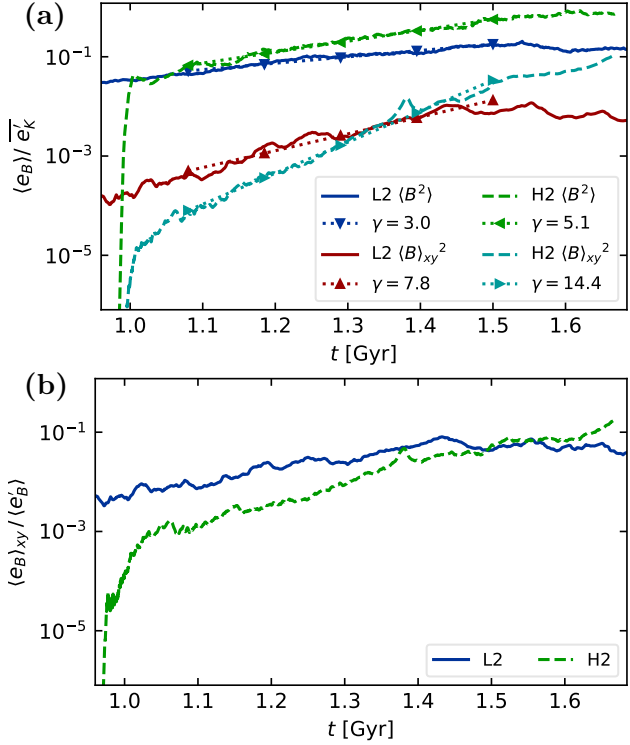
\begin{figure}
\centering
\includegraphics[trim=0.5cm 0.5cm 0.0cm 0.2cm,clip=true,width=\columnwidth]{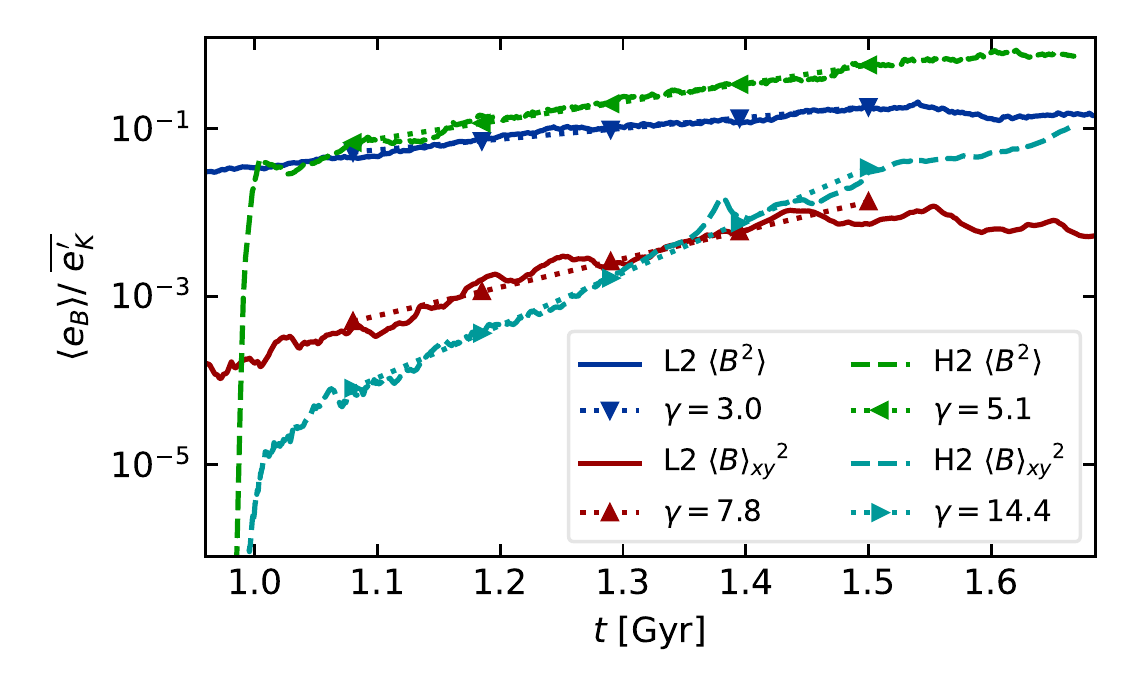}
\includegraphics[trim=0.5cm 0.5cm 0.0cm 0.2cm,clip=true,width=\columnwidth]{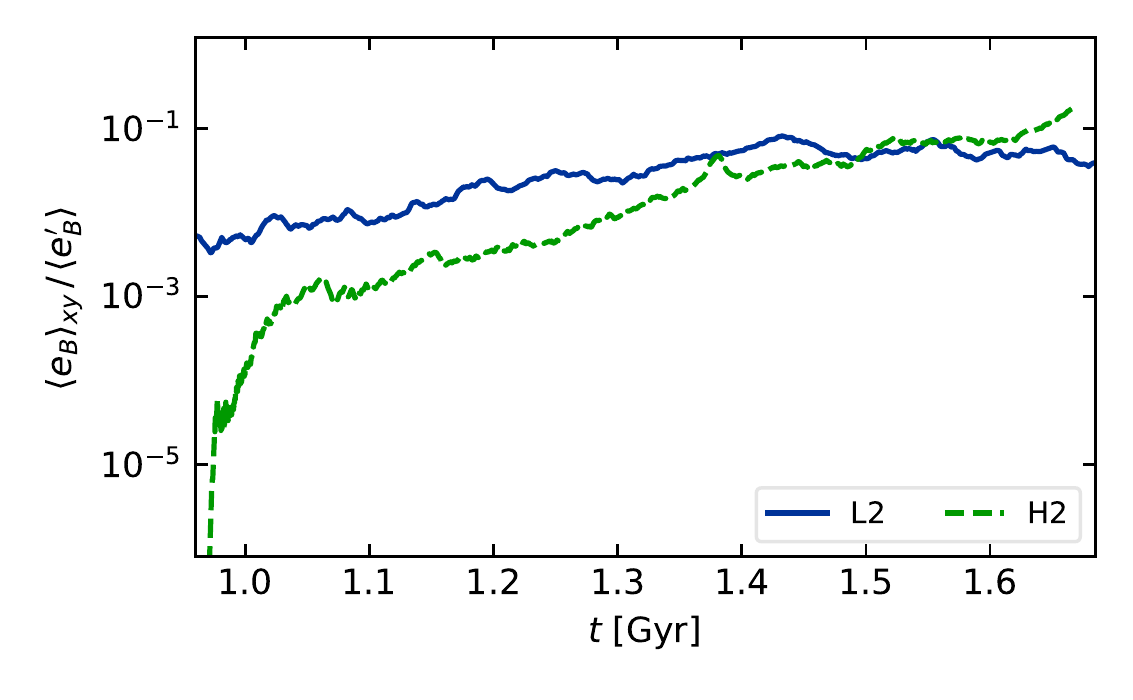}
 \begin{picture}(1,0)
    \put(-113,285){{\sf\bf{(a)}}}
    \put(-113,145){{\sf\bf{(b)}}}
  \end{picture}
\caption{
Volume averaged energy density \emph{(a)} of total magnetic energy
$\langle B^2\rangle(8\uppi)^{-1}$ and horizontally averaged mean-field energy
${\langle B \rangle_{xy}}^2(8\uppi)^{-1}$, normalised by time-averaged
turbulent kinetic energy density $\eKt$ for Models~\LA\ and \HA\ (with the
latter shifted 0.97~Gyr to align saturation of their respective SSDs).
Exponential fits of growth rates, as listed in the legend, apply to the
interval $1.08\Gyr<t<1.5\Gyr$.  \emph{(b)} Ratio of horizontally-averaged
mean-field energy to turbulent magnetic field energy for same models and
intervals as \emph{(a)}.
\label{fig:H2OvL2O}
}
\end{figure}

Figure~\ref{fig:H2OvL2O}\emph{(a)} compares the total and horizontally-averaged
magnetic energy of Model~\LA\ to the time-shifted solution of Model~\HA.  The
growth rate at four times higher linear resolution is less than a factor of 1.5
higher. The amplitude of the magnetic energy, 500 Myr following saturation of
the SSD, differs by a factor of 3.8.

By shifting the solution displayed in Figure~\ref{fig:eBall} for \LB\ by
420~Myr the stage immediately following the saturation of the SSD aligns with
Model~\LA.  Taking this comparative interval of the LSD for Model~\LB\ as that
fitted for Model~\LA\ in Figure~\ref{fig:H2OvL2O} yields $\gamma=4\Gyr^{-1}$ vs
3 for the total magnetic energy and 11.1 vs 7.8 for the mean magnetic energy,
respectively (See also Appendix~\ref{subsec:appen-density}).

Figure~\ref{fig:H2OvL2O}\emph{(b)} shows the ratios obtained from the values in
panel \emph{(a)}, which are $\lesssim0.1$ and depend more on the evolution of
the mean-field energy than the resolution of the SSD. Estimates from
observation place the ratio of mean to turbulent magnetic energy at 0.1--1
\citep[e.g.,][]{TKFB08, Houde13, Haverkorn15, Beck15, BCEB19}. In the later
stages the proportion of turbulent energy is even higher for Model \LA\ than
\HA. However, this is explained by the increased efficiency of the LSD in Model
\HA\, which remains well below saturation, so the system is not yet in a
statistical steady state.

In Appendix~\ref{subsec:appen-conv} comparisons of spectra from equivalent
times and time-latitude diagrams of the magnetic fields during this
corresponding interval illustrate the convergence in the structure and topology
of the magnetic field.  At this early stage of the LSD the azimuthal field is
predominantly negative and has similar scale height independent of resolution,
although it later evolves, at least in the low-resolution models, to be
positive and symmetric.  Computationally, the LSD solution for \HA\ requires a
long enough integration time that it has not yet been determined whether the
similarity in the LSD solution  to the low-resolution model extends to the
later stages of the dynamo.

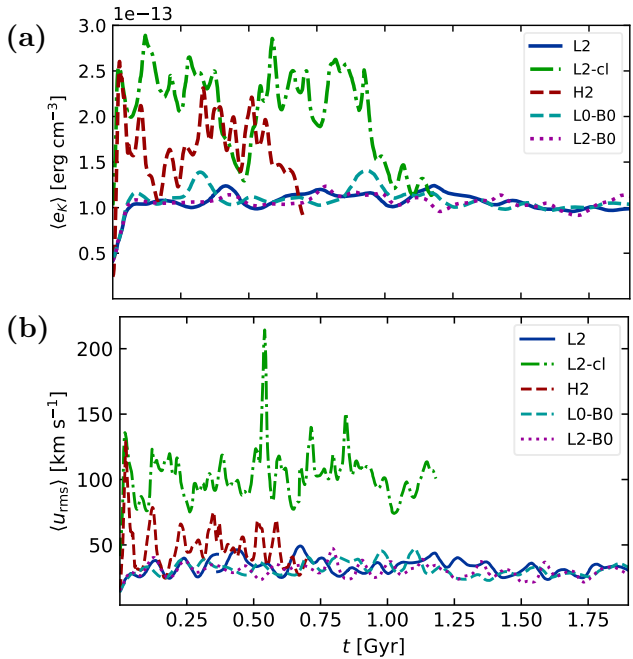
\begin{figure}
\centering
\includegraphics[trim=-0.3cm 1.35cm 0.0cm 0.2cm,clip=true,width=\columnwidth]{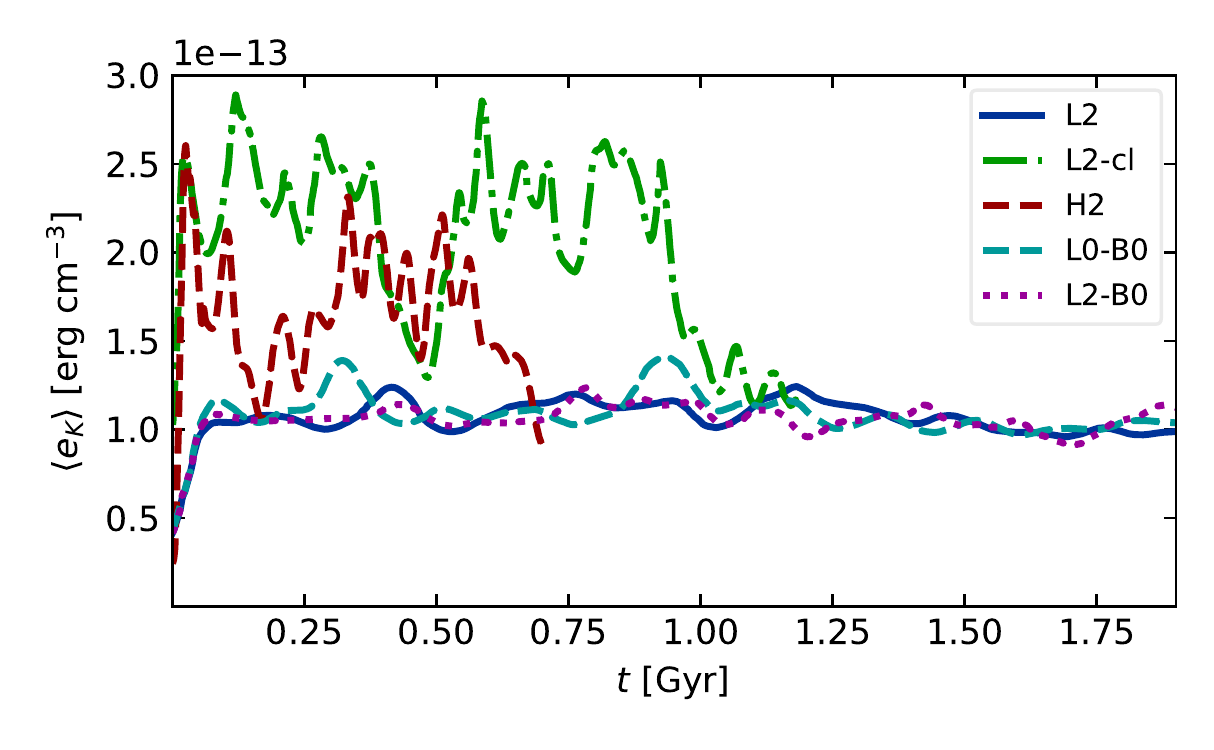}
\includegraphics[trim=-0.2cm 0.50cm 0.0cm 0.2cm,clip=true,width=\columnwidth]{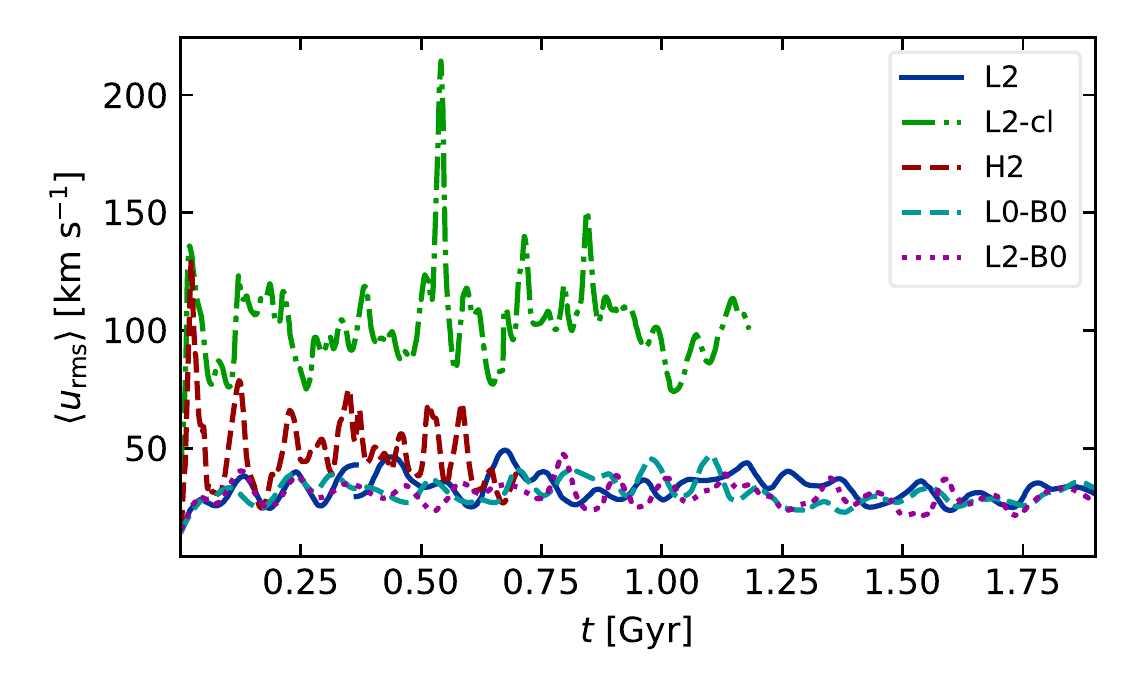}
 \begin{picture}(1,0)
    \put(-120,245){{\sf\bf{(a)}}}
    \put(-120,135){{\sf\bf{(b)}}}
  \end{picture}
\caption{
Volume-averaged total kinetic energy density $\eKt$ \emph{(a)} for all models
as listed in the legend and \emph{(b)} volume-averaged root-mean-square total
velocity $u$ for the same models.
\label{fig:eKt}
}
\end{figure}

\subsection{{Hydrodynamics}} \label{sec:hydro}
In Figure~\ref{fig:eKt}(a) we see that OB clustering supports higher total
kinetic energy, but that this primarily results from stronger large-scale
motions as the turbulent contribution to the kinetic energy, plotted separately
in Figure~\ref{fig:eBall}(b), is actually weaker than for Model~\LA.  By
contrast the turbulent velocity in Figure~\ref{fig:eBall}\emph{(c)} has a
higher magnitude (Fig.~\ref{fig:eKt}\emph{(b)}) when OB clustering is included,
because the high turbulent velocities are in regions of low density and thus
contribute little to the kinetic energy.  \citet{GMKS22} have demonstrated that
it is the flow velocity, rather than the kinetic energy, that is critical to
the SSD growth rate in the multiphase ISM, while saturation strength is related
to the kinetic energy and is thus independent of the inclusion of clustering.

For Model~\LB\ dips in total kinetic energy can be traced to the blowout of gas
occurring in the northern latitudes at these times.  We include in
Appendix~\ref{subsec:appen-density} time-latitude diagrams of the gas density
and azimuthal magnetic field to contrast the topology of the gas and resultant
field between Models~\LA\ and \LB.

\section{{Discussion}}\label{sec:disc}

\subsection{Mean-field decomposition}\label{sec:MFdecomp}

When considering mean-field dynamo theory it can be helpful to decompose the
field into the mean and fluctuating fields subject to a large separation of
scales. In practice, however, it is instead common in simulations, experiments,
and observations to find a continuum of power across scales without any clear
minimum in the spectrum between large and small scales.  Horizontal averages
have an arbitrary dependence on the choice of domain, while kernel averaging
can rely on an arbitrary smoothing scale.  Following \citet{Gent:2013a,HSSFG17}
we adopted a Gaussian smoothing scale of $\ell=50\pc$ to separate the mean and
random magnetic fields for $\langle B\rangle_\ell$ in
Figure~\ref{fig:eB-gamma}.

Our analysis of the temporal evolution of the spectrum provides a physically
motivated criterion for separating the mean field from the turbulent field.
This depends on their sensitivity, or lack thereof, to the action of the LSD.
Truncating the spectrum of the magnetic field at $k\geq17k_1$ allocates to the
mean-field only those scales sensitive to the LSD, while masking the spectrum
at $k<17k_1$ allocates to the random field those scales insensitive to the LSD.

\subsection{OB clustering}\label{sec:OB}

In a horizontal domain of 1~kpc we find that simulations including clustering
of OB stars in associations can rapidly develop thermal runaway \citep{KO15} on
either side of the midplane. {Asymmetry arises in SN-driven turbulence due
to sporadic random clustering of explosions on one side of the midplane.  This
is enforced by OB-clustering, which ensures an extended period when subsequent
explosions will reoccur on that side. This is a physical behaviour, but across
the span of the disc would tend to average out, leaving a pocklike imprint over
its gas surface.}

The expanded fractional volume of hot gas{, in this particular instance
above the midplane for Model~\LB,} is beneficial to the SSD
\citep{GMKS22}{. However, it} does not favour the LSD, as can be seen by
comparing {the spread of the mean magnetic field to large $|z|$ in}
Figure~\ref{fig:tav-L2O}(c), while it is confined to the {midplane with the
warm gas for Model~\LB\ (see Appendix~\ref{subsec:appen-density})}.
Figure~\ref{fig:eBall}(a){, and comparison of their mean fields in
Appendix~\ref{subsec:appen-density}, show that after saturation of the SSD} the
{mean field} actually grows more efficiently in the thinner, more dense
disc of the OB-clustered model {\LB\ }than {\LA\ }without.

{\citet{GMKS22} show sporadic rapid growth in  the diffuse hot gas during
the kinematic SSD, but upon saturating the turbulent field becomes increasingly
correlated with the gas density.} Both {absence of mean field in the hot
gas and increased LSD growth rate associated with a more uniform warm phase}
support the hypothesis that the mean-field dynamo grows mainly in the warm
rather than the hot gas{, but could also indicate correlation of the mean
field with mean gas density}.  {\citet{EGSFB16} demonstrate that the
turbulent field is proportionate to gas density in both warm and hot phases,
but the mean field preferentially resides in the warm gas.}

\subsection{Quenching of the \texorpdfstring{$\alpha$-effect}{alpha effect}}\label{sec:aq}

We have used kinetic and current helicities as proxies for measuring the
collective inductive action of SN turbulence, called the  $\alpha$-effect. A
more thorough analysis, yielding also other important turbulent transport
coefficients, would require the usage of a suitable test field method
\citep[see][]{KBR22}, but due to significantly increased computational costs,
such analysis is not performed here. {Singular value decomposition
\citep{RCGBS11,BSEG20} may also be applied retrospectively, although the
presence of SSD in our models needs to be taken into account.}

As Model~\LA\ approaches equipartition in Figure~\ref{fig:av-alpha}(a) we
observe $\alpha$-quenching near the midplane, with magnetic helicity of
opposite sign suppressing the dynamo.  While the LSD is active, small-scale
opposite sign helicity is removed from the disc by the outflows. At large
$|z|$, the $\alpha$-effect remains non-zero, with magnetic and kinetic helicity
typically of the same sign. However, both magnetic and kinetic energy density
are extremely weak relative to the disc.

It is possible that the reversal of the midplane field at around 2~Gyr is an
indication of the nonlinear magnetic buoyancy instability. The coincident
change of sign of $\alpha$ in the northern latitudes, which is a magnetic
$\alpha$-effect, is a signature of this behaviour as identified by
\citet{Machida13,DT2022b}, and \citet{QSTGB23}.  Subsequent reversals do not
occur, as the scale height of the magnetic field expands, reducing the
vulnerability of the system to magnetic buoyancy instability.

\section{Conclusions}\label{sec:conc}

We have verified that the large-scale dynamo (LSD) in disc galaxies is robust
in the presence of SSD at high magnetic Reynolds {number }Rm{. (This
varies strongly across the domain, reaching values as high as $\sim10^5$ in hot
regions \citep[see][]{GMKS22}.)  The LSD} saturates in approximate
equipartition with the kinetic energy density.  However, the LSD growth rate
appears to be impeded by the presence of a vigorous small-scale dynamo (SSD),
in comparison with the LSD in cases where the SSD is weaker or absent
altogether \citep{Korpi:1999b, Gressel:2008, Gent:2013a, Bendre2018}.

{\citet[][see their Figure~2]{CS17} seek to relate the growth of the mean
galactic field to various {ratios of small-scale magnetic to turbulent
kinetic energy density.} Our normalised growth rate for model \LA\ would be
$\sim3.5$ vs 4.5 \citep{Gent:2013a}, which are slightly higher values than the
upper estimate from \citet{CS17} of $\sim2.5$ for $\Omega=2\OSN$. The increased
mean field growth with reduced {ratios} is however consistent with their
model.}

In contrast, the SSD appears quite insensitive to the presence of an LSD.
{Given that the peak in the power spectrum of Model \LA\ remains at a scale
below the range of forcing scales provided by SNe, we can be confident the
magnetic energy growth is attibutable to SSD rather than tangling of the
mean-field \citep[][their Figure~1]{GMKS21}. Tangling injects peak energy at
the scale of the forcing.} The presence of large-scale rotation and shear does,
however, inject more vorticity, increasing the growth rate of the SSD
\citep{SRB17,GMKS22}. The increased growth rates apply primarily to larger
scales.

Unlike the case lacking LSD, the small scales of the magnetic field do continue
to grow after the SSD has saturated, but much more slowly than the large
scales, which might be explained by tangling of the growing large-scale field.
For $k\lesssim17k_1$, before the SSD saturates the self-similar growth ceases
and growth at the largest scales slows down.  After the LSD becomes dominant,
the larger scales overtake the energy at $k\simeq17k_1$ and the growth at these
smaller scales slows even further, until eventually the largest scale contains
the peak energy and grows faster than any other scale.  This indicates that a
horizontal domain of 1~kpc is insufficient to identify the peak energy scale of
the mean magnetic field in a disc galaxy with an active LSD.

In the kinematic stage of the SSD, all scales in the magnetic field grow
self-similarly.  Upon saturation and in the absence of SSD, the hierarchy of
the magnetic field spectrum is preserved. The power peaks at a wavenumber
$k\simeq17k_1${, or 30~pc scale,} independent of resolution (or effective
Rm){. This} remains the case in the nonlinear phase of the {pure} SSD,
{even though the portion of total magnetic energy at scales larger than 30
pc increases at late times}.  When the LSD is present, the self similar growth
of the magnetic energy spectra continues in the same manner for all
$k\gtrsim17k_1$, through to saturation of the SSD and beyond.  This identifies
the turbulent field of the SN driven ISM as occupying structures organised on
length scales $\ell \lesssim 30\pc$.

{We have shown that the SSD in the ISM can be expected to saturate at a few
percent of equipartition within a few megayears, depending on the SN rate and
disc mass.  During the kinematic stage of the SSD the mean field is also
amplified at slightly above the self-similar growth rate of the SSD up to
$\sim10^{-5}$--$10^{-4}$ of equipartition.  The LSD in the presence of high Rm
SSD is slower relative to LSD at lower Rm, although its dependence across a
range of turbulent Rm remains to be explored.  The mean field in spiral
galaxies is likely to require more than 2 Gyr to saturate, depending on SN
rate, rate of shear and scale height, which also remains to be explored, and
its saturation strength depends on the sum of turbulent kinetic energy and
energy from the galactic shear. The properties of the saturated magnetic field,
including its mean-field fraction, topology, and sensitivity to parameters are
pending further investigation.}

{The total magnetic field already approaches equipartition with the
turbulent kinetic energy within 1 Gyr.  The turbulent magnetic field continues
to grow in the presence of the LSD at relative rates below that of the LSD, but
can increment magnetic energy orders of magnitude larger than the mean field
itself.  The alpha effect appears to support the LSD by transporting helicity
on small scales away from the midplane, and the magnetic alpha, which can
acquire magnitude comparable with the kinetic alpha within 400 Myr, initially
has the same sign and supports the LSD. In the later stages, as the mean field
continues to grow, the magnetic alpha reverses sign and quenches the LSD. A
more robust analysis of the turbulent transport processes is required.}

\begin{acknowledgments}
{We thank the anonymous referee for a thorough, constructive, and
intelligent review of the work, which has inspired substantial improvements to
the quality of the presentation.} F.A.G. and M.J.K.-L. acknowledge support from
the Academy of Finland ReSoLVE Centre of Excellence (grant 307411), the
Ministry of Education and Culture Global Programme USA Pilot 9758121 and the
ERC under the EU's Horizon 2020 research and innovation programme (Project
UniSDyn, grant 818665) and generous computational resources from CSC -- IT
Center for Science, Finland, under Grand Challenge GDYNS Project 2001062.
F.A.G. benefited from in-depth discussions at ``Towards a Comprehensive Model
of the Galactic Magentic Field'' at Nordita 2023 supported by NordForsk.
M-M.M.L. was partly supported by US NSF grants AST18-15461 and AST23-07950, and
thanks the Inst.\ f\"ur Theoretische Astrophysik der Uni.\ Heidelberg for
hospitality.
\software{Pencil Code \citep{brandenburg2002,Pencil-JOSS}}
\end{acknowledgments}

\bibliographystyle{aasjournal}
\bibliography{refs}{}
\appendix

\section{Supplementary material}\label{sec:supp}
\subsection{{Spectral comparison of SSD and LSD}}\label{subsec:SSD-spectra}

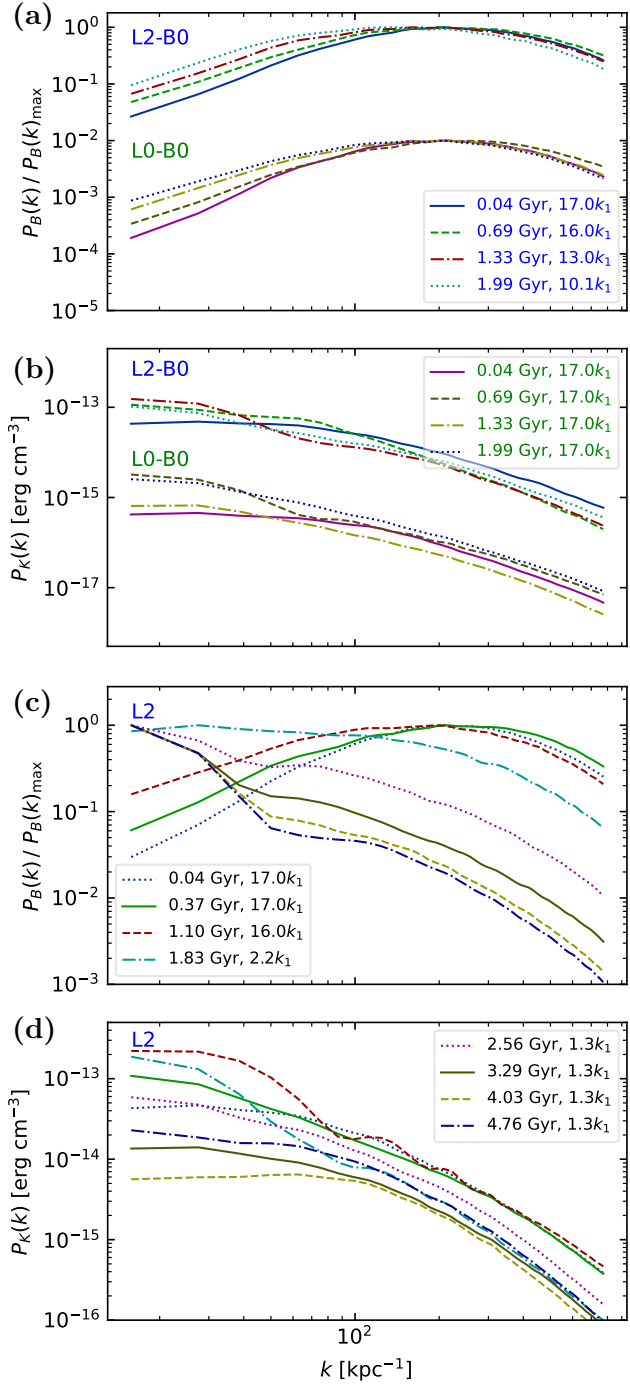
\begin{figure}
\centering
\includegraphics[trim=0.5cm 1.6cm 0.0cm  0.0cm,clip=true,width=\columnwidth]{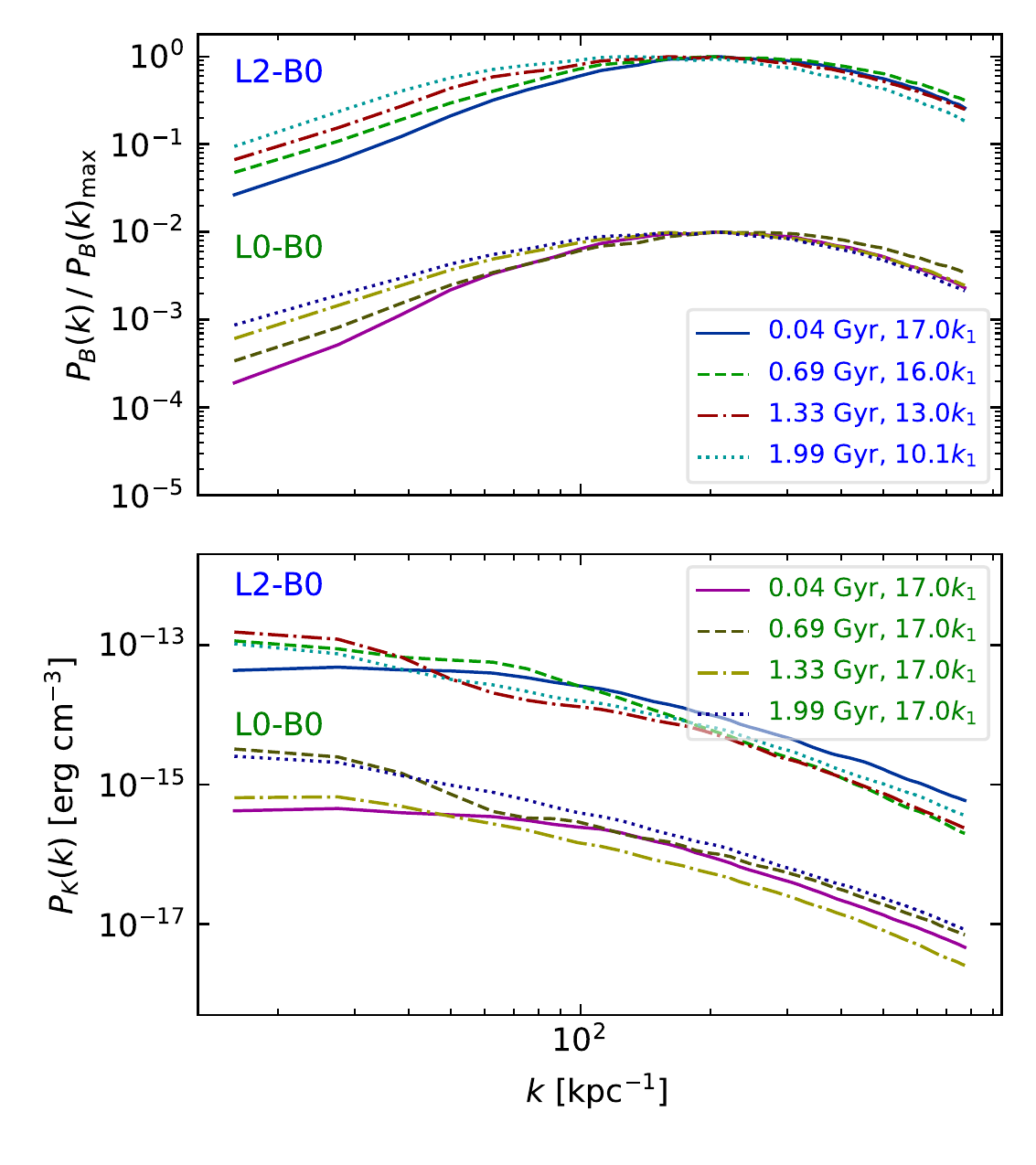}
\includegraphics[trim=0.5cm 0.5cm 0.0cm -0.2cm,clip=true,width=\columnwidth]{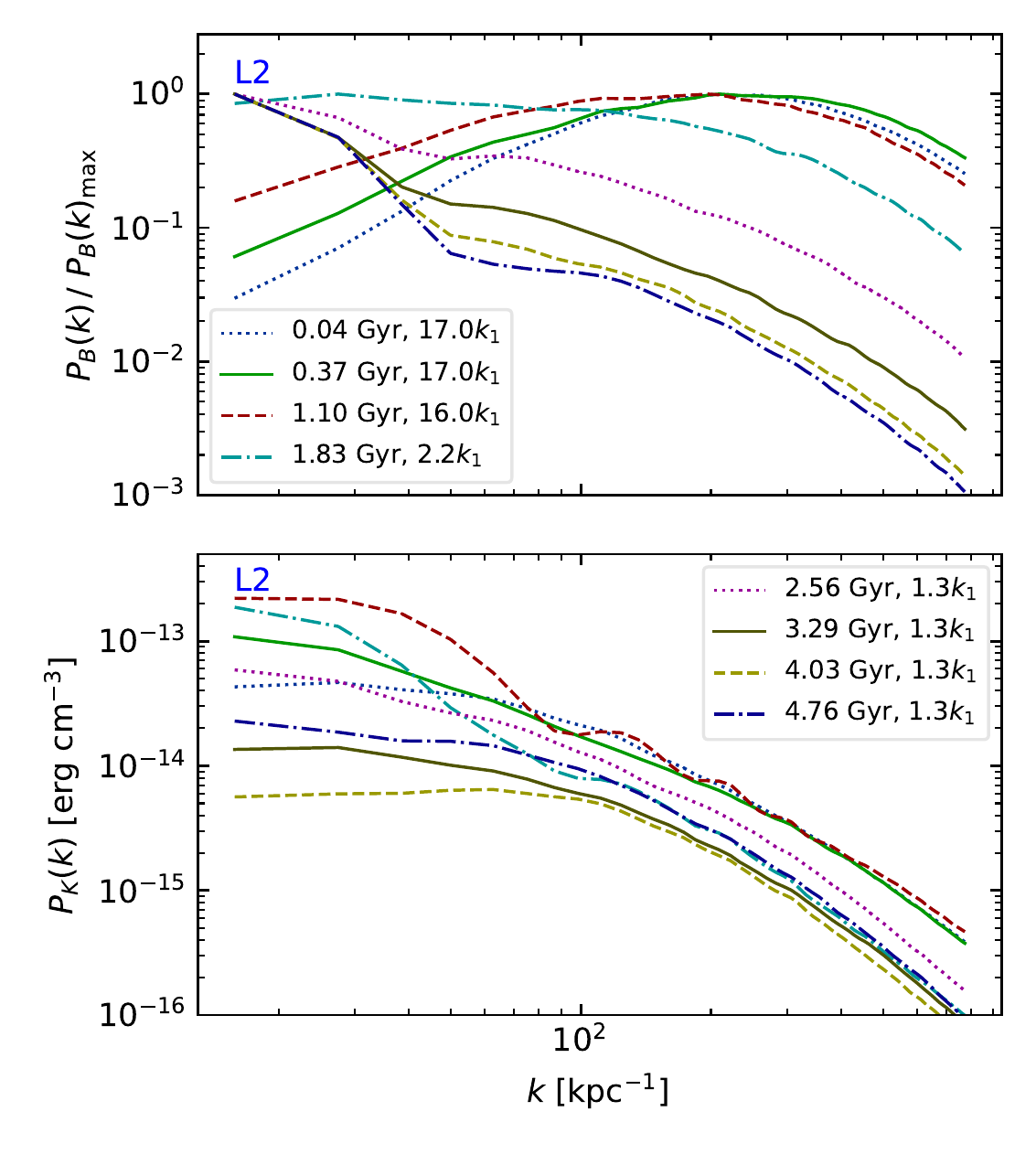}
 \begin{picture}(0,0)
    \put(-118,530){{\sf\bf{(a)}}}
    \put(-118,396){{\sf\bf{(b)}}}
    \put(-118,270){{\sf\bf{(c)}}}
    \put(-118,146){{\sf\bf{(d)}}}
  \end{picture}
\caption{
{Magnetic energy spectra \emph{(a)} in pure SSD Models~\LC\ and \LD\ and
\emph{(c)} the fiducial Model~\LA, each normalised by its maximum power.  The
spectra for Model~\LC\ are shifted $10^{-2}$ to differentiate from Model \LD.
Kinetic energy spectra \emph{(b)} and \emph{(d)} with matching models, line
styles, times (and axis shift) to panels \emph{(a)} and \emph{(c)},
respectively.} Listed wavenumbers $k$, in multiples of $k_1=2\uppi{L_x}^{-1}$,
locate the maximal magnetic power at each time.  Spectra have been smoothed
with a Gaussian kernel of length 5\,Myr {about time $t$} and 1.5\,$k_1$
{about wavenumber $k$}.
\label{fig:power-LSD}
\label{fig:ssd-kspec}
}
\end{figure}

Samples of the spectral properties of the magnetic energy and kinetic energy of
both pure SSD models are displayed in Figure~\ref{fig:ssd-kspec}{\emph{(a)}
and \emph{(b)}}. Spectra have been smoothed with a Gaussian kernel of length
5\,Myr and 1.5\,$k_1$ in time and wavelength{, with smoothing both forward
and backward in time}. The magnetic energy spectra, which grow many orders of
magnitude over time, are normalised by maximum energy density to ease
comparison between times.  {The equivalent spectra for Model \LA\ are
plotted in panels \emph{(c)} and \emph{(d)}.} In {all} models the peak
wavenumber of the magnetic energy during the kinematic stage of the SSD is at
$17 k_1$, with $k_1=2\uppi {L_x}^{-1}\simeq12.3\kpc^{-1}$.  This approaches the
resistive scale for these models, as expected for the SSD.

The statistically steady kinetic spectra are derived only from the velocity
$\vect{u}$, but are multiplied by the conserved volume-averaged mean gas
density to approximate the kinetic energy density.  The spectral properties of
the flow appear quite insensitive to the large-scale rotation and shear, but in
the saturated state the shear appears to excite a peak magnetic energy with
slightly larger scales than without.  After saturation of the SSD and
transition to the LSD, the magnetic energy shifts to larger scales
($k\rightarrow k_1$). The dominance of the largest scale increases over time
until the LSD saturates around 5\,Gyr.  The kinetic energy varies strongly over
time due to the sporadic and heterogeneous nature of the SN forcing, but the
dominant energy persistently remains at large scales $k<5k_1$.  {Later, as
the mean field becomes significant beyond 3~Gyr, energy is lost at large scales
with the spectrum at $k\lesssim5k_1$ becoming more shallow.}

\begin{figure}
\centering
\includegraphics[trim=0.5cm 1.34cm 0.0cm 0.0cm,clip=true,width=\columnwidth]{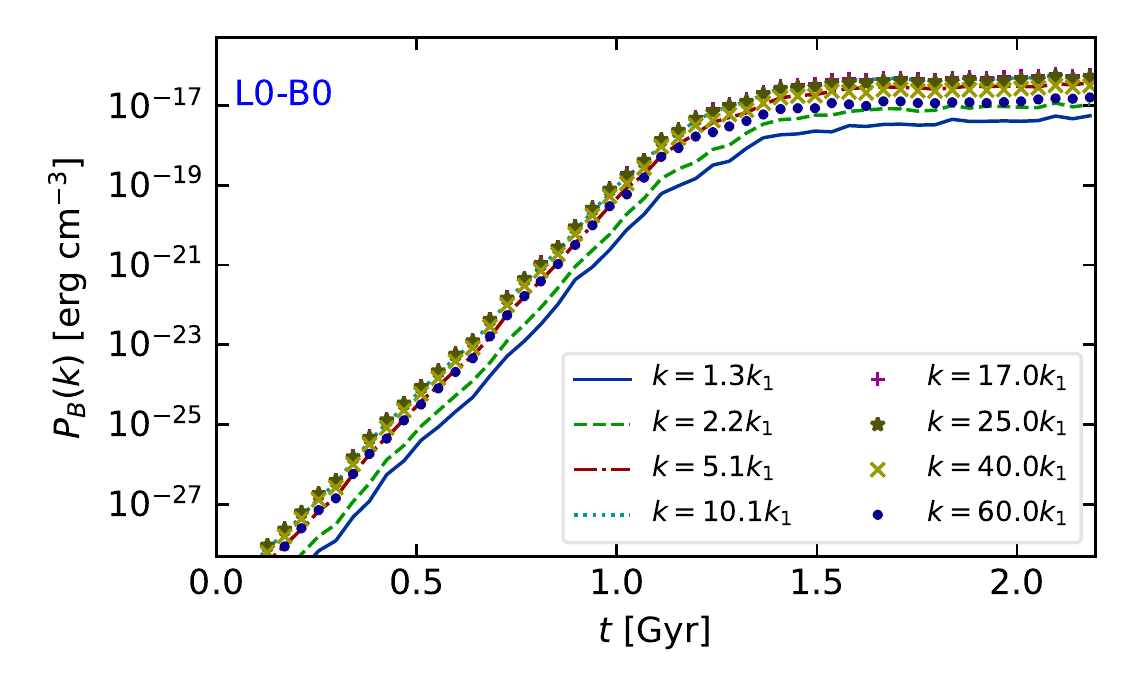}
\includegraphics[trim=0.5cm 0.5cm 0.0cm 0.0cm,clip=true,width=\columnwidth]{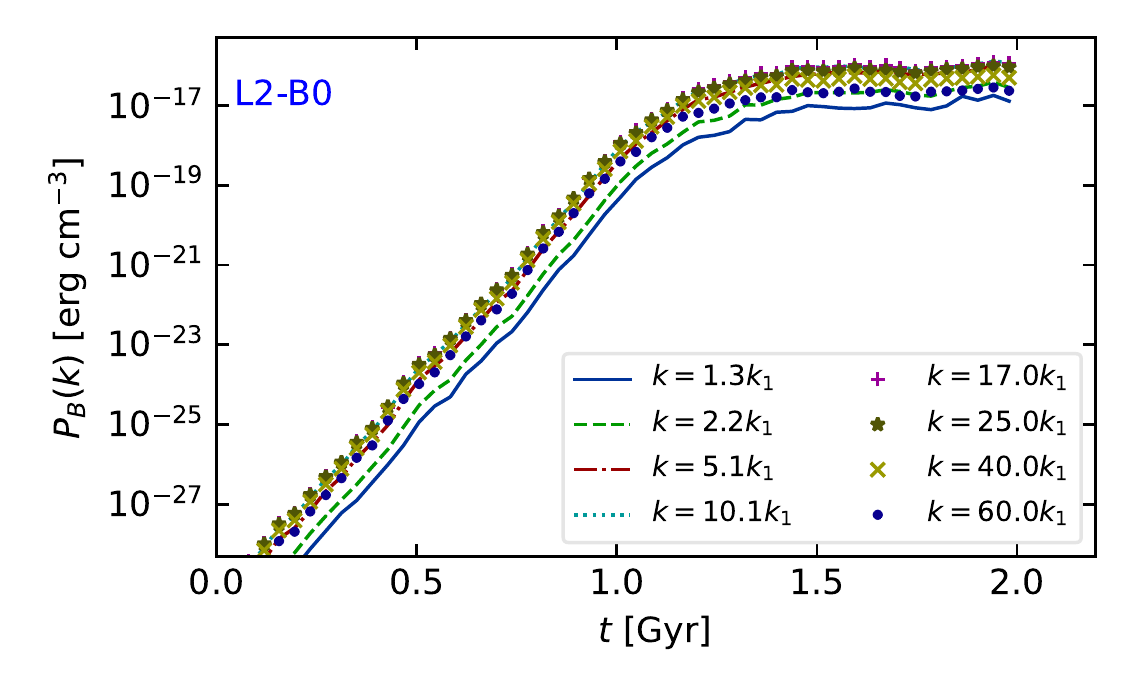}
\includegraphics[trim=0.5cm 0.5cm 0.0cm 0.0cm,clip=true,width=\columnwidth]{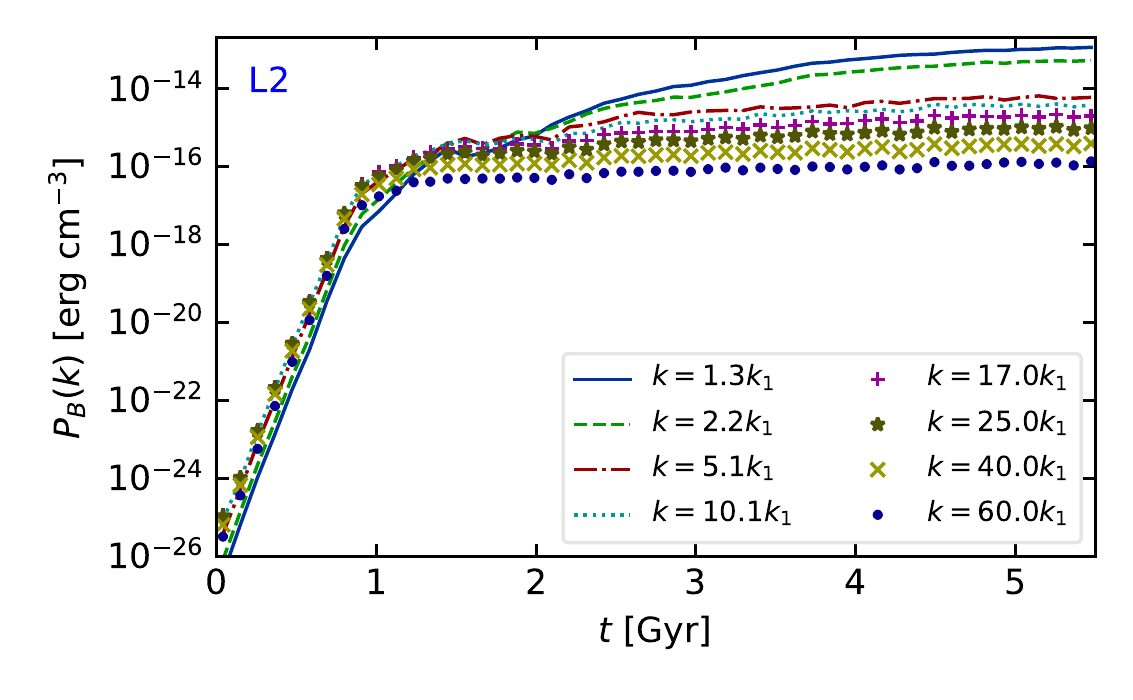}
 \begin{picture}(0,0)
    \put(-125,415){{\sf\bf{(a)}}}
    \put(-125,282){{\sf\bf{(b)}}}
    \put(-125,143){{\sf\bf{(c)}}}
  \end{picture}
\caption{
Time evolution of magnetic energy for SSD models \emph{(a)} \LC{,} \emph{(b)}
\LD\ {and \emph{(c)} \LA\ }at sampled scales of $k$, as noted in the legends in multiples of
$k_1=2\uppi {L_x}^{-1}$.
\label{fig:L2Otpower}
\label{fig:ssd-tspec}
}
\end{figure}
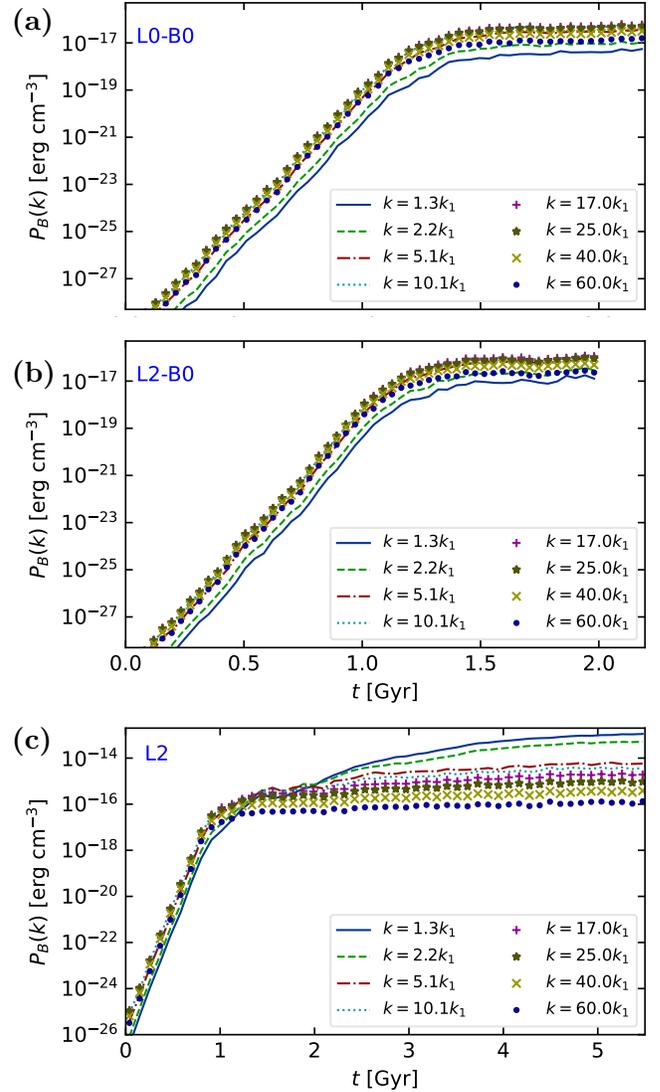

To examine the evolution of the magnetic field {energy spectra} over time,
we plot the power in selected wavenumbers of {the pure SSD models} in
Figure~\ref{fig:ssd-tspec}{\emph{(a)} and \emph{(b)}}.  Both models follow
an eigenmode growing at the same rate at all wavelengths.  The ranking of the
power in each wavenumber $k$ is quite rigid throughout for both models.  These
results show that the SSD behaviour and structure is {quite insensitive to}
the large-scale rotation and shear, other than to increase the growth rate and
saturation energy of the SSD.

In Figure~\ref{fig:L2Otpower}{\emph{(c)}} the evolution of the energy in
wavenumber bins with $k\geq17k_1$ is insensitive to the LSD and retains the
same {ranking} as for the SSD in {panels \emph{(a)} and \emph{(b)}}.
In contrast, wavenumber bins with $k<17k_1$, which had lower energies during
the SSD, {first detach from the eigenmode shortly before 1~Gyr, slowing
down. However, when the SSD saturates shortly after 1~Gyr these wavenumbers
continue to gain energy, eventually} becoming dominant over the larger
wavenumber bins with $k>17k_1${. During this transition up to about 2~Gyr
the ranking within $k<17k_1$ becomes inverted, while at smaller scales the
ranking from the kinematic stage is conserved.} Thus, the LSD can be identified
as acting primarily on scales $k\lesssim17k_1$.  However, in the nonlinear
stage of the LSD, even the smallest scales of the magnetic field increase in
energy, tenfold compared to the pure SSD models, with {even} $k=60k_1$
exceeding $10^{-16}$ erg cm$^{-3}$.

{Determining how much the pivot point of $k=17k_1$, corresponding to a
scale of 30 pc, depends on parameters such as supernova rate, galactic shear,
the mass of the disc, or the Reynolds numbers would require a parameter sweep
beyond the scope of this study. We hypothesise that this qualitative behaviour
will provide a useful condition to identify the separation between mean and
fluctuation scales, given that spectra from simulations typically do not
exhibit minima that could identify separation between mean and fluctuation
scales.}

{The inversion of rank in energy for $k<17k_1$} mirrors contrasting growth
in Figure~\ref{fig:eB-gamma} between $\langle e_B\rangle_{xy}$, consisting of
only the longest length scale, and $\langle e_B \rangle_\ell$, comprising a
range of intermediate scales.  The growth of $\langle e_B\rangle_{xy}$ slows
down earliest, but it subsequently grows more rapidly than $\langle e_B
\rangle_\ell$.

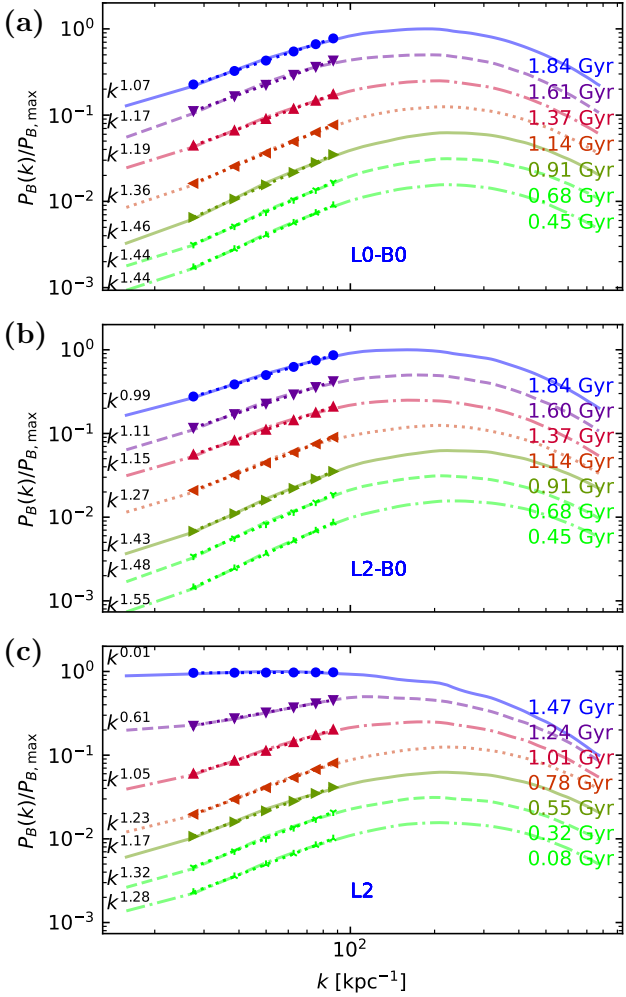
\begin{figure}
\centering
\includegraphics[trim=0.2cm 0.6cm 0.0cm 0.2cm,clip=true,width=\columnwidth]{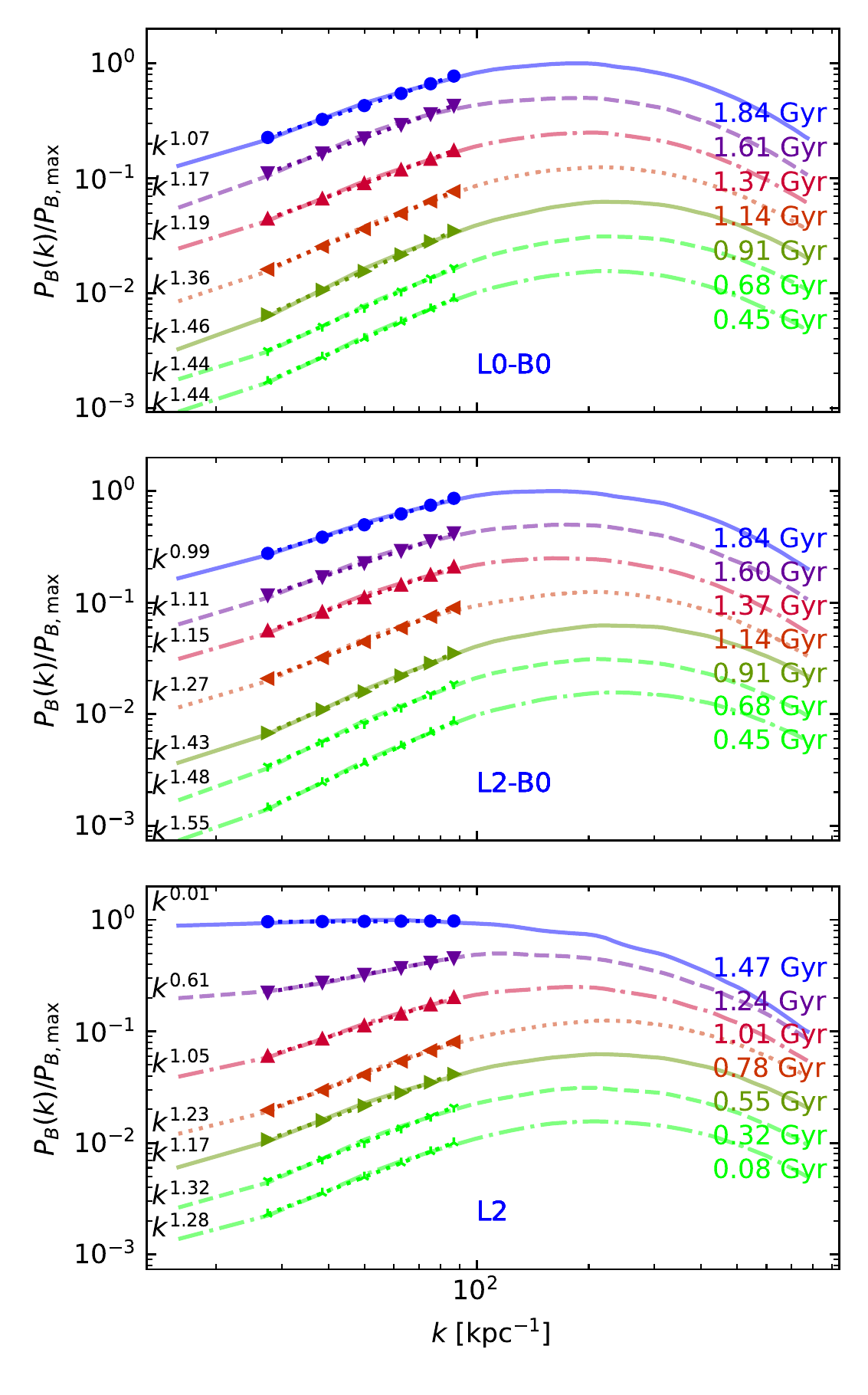}
 \begin{picture}(1,0)
    \put(-119,378){{\sf\bf{(a)}}}
    \put(-119,260){{\sf\bf{(b)}}}
    \put(-119,140){{\sf\bf{(c)}}}
  \end{picture}
\caption{
Energy spectra  for\emph{(a)} Model~\LC, \emph{(b)} \LD\ and \emph{(c)} \LA.
Power law fits are shown for times indicated (not time shifted).  Each
spectrum is normalised to its own maximum, but shifted by a small factor to
differentiate each time, while adequately comparing spectral changes that
differ by a factor of $10^{10}$ over time.
{Spectra have been smoothed with a Gaussian kernel of length 5\,Myr {about time
$t$} and 1.5\,$k_1$ {about wavenumber $k$}.}
\label{fig:comp-powerB0}
}
\end{figure}

Figure~\ref{fig:comp-powerB0} displays evolving spectral structure for Models
\LC, \LD\ and \LA\ from the kinematic stage to saturation of the SSD.  Even in
the presence of the LSD, evolution is predominantly self-similar until the SSD
saturates, with scaling $\simeq k^{5/4}$ for Model~\LA\ between 0.08 and 0.78
Gyr in Figure~\ref{fig:comp-powerB0}\emph{(c)}, and $\simeq k^{3/2}$ for the
equivalent stages in \emph{(a)} and \emph{(b)}.  The steepening resistive tail
in Figure~\ref{fig:comp-powerB0}\emph{(c)} after the SSD saturates indicates
that LSD continues to pile energy into larger scales within the intertial
range, which is not so for \LC\ \emph{(a)} or \LD\ \emph{(b)}.  {Given the
mean field in the form of horizontal averages is removed from Model \LD\ and
yet the shear adds energy preferentially to scales at $k<17k_1$, this indicates
the mean field is \emph{not} fully expressed in this form.}

\subsection{Vertical density structure and effect of clustering}\label{subsec:appen-density}

\begin{figure}[h]
\centering
\includegraphics[trim=0.32cm 0.45cm 0.54cm 0.25cm,clip=true,width=1.05\columnwidth]{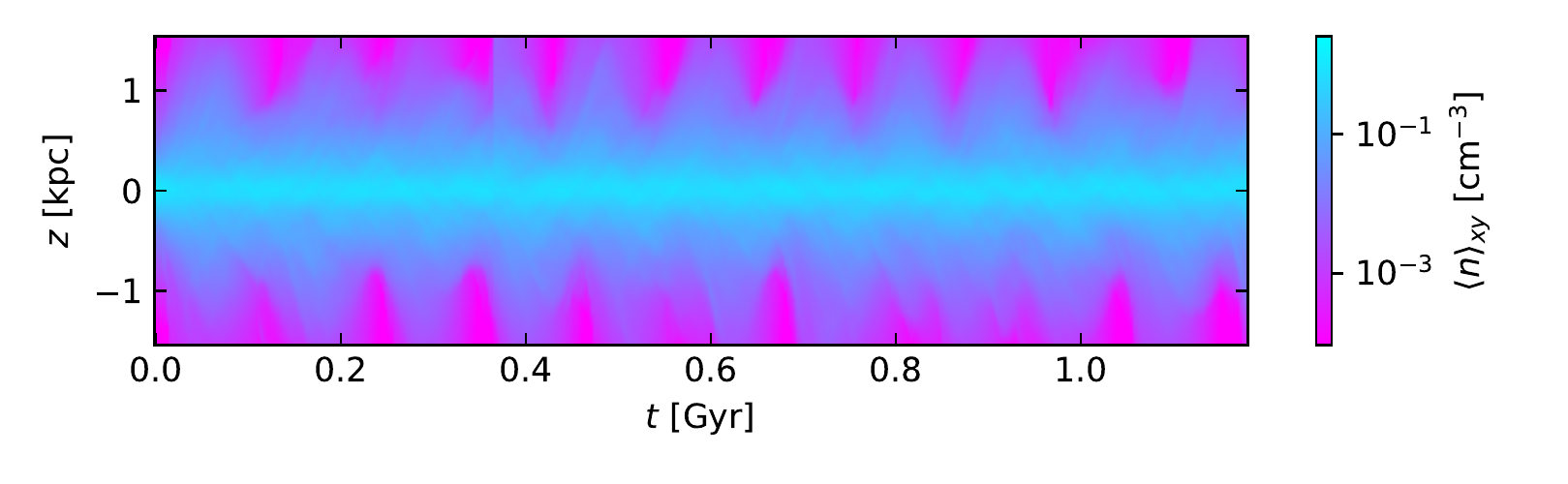}
\includegraphics[trim=0.32cm 0.45cm 0.54cm 0.25cm,clip=true,width=1.05\columnwidth]{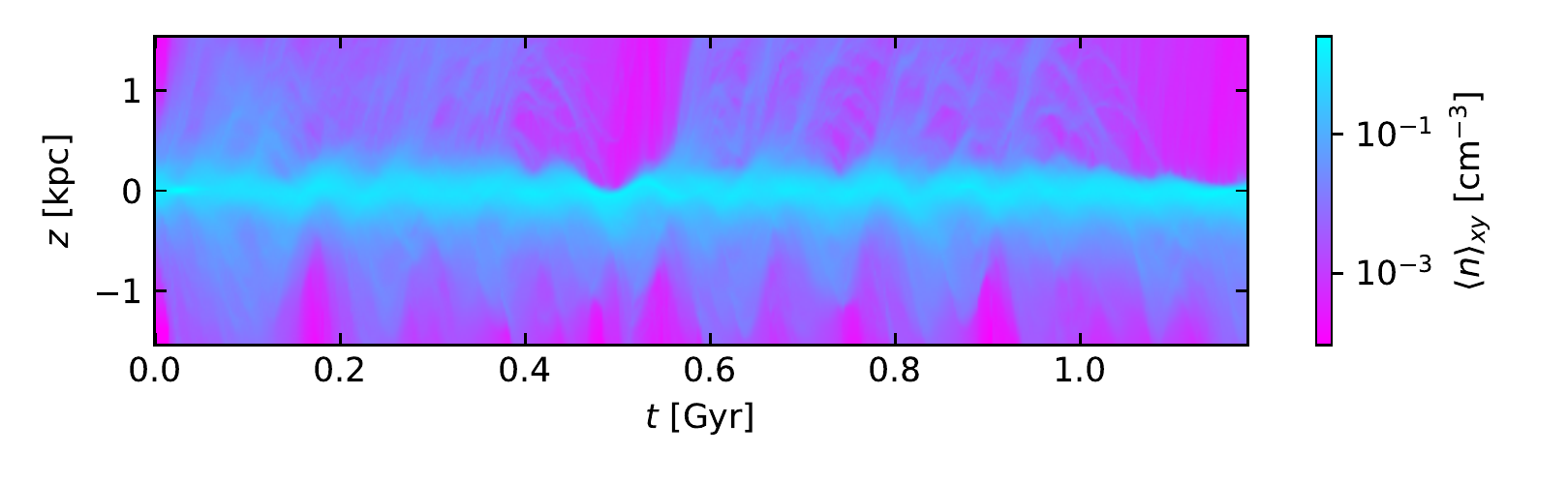}
\includegraphics[trim=0.28cm 0.45cm 0.74cm 0.25cm,clip=true,width=1.05\columnwidth]{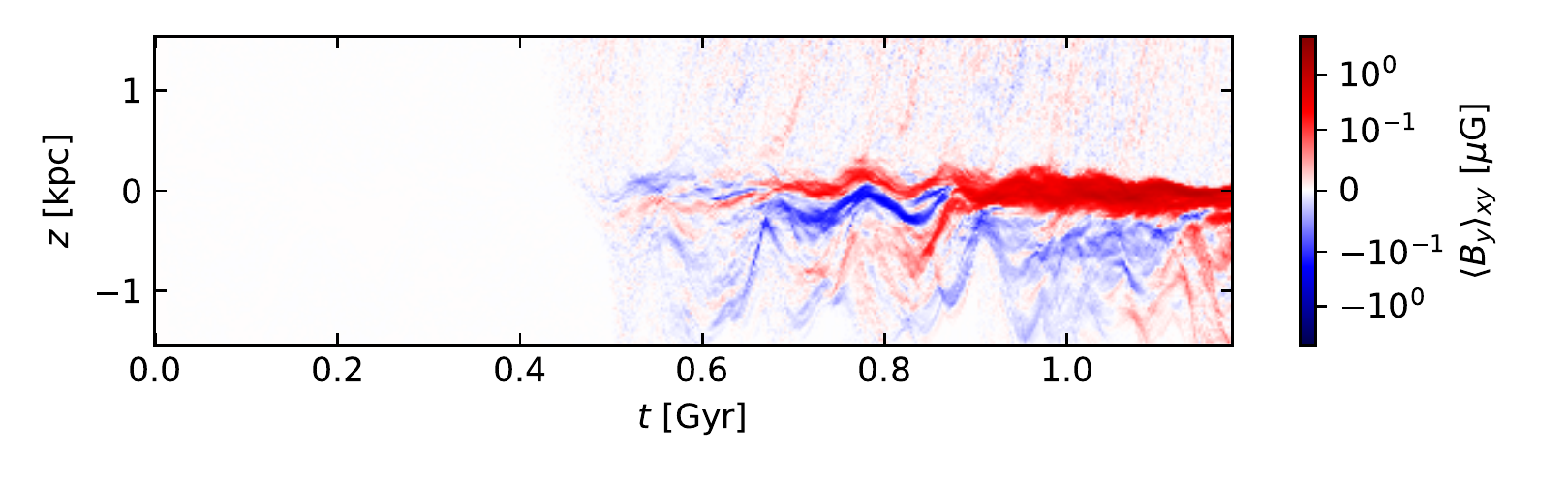}
 \begin{picture}(0,0)
    \put( -90,220){{\sf\bf{\LA}}}
    \put( -90,145){{\sf\bf{\LB}}}
    \put( -90, 75){{\sf\bf{\LB}}}
    \put(-124,220){{\sf\bf{(a)}}}
    \put(-124,145){{\sf\bf{(b)}}}
    \put(-124, 75){{\sf\bf{(c)}}}
  \end{picture}
\caption{
Time-latitude diagrams of horizontally averaged gas number density for
\emph{(a)} Model~\LA\ and \emph{(b)} \LB\ and \emph{(c)} {$\langle
B_y\rangle_{xy}$} for Model~\LB.
\label{fig:OB-av}
}
\end{figure}
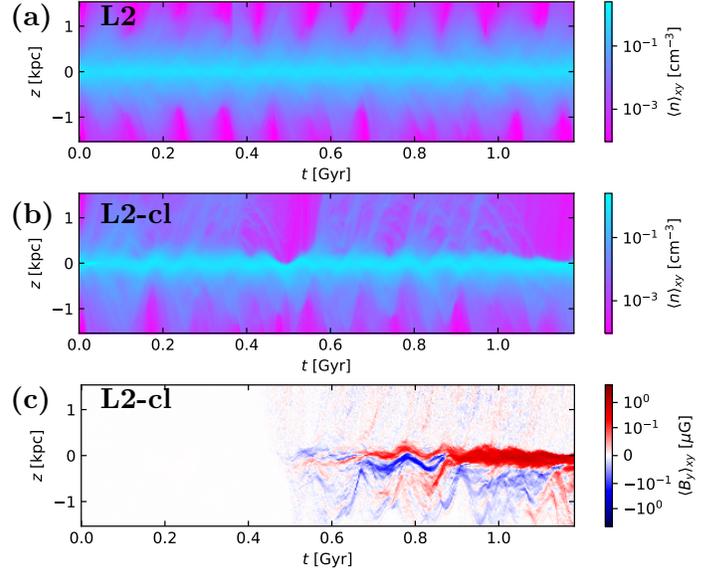

In Figure~\ref{fig:OB-av} the time-latitude diagrams of gas density are shown
for Model \LA\ without and Model \LB\ with OB clustering. The effect of
clustering is to more efficiently evacuate the gas from the lower halo. The
secondary effect on the topology of the magnetic field shown in panel
\emph{(c)} is to confine the mean field to the disc. Compare panel \emph{(c)}
to Figure~\ref{fig:tav-L2O} for Model~\LA.

\begin{figure}
\centering
\includegraphics[trim=0.5cm 0.5cm 0.2cm 0.2cm,clip=true,width=\columnwidth]{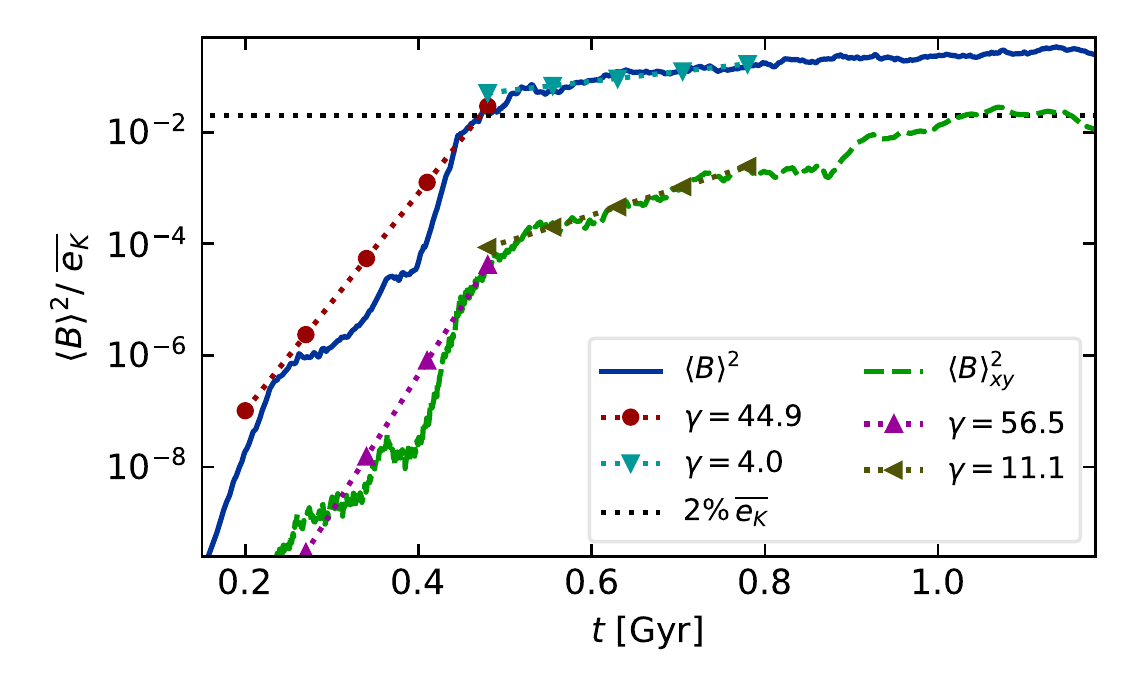}
\caption{
Model \LB: magnetic energy density normalised by time-averaged turbulent
kinetic energy $\eKt$.  Total magnetic energy $\langle B^2\rangle(8\uppi)^{-1}$
(\emph{blue}) and horizontally-averaged mean-field energy ${\langle B
\rangle_{xy}}^2(8\uppi)^{-1}$ (\emph{green}) have fits for exponential growth
rate $\gamma$ as listed in the legend, spanning $0.2\Gyr<t<0.48\Gyr$ and
$0.48\Gyr<t<0.78\Gyr$, respectively.
\label{fig:eBO}
}
\end{figure}

Figure~\ref{fig:eBO} shows SSD and LSD growth rates for Model~\LB, for
comparison with the growth rates in Figure~\ref{fig:eB-gamma} for Model~\LA.
The latter 300~Myr interval {when} shifted 0.42~Myr corresponds to the LSD
stage fitted for Model~\LA\ in Figure~\ref{fig:H2OvL2O}.

\subsection{Convergence}\label{subsec:appen-conv}

\begin{figure}[h]
\centering
\includegraphics[trim=0.32cm 0.45cm 0.59cm 0.25cm,clip=true,width=1.05\columnwidth]{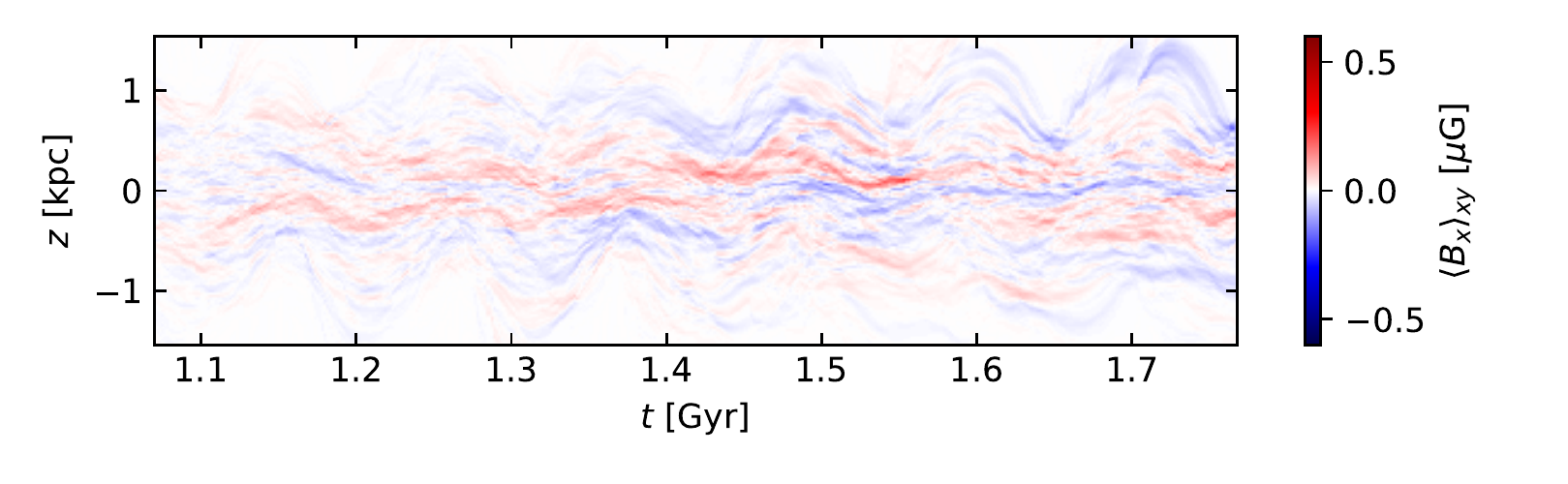}
\includegraphics[trim=0.32cm 0.45cm 0.59cm 0.25cm,clip=true,width=1.05\columnwidth]{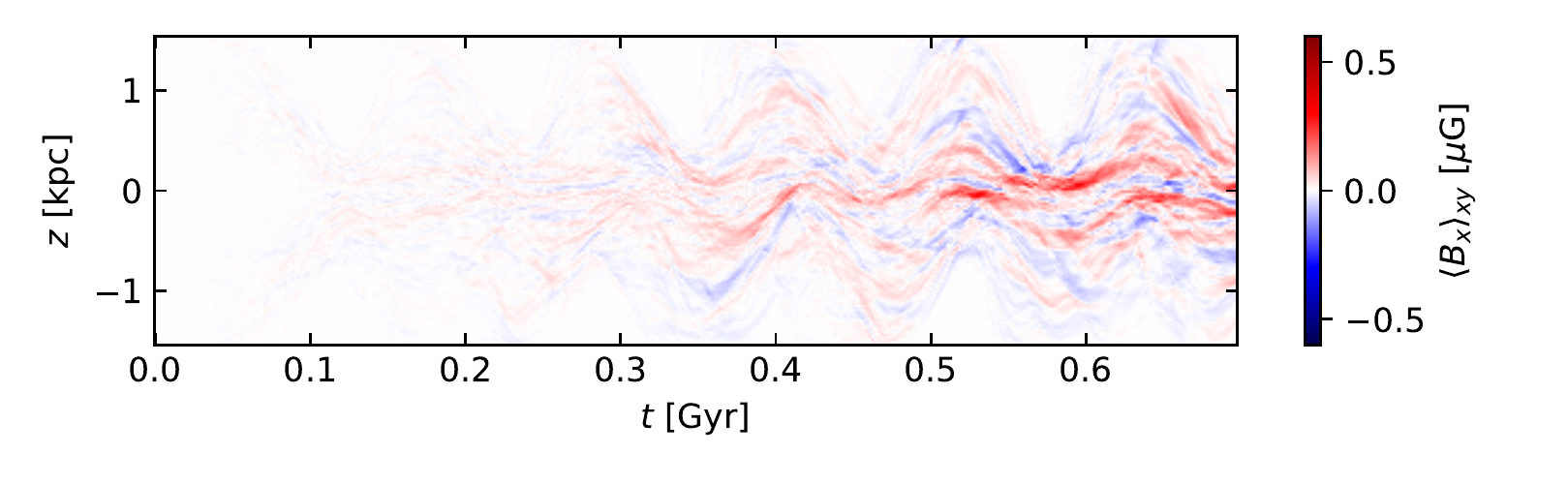}
\includegraphics[trim=0.32cm 0.45cm 0.59cm 0.25cm,clip=true,width=1.05\columnwidth]{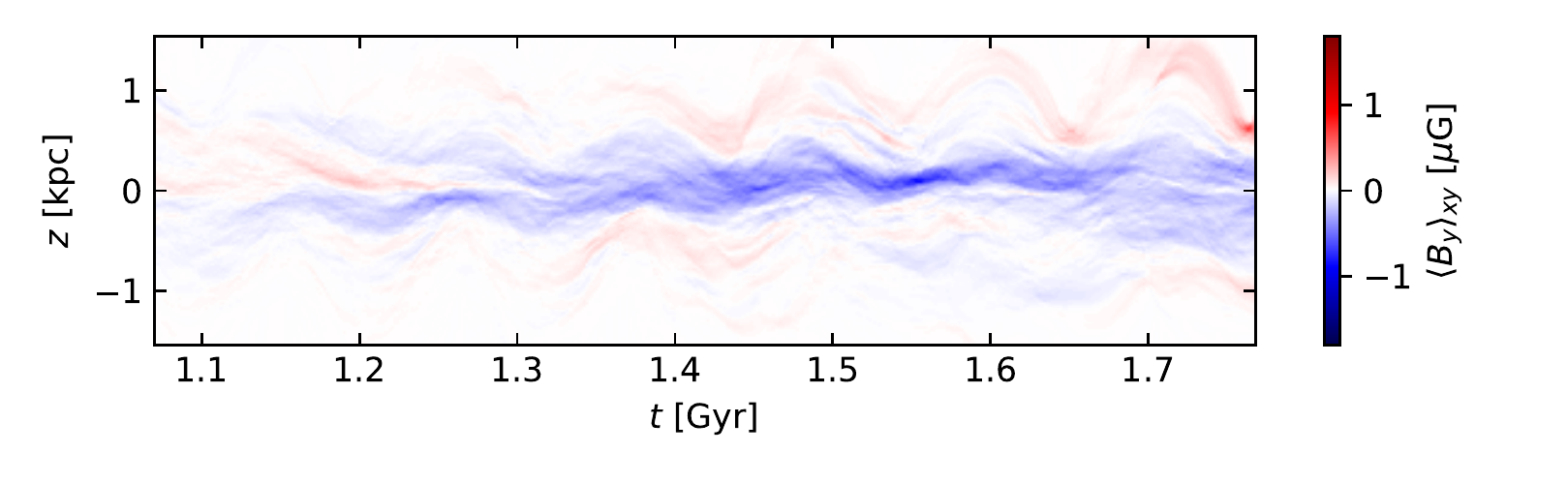}
\includegraphics[trim=0.22cm 0.45cm 0.69cm 0.25cm,clip=true,width=1.05\columnwidth]{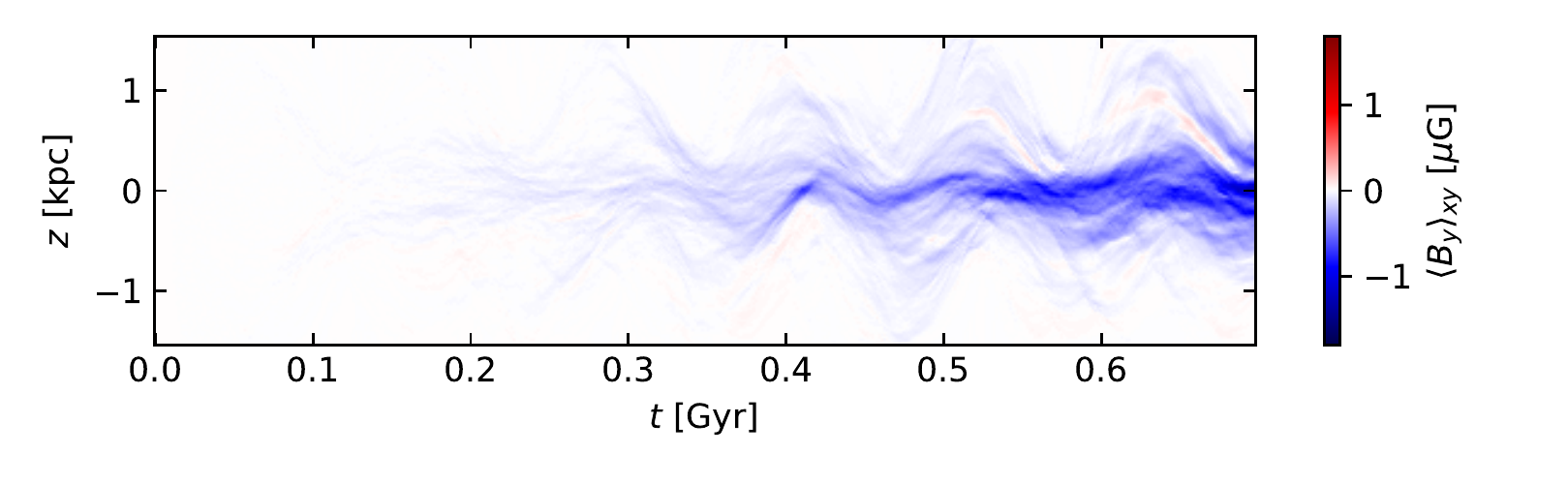}
 \begin{picture}(0,0)
    \put(-90,290){{\sf\bf{\LA}}}
    \put(-90,218){{\sf\bf{\HA}}}
    \put(-90,145){{\sf\bf{\LA}}}
    \put(-90, 73){{\sf\bf{\HA}}}
  \end{picture}
\caption{
Time-latitude diagrams of horizontally averaged $B_x$ and $B_y$ immediately
after the SSD has initially saturated for Models \LA\ and \HA, as indicated by
the labels in each panel. Note the different starting time for the two models,
chosen to capture the saturated SSD in each case.
\label{fig:H2OvL2O-av}
}
\end{figure}
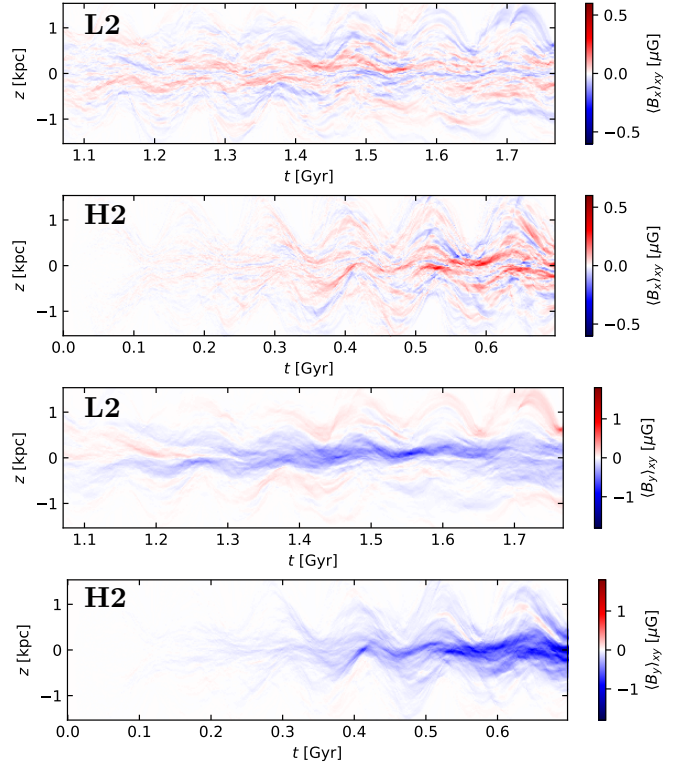

A further indication of the level of convergence of the solution (see
Sect.~\ref{sec:convergence}) is a comparison of the topology of the mean
magnetic field produced between Models~\HA\ and \LA. In
Figure~\ref{fig:H2OvL2O-av} the time-latitude diagrams for the
horizontally-averaged magnetic field components are plotted for both models.
During this stage of the LSD, the field strength, orientation, and topologies
are qualitatively similar prior to the later reversal in the sign of the
azimuthal field in Model~\LA. Model~\HA\ has not yet progressed far enough to
determine whether such a field reversal occurs at high resolution.

\begin{figure}[h]
\centering
\includegraphics[trim=0.45cm 0.5cm 0.0cm 0.2cm,clip=true,width=\columnwidth]{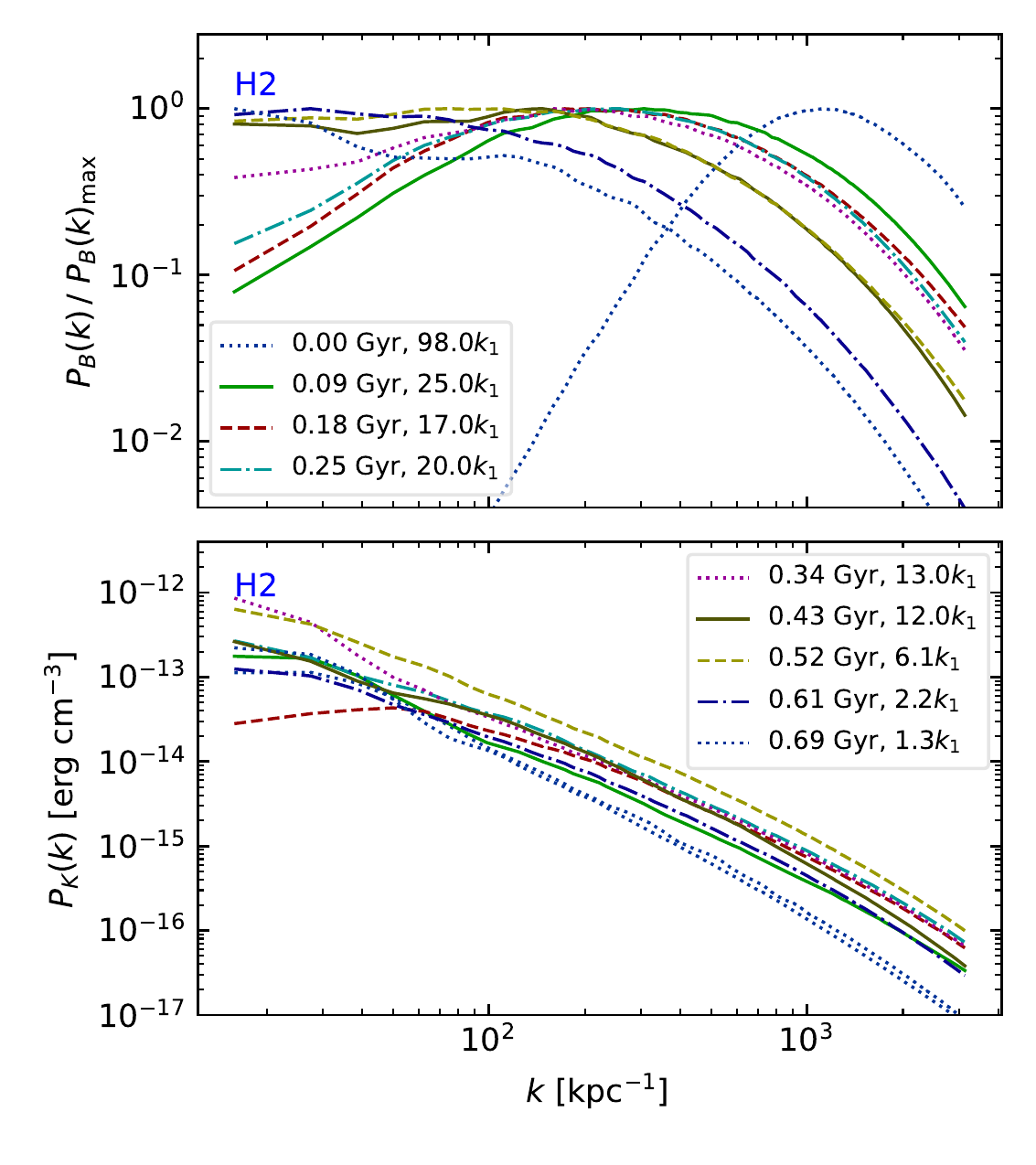}\\
\includegraphics[trim=0.40cm 0.5cm 0.2cm 0.2cm,clip=true,width=\columnwidth]{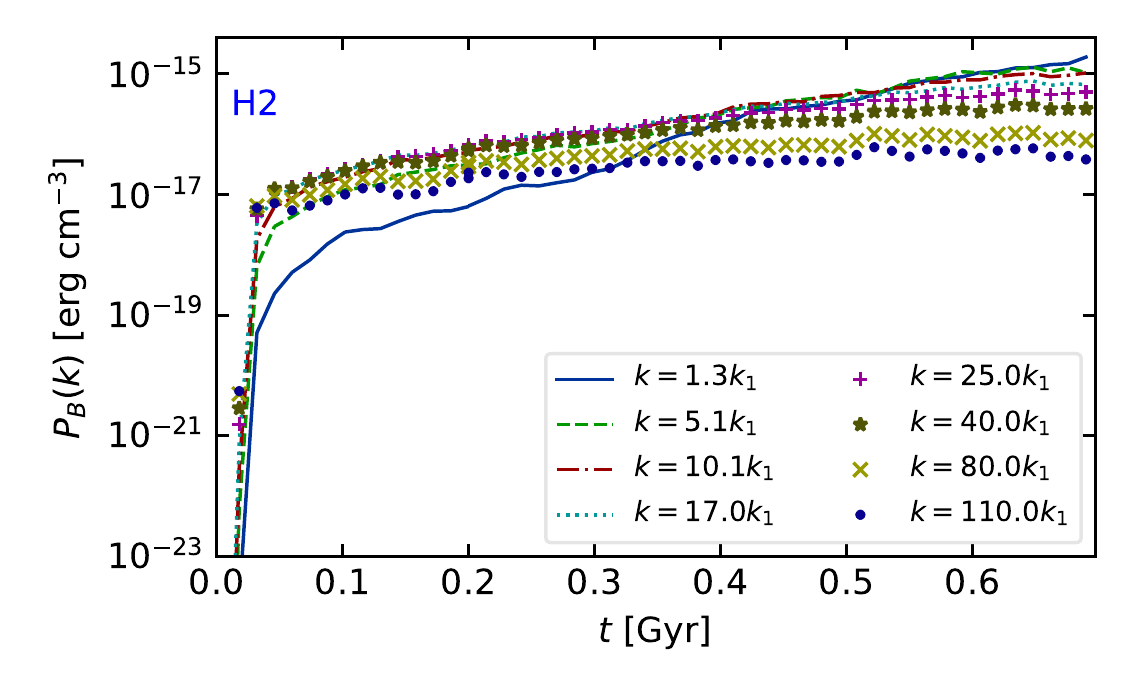}
 \begin{picture}(1,0)
    \put(-123,410){{\sf\bf{(a)}}}
    \put(-123,282){{\sf\bf{(b)}}}
    \put(-123,140){{\sf\bf{(c)}}}
  \end{picture}
\caption{
\emph{(a)} Magnetic energy spectra normalised by maximum power in each
snapshot. Wavenumber $k$ corresponding to maximal power for each snapshot are
listed in the legend.  \emph{(b)} Kinetic energy spectra.  The snapshot time in
Gyr is listed in the legend.  Both panels show the matching snapshots for
matching line style and colour.  Spectra have been smoothed with a Gaussian
kernel of length 5\,Myr and 1.5\,$k_1$ in time and wavelength.  \emph{(c)} Time
evolution of magnetic energy over sampled scales of $k$, as noted in the
legends in multiples of $k_1=2\uppi {L_x}^{-1}$.
\label{fig:H2Ospec}
}
\end{figure}
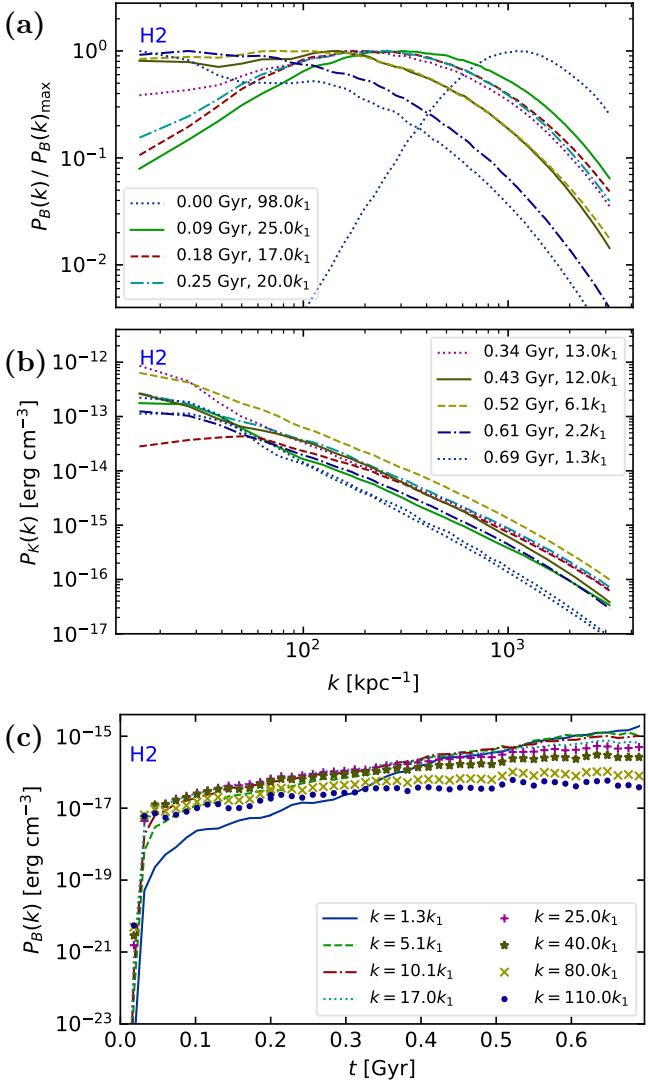

In Figure~\ref{fig:H2Ospec} from Model~\HA\ magnetic (a) and kinetic (b) energy
spectra are shown for comparison to the equivalent spectra from Model~\LA\ in
Figure~\ref{fig:power-LSD}.  The peak wavenumber for the magnetic spectra
during the kinematic stage of the SSD are at $k\simeq17k_1$ for \LA\, but vary
$16k_1\leq k \leq 26k_1$ at high resolution contributing to the greater
efficiency of the SSD.  Although both models fluctuate over time, the peak
energy for the kinetic spectra at low $k$ are almost an order of magnitude
higher for \HA\ than for \LA, which has previously been shown for SSD in
\citet{GMKS21}.  The inertial range for both models follows a consistent
$k$-scaling, although  extending much further at high resolution, as would be
expected.  Critically, lower resolution affects not only the energetics of the
small-scale turbulence, but also the systemic flows.  Nevertheless, as can be
seen from Figure~\ref{fig:H2OvL2O-av} these differences do not appear to affect
the horizontally-averaged structure of the large-scale field.

Figure~\ref{fig:H2Ospec}(c) shows how different scales $k$ within the magnetic
field grow over time for Model~\HA, to compare with that of Model~\LA\ in
Figure~\ref{fig:L2Otpower}(b).  Saturation of the SSD occurs at about
$10^{-17}\erg\cmcube$ for all $k\gtrsim17k_1$ in both models, after which the
hierarchy between these scales is conserved in the nonlinear stage of the SSD.
For $k\lesssim17k_1$ the energy continues to grow, and the hierarchy between
these scales are eventually inverted relative to their hierarchy during the
kinematic stage of the SSD.  This spectral evolution is not only qualitatively
consistent, but quantitatively compatible independent of resolution subsequent
to the kinematic stage of the SSD.

\citet{GMKS21} demonstrate that SSD solutions converge for resolution $\delta
\vect{x}\lesssim 1\pc$.  The magnetic energy spectra plotted in
Figure~\ref{fig:H2Ospec}\emph{(a)} show that the peak energy during the SSD
varies around slightly smaller scale, $17k_1\lesssim k\lesssim27k_1$ rather
than more consistently near $17k_1$.  Comparing the kinetic energy spectra
between Figure~\ref{fig:power-LSD}\emph{(b)} and
Figure~\ref{fig:H2Ospec}\emph{(b)} it is evident that tail of the inertial
range occurs $k$ just below $17k_1$ for Model \LA, but continues as far as
$25k_1$ for Model \HA.  This extends the scales at which kinetic energy can be
transferred to the magnetic field, but in addition the growth of the magnetic
field is more rapid at higher resolution at all scales up to $k=k_1$
\citep[Figure~4 of][]{GMKS21}.  The primary driver of this enhanced SSD is the
increased vorticity, lower sound speed, and higher fractional volume of the hot
gas available at higher resolution \citep{GMKS22}.

\begin{figure}[h]
\centering
\includegraphics[trim=0.38cm 0.45cm 0.61cm 0.25cm,clip=true,width=1.05\columnwidth]{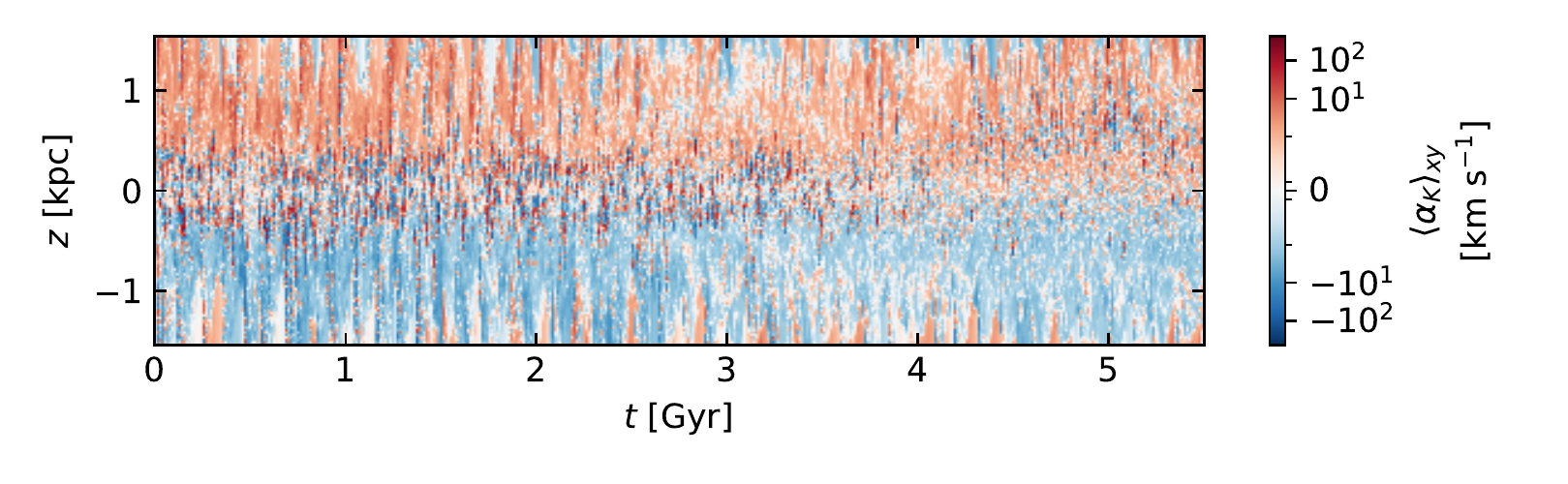}
\includegraphics[trim=0.30cm 0.45cm 0.69cm 0.25cm,clip=true,width=1.05\columnwidth]{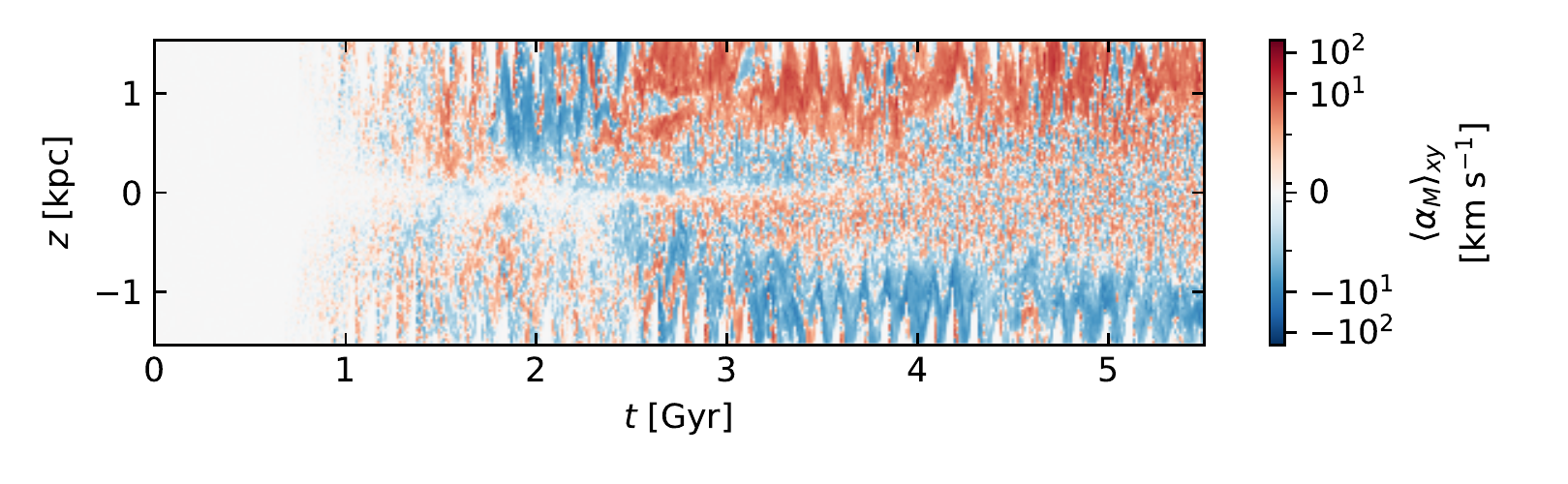}
 \begin{picture}(0,0)
    \put(-124,145){{\sf\bf{(a)}}}
    \put(-124, 73){{\sf\bf{(b)}}}
    \put( -90,145){{\sf\bf{\LA}}}
    \put( -90, 73){{\sf\bf{\LA}}}
  \end{picture}
\caption{
Time-latitude diagrams of horizontally averaged $\alpha_K$-effect \emph{(a)}
and $\alpha_M$-effect \emph{(b)} for Model~\LA.  A linear colour-scale is
applied $-5\kms<\alpha<5\kms$ and symmetric logarithmic colour-scale otherwise.
\label{fig:av-Lalpha}
}
\end{figure}
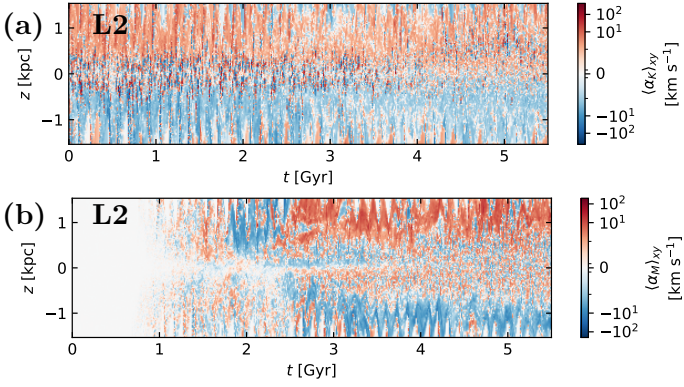

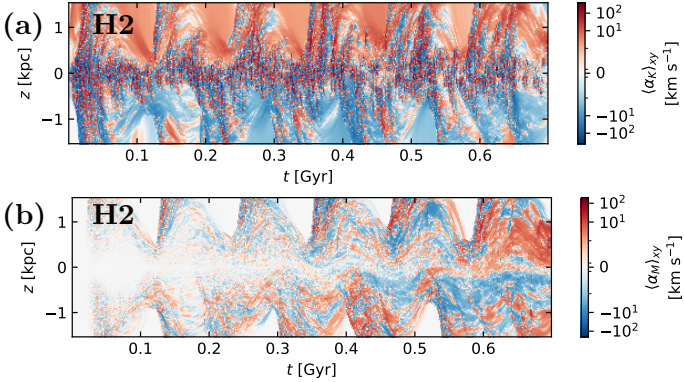
\begin{figure}[h]
\centering
\includegraphics[trim=0.38cm 0.45cm 0.61cm 0.25cm,clip=true,width=1.05\columnwidth]{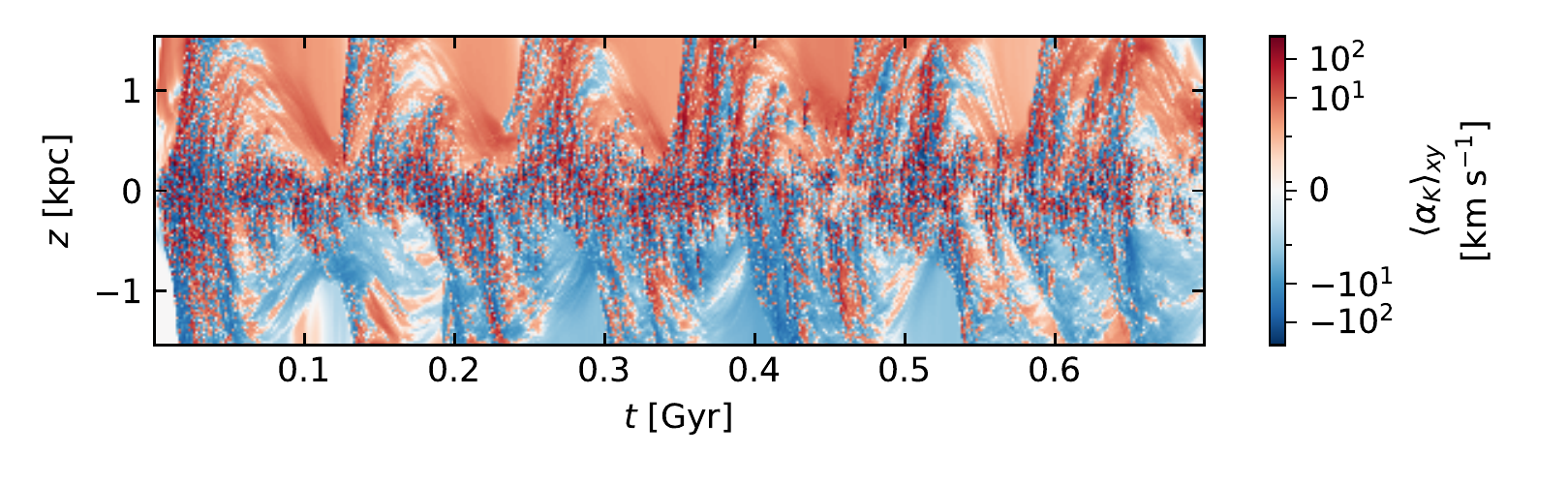}
\includegraphics[trim=0.30cm 0.45cm 0.69cm 0.25cm,clip=true,width=1.05\columnwidth]{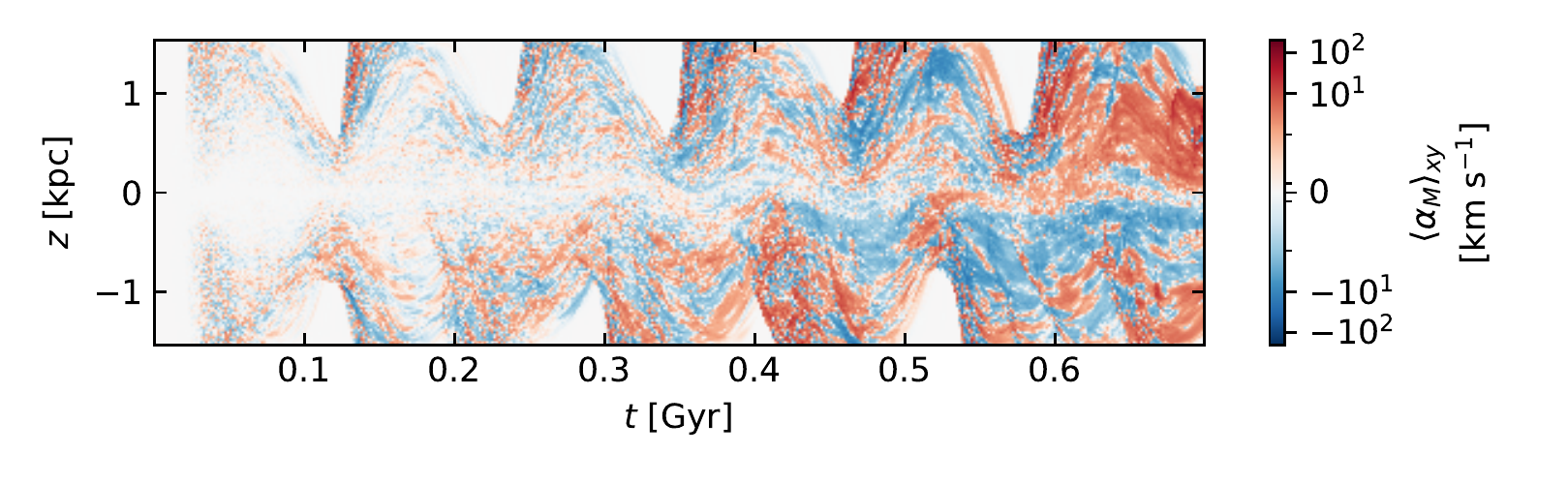}
 \begin{picture}(0,0)
    \put(-124,145){{\sf\bf{(a)}}}
    \put(-124, 73){{\sf\bf{(b)}}}
    \put( -90,145){{\sf\bf{\HA}}}
    \put( -90, 73){{\sf\bf{\HA}}}
  \end{picture}
\caption{
Time-latitude diagrams of horizontally averaged $\alpha_K$-effect \emph{(a)}
and $\alpha_M$-effect \emph{(b)} for Model~\HA.  A linear colour-scale is
applied $-5\kms<\alpha<5\kms$ and symmetric logarithmic colour-scale otherwise.
\label{fig:av-Halpha}
}
\end{figure}

\subsection{\texorpdfstring{$\alpha$}{Alpha} coefficients}\label{subsec:appen-alpha}

In Figures~\ref{fig:av-Lalpha} and \ref{fig:av-Halpha} for Models~\LA\ and \HA,
respectively, we show the separate contributions of kinetic $\alpha_K$ and
magnetic $\alpha_M$.  For Model \LA\ the systemic sign of the $\alpha_K$ either
side of the midplane is permeated by small-scale reversals, especially in the
SN active region nearest the midplane. The strength of the kinetic helicity and
the reversals dissipate over time.  The magnetic $\alpha_M$ is negligible
almost to 2 Gyr. Subsequently it has large-scale structure away from the
midplane, typically with same sign to the kinetic $\alpha_K$. These correspond
to the intersections between outflows and inflows and are dynamically weak,
occuring in diffuse gas with weak magnetic energy.

For Model~\HA\ we see that $\alpha_M$ at the midplane is initially negligible.
When it becomes dynamically important after about 400~Myr it has the same sign
as that typical of $\alpha_K$, thus enhancing the LSD. {This is remarkably
consistent with the analytic prediction by \citet{GS23} of the timescale over
which $\alpha_M$ would become dynamically effective. At this stage it has the
same sign as $\alpha_K$ and may enhance the LSD. For Model \LA\ this is delayed
to almost 2~Gyr, and soon exhibits sign opposite to $\alpha_K$.} In the
outflows we see that $\alpha_K$ {exhibits fluctuating sign on short time
and spatial scales, which may also have} a positive effect on the LSD by
supporting the removal of small-scale {kinetic} helicity {of opposite sign
to the mean field}.

\begin{table*}\caption{\label{tab:notation} Meanings of variables}
\begin{tabular}{lll}
\hline
\hline
 {Symbol}                           & {Denoting}                          & {Units/Definition}\\
\hline
$\vect{U}$                          & galactic shear velocity {about our radial centre}              & [cm s$^{-1}$] \\
$\Omega=-S$                         & angular velocity, rate of shear      & [s$^{-1}$] \\
$\vect{u}$                          &deviation of gas velocity from shear  & [cm s$^{-1}$] \\
$\vect{u}^\prime$                   & gas turbulent velocity               & [cm s$^{-1}$] \\
{$\ell^\prime$}                 & turbulent driving scale              & [cm] \\
$\rho$                              & gas density                          & [g $\cmcube$]  \\
$t$                                 & time                                 & [s] \\
$T$                                 & gas temperature                      & [K] \\
$s$                                 & specific entropy                     & [cm$^2$ s$^{-2}$ K$^{-1}$] \\
$\vect{A}$                          & magnetic vector potential            & [G cm] \\
$\vect{B}$                          & magnetic field {$\nabla\times\vect{A}$} & [G] \\
$\dfrac{\text{D}~}{\text{D}t}$      & advective (material) derivative      & \footnotesize{$\dfrac{\partial~ }{\partial t}+(\vect{U}+\vect{u})\cdot \nabla$} \\
$\zeta_{D},\zeta_{\nu},\zeta_{\chi}$& shock diffusion coefficients         & $\propto {\cal C}=-\left(\nabla\cdot\vect u|_{-\rm ve}\right)$\\
$\nu,\eta$                          & {diffusivity}
& [cm$^{2}$ s$^{-1}$]\\
$\nu_6,\chi_6,\eta_6$               & hyperdiffusion coefficients          & [cm$^{6}$ s$^{-1}$]\\
$\mu_0$                             & vacuum magnetic permeability         & 1 \\
$\cs$                               & sound speed                          & [cm s$^{-1}$]\\
$\cplocal$                          & specific heat at constant pressure   & [cm$^2$ s$^{-2}$ K$^{-1}$]\\
$\cv$                               & specific heat at constant volume     & [cm$^2$ s$^{-2}$ K$^{-1}$]\\
$\dot\sigma$                        & SN explosion rate                    & [cm$^{-2}$ s$^{-1}$]\\
$\SNr$                              & Solar neighbourhood SN rate          & [cm$^{-2}$ s$^{-1}$]\\
$\Gamma$, $\Gamma_{\rm UV}$         & far ultraviolet heating function     & [cm$^2$ s$^{-3}$]\\
$\Lambda$                           & radiative cooling                    & [cm$^5$ s$^{-3}$ g$^{-1}$]\\
$\EST$                              & SN explosion energy                  & 10$^{51}$ [erg]\\
$\ESK$                              & SN explosion momentum injection      & $\sum_{\rm Sn}\rho\Delta\vect{u}$ [g cm s$^{-1}$]\\
$h$                                 & vertical domain size                 & $\leq$3.072{~kpc}\\
$h_{\rm II}\,,h_{\rm I}$            & scale height for SN Type II, I       & 0.09, 0.325 kpc\\
$e_{ B}$                            & magnetic energy density              & [erg cm$^{-3}$]\\
$e_K$                               & total kinetic energy density         & [erg cm$^{-3}$]\\
$e_K^\prime$                        & turbulent kinetic energy density     & [erg cm$^{-3}$]\\
$\gamma$                            & volume-averaged $\eB$ growth rate    & [s$^{-1}$]\\
$k_1$                               & reference spectral wavenumber        & $\sim12.3$ kpc$^{-1}$\\
$\alpha$                            & inductive term in EMF                & [cm s$^{-1}$] \\
$\nabla$                            & gradient vector                      & e.g., \footnotesize{$\left(\dfrac{\partial~}{\partial x},\dfrac{\partial~}{\partial y},\dfrac{\partial~}{\partial z}\right)$} \\
$\nabla^{2n}$                       & 2nd or 6th order Lapacian            & e.g., \footnotesize{$\dfrac{\partial^{2n}~}{\partial x^{2n}}+\dfrac{\partial^{2n}~}{\partial y^{2n}}+\dfrac{\partial^{2n}~}{\partial z^{2n}}$}, $n=1,3$ \\
$\mathbfss W$                       & traceless rate of strain tensor      & \footnotesize{${\mathsf W}_{ij} = \dfrac{1}{2}\left(\dfrac{\partial u_i}{\partial x_j}+\dfrac{\partial u_j}{\partial x_i}-\dfrac{2}{3} \delta_{ij}\nabla\cdot \vect u\right)$} \\
$|\mathbfss W|^2$                   & contraction of $\mathbfss W$         &    $|\mathbfss W|^2={\mathsf W}_{ij}{\mathsf W}_{ij}$\\
$\mathbfss W^{(5)}$                 & 5th order rate of strain tensor      &    \footnotesize{${\mathsf W}_{ij}^{(5)} = \dfrac{1}{2}\left(\dfrac{\partial^5 u_j}{\partial x_i^5}+\dfrac{\partial^4}{\partial x_i^4}\left(\dfrac{\partial u_i}{\partial x_j}\right)-\dfrac{1}{3}\dfrac{\partial^4}{\partial x_i^4}\left(\nabla\cdot \vect u\right)\right)$} \\
{$\epsilon_{ijk}$}              & Levi-Cevita symbol, $i,j,k\in(1,2,3)$& $+(-)1$ if permutation of 1,2,3 is even(odd) or 0\\&&if an index repeats.  \\[0.4cm]
\hline
\end{tabular}
\end{table*}

%
\section{Code and parameter evolution during run time}\label{sec:code}

\begin{table*}
\caption{ \label{tab:params}
Tracking parameter changes over duration of simulations: For the models listed
the parameters applied commencing at time $t$ are shown:  $h_{\rm adj}$
indicates the scale height of the SN vertical distribution varies with the
scale height of the gas; $\Gamma_{\rm UV}$ is the photoelectric heating factor
relative to the the Habing rate $\Gamma_0$. $T_{\rm max}$ is the maximum
temperature allowed within $r_0${, the SN radial scale as explained in
Section~\ref{sec:SN}} of the SN explosion, and $T_{\rm ratio}\times T_{\rm
max}$ is the maximum allowed within 2.25$r_0$ of the explosion. $n_{\rm cool}$
indicates whether cooling mass is used to reduce the temperature, otherwise the
SN site is rejected and an alternative selected.  $n_{\rm max}$ is the maximum
gas number density at the grid location of an explosion, to avoid locations
where SN evolution is not well resolved at $\dx=4\pc$ $n_{\rm ratio}$ is the
maximum ratio between maximum and minimum gas density within $r_0$ of an
explosion, to avoid excessive viscous forces.
}
\centering
{\footnotesize{
\begin{tabular}{lllllllllll}
\hline\hline
Model             &$t$   & $h_{\rm adj}$&$\Gamma_{\rm UV}$&$T_{\rm max}$&$n_{\rm max}$&$n_{\rm ratio}$                 &$T_{\rm ratio}$       & $E_{m}$  & $n_{\rm cool}$ & $\nu_6\,\chi_6\,\eta_6$\\
                  &[Gyr] &              &[$\Gamma_0$]     &[$10^6$ K]   &[cm$^{-3}$]  &[$n_{\rm max}{n_{\rm min}}^{-1}$]&[${T_{\rm max}}^{-1}$]&          &                &[kpc$^{-5}$ km s$^{-1}$]\\ \hline
\HA               &0     &F             & 1               & 25          & 50          &2e4                             & 8e2                  &  T, 7.5\%&  F             &  1e-15                 \\
\HA               &0.02  &F             & 1               & 25          & 150         &3e4                             & 9e2                  &  T, 7.5\%&  F             &  1e-15                 \\
\HA               &0.028 &T             & 2.5             & 25          & 150         &3e4                             & 9e2                  &  T, 7.5\%&  F             &  1e-15                 \\
\HA               &0.037 &T             & 3.5             & 25          & 100         &1e4                             & 7                    &  T, 7.5\%&  T             &  1e-15                 \\
\HA               &0.465 &T             & 3.5             & 25          & 100         &1e4                             & 7                    &  T, 7.5\%&  T             &  1.1e-15               \\
\HA               &0.652 &T             & 3.5             & 7.5         & 100         &1e4                             & 12                   &  F       &  F             &  1.2e-15               \\
\HA               &0.672 &T             & 3.5             & 50          & 100         &1e4                             & 10                   &  F       &  F             &  1.2e-15               \\
\LA, \LB, \\
\LC, \LD          &0     &F             & 3.5             & 12.5        & 50          &1e4                             & 7                    &  T, 7.5\%&  T             &  5e-12                 \\
\LA               &0.365 &F             & 3.5             & 12.5        & 50          &1e4                             & 7                    &  T, 7.5\%&  F             &  5e-12                 \\
\LA               &0.461 &F             & 3.5             & 12.5        & 50          &1e4                             & 7                    &  T, 7.5\%&  T             &  5e-12                 \\
\LA               &3.272 &T             & 3.5             & 25          & 100         &1e4                             & 7                    &  T, 7.5\%&  T             &  5.2e-12               \\
\LA               &3.435 &T             & 3.5             & 25          & 100         &1e4                             & 7                    &  T, 7.5\%&  T             &  5.25e-12              \\
\LB               &0.1   &T             & 3.5             & 25          & 50          &1e4                             & 4                    &  T, 7.5\%&  T             &  5e-12                 \\
\LB               &0.128 &T             & 3.5             & 20          & 50          &1e4                             & 5                    &  T, 7.5\%&  T             &  5e-12                 \\
\LB               &0.349 &T             & 3.5             & 25          & 50          &1e4                             & 4                    &  T, 7.5\%&  T             &  5e-12                 \\
\LB               &0.375 &T             & 3.5             & 12.5        & 50          &1e4                             & 7                    &  T, 7.5\%&  T             &  5e-12                 \\\hline
\end{tabular}
}}
\end{table*}

A summary of the symbols used in the manuscript are listed in
Table~\ref{tab:notation}.  It is in the nature of such simulations, which
require long integration time, spanning several months and longer in real time,
that new code adaptations to stability issues or improvements to algorithms are
discovered and applied while the simulations are progressing. Such is the
expense in terms of computation and storage that it is not feasible to restart
the simulations with the new implementation and repeat through to saturation of
the LSD. This is particularly the case for Model \HA, which has used tens of
millions of computational billing units.

Prior to the suite of simulations presented in this study, we explored a
variety of lower resolution runs with larger and smaller domains, and a variety
of far ultraviolet heating rates and vertical distributions of SNe, so that the
parameters chosen during this suite of experiments has been fairly robust and
consistent. Parameter adjustments are listed for each Model in
Table~\ref{tab:params}. Nevertheless, some updates, as detailed in
Table~\ref{tab:pcode}, needed to be applied during the course of the
experiment. When these updates have been committed to address our specific
issues, we have also acquired any updates applied by other developers between
our commits, which are not particular to SN-driven turbulence, but nevertheless
may affect code performance.

\begin{table*}
\caption{
Tracking code version over duration of simulations: date, repository hash code
at {\tt github.com/pencil-code}, model, and final column summary of update.
The start time in Gyr when each version applies is listed for each model.\\
$^\dagger$: Model \LA\ originally initiated with $\dot{\sigma}=\SNr$, inducing
a higher SSD growth rate. Beyond 0.3~Gyr continued with $\dot{\sigma}=0.5\SNr$.
The period 0--0.365~Gyr runs retrospectively with higher initial field strength
to replace the original data.\\ **: Update directly affects the included
models. The cooling ejecta is absent only for a small span of three models:
\HA\ $t<37$ Myr; \LA\ $365<t<461$ Myr and \LB\ $t<208$ Myr, and will have
little effect on the properties of the turbulence.\\ General changes between
versions could affect the timestep or precise realisations of the inherently
chaotic calculations.  The effect does not alter the statistical properties of
the simulations.
\label{tab:pcode}
}
{\footnotesize{
\begin{tabular}{llllllll}
\hline\hline
Date      & hash & \HA\ & \LA\ & \LB\ & \LC\ & \LD\ & \\\hline
2021-01-22& 8b05b3884      &0     &      &      &      &      &Machine specific \\
2021-04-13& 163e3e15d      &0.025 &      &0     &      &      &Citations update \\
2021-10-14& f7bb7be7a      &      &0$^\dagger$& &      &      &Test solutions \\
2021-06-01& 92794628c      &0.033 &0.365 &0.1   &      &      &Complex numbers\\
2021-06-03& b482858d2      &      &0.461 &0.208 &      &      &Added cooling ejecta** \\
2021-06-04& 533ad9dad      &      &0.584 &0.334 &      &      &Add no mass in dense gas\\
2021-06-15& 9426d1a5f      &      &1.017 &0.513 &      &      & \\
2021-06-02& f7bb7be7a      &0.037 &1.237 &      &0     &0     &Test fields initialized\\
2022-10-18& 6b8c45449      &0.192 &      &      &      &      &Streamlined SN location\\
2022-10-19& 524ed6ca2      &0.196 &      &      &      &      &Omit redundant MPI call\\
2022-11-08& 81cc4c1d0      &0.205 &      &      &      &      & \\
2022-12-12& 2b0563676      &0.216 &      &      &      &      &Updated SN list length\\
2023-02-15& c4b34885f      &0.368 &      &      &      &      &Use log $T$ if calculated\\
2023-04-11& 86e11c9c6      &0.651 &      &      &      &      & Added diagnostics\\\hline
end time  &                &0.672 &5.500 &1.181 &2.195 &2.008 & \\\hline
\end{tabular}
}}
\end{table*}

Here, we seek to explain what issues have arisen in our experiments, how these
have been addressed and apply to each model, and list which version of the code
has been used during each experiment.  The primary causes of code instability
arise from the location of specific SN remnants, which are difficult to stably
resolve. This is most likely when applied to a high density region, in which
momentum injection is required, but may also include neigbourhoods of low
density and therefore high temperatures and high viscous stresses. Usually
restarting from a recent snapshot with an updated random seed for SN locations
avoids this location and advances the simulation with a statistically
equivalent realisation. High density sites that do not cause stability crises
continue to be included.  This is more common at low resolution. At high
resolution it is easier to resolve more challenging SN remnant structures,
although occasionally a seed reset is needed. This does not substantially
affect the comparability of the simulations over time or between models.

One update to the code that was included during the experiments is to permit
the use of additional SN mass ejecta, where the temperature exceeds a limit of
75 MK to cool the site to less than this temperature and therefore maintain a
longer timestep.  Prior to this change these locations would be avoided and an
alternate site without such high temperatures arising randomly selected, so
that remnants in diffuse locations may have been less frequent. This change
should have a minor effect on the statistics of the dynamo, as these locations
are in any case selected rarely and higher temperatures continue to arise, when
remnants expand into adjacent diffuse ISM, rapidly heating the gas outside the
initial SN remnant radius.

Another occasional cause of code instability is too low hyperdiffusivity
coefficients. The values for $\nu_6$, $\chi_6$ and $\eta_6$ reported in
Section~\ref{sec:shocks} were adopted following experiments in periodic
unstratified simulations, where they adequately resolved grid-scale
instabilities during the linear and nonlinear stages of the SSD. As our
simulations approached the nonlinear stage of the LSD, we discovered some
grid-scale instabilities that could not be attributed to remnant location, as
they occurred far away from any explosions. These were solved by increasing the
hyperdiffusivity coefficients modestly. It is unlikely that these affect the
statistics of the dynamo, because they still apply only at the very largest
wavenumbers in the model, where the energies are very weak compared to the
energies at even the tail end of our inertial range, let alone those in the LSD
range of wavenumbers, as is evident in the later spectra of Model \LA, which
are displayed in Figure~\ref{fig:power-LSD} and Model \HA, in
Figure~\ref{fig:H2Ospec}(a). In Model \HA\ these coefficients were increased to
$1.2\times10^{-15}$ after 652~Myr.  In this study, $5\times10^{-12}$ was
applied throughout for $\dxx=4\pc$, except as listed in Table~\ref{tab:params}
at late times for Model \LA.

\end{document}